\documentclass[prd,preprintnumbers,nofootinbib,superscriptaddress,onecolumn]{revtex4}

\pdfoutput=1
\usepackage{graphicx}
\usepackage{amsmath}
\usepackage{amssymb}
\usepackage{slashed}
\usepackage{mathtools}
\usepackage[utf8]{inputenc}
\usepackage[colorlinks=true,linkcolor=blue,bookmarksopen,bookmarksnumbered]{hyperref}
\usepackage{color}
\usepackage[usenames,dvipsnames]{xcolor}
\usepackage[normalem]{ulem}
\usepackage{hhline,multirow,tabularx}
\usepackage{epstopdf}

\usepackage{physics}
\usepackage{caption}
\usepackage{subcaption}
\usepackage{qcd}

\begin{document}

\title{Basics of factorization in a scalar Yukawa field theory}

\preprint{JLAB-THY-22-3760}

\author{F.~Aslan}
\email{fpaslan@jlab.org}
 \affiliation{Department of Physics, University of Connecticut, Storrs, CT 06269, U.S.A.}
 \affiliation{Jefferson Lab, 12000 Jefferson Avenue, Newport News, VA 23606, USA}
\author{L.~Gamberg}
\email{lpg10@psu.edu}
\affiliation{Division of Science, Penn State Berks, Reading, Pennsylvania 19610, USA}
\author{J.~O.~Gonzalez-Hernandez}
\email{joseosvaldo.gonzalezhernandez@unito.it}
\affiliation{Dipartimento di Fisica, Universit\`a degli Studi di Torino, Via P. Giuria 1, I-10125, Torino, Italy}
\affiliation{INFN, Sezione di Torino, Via P. Giuria 1, Torino, I-10125, Italy}
%
%
\author{T.~Rainaldi}
\email{train005@odu.edu}
\affiliation{Department of Physics, Old Dominion University, Norfolk, VA 23529, USA}
\author{T.~C.~Rogers}
\email{tedconantrogers@gmail.com}
\affiliation{Department of Physics, Old Dominion University, Norfolk, VA 23529, USA}
\affiliation{Jefferson Lab, 12000 Jefferson Avenue, Newport News, VA 23606, USA}

\date{December 1, 2022}


\begin{abstract}
The factorization theorems of quantum chromodynamics (QCD) apply equally well to most simple quantum field theories that require renormalization but where direct calculations are much more straightforward. Working with these simpler theories is convenient for stress-testing the limits of the factorization program and for examining general properties of the parton density functions (pdfs) or other correlation functions that might be necessary for a factorized description of a process. With this view in mind, we review the steps of factorization in a real scalar Yukawa field theory for both deep inelastic scattering (DIS) and semi-inclusive deep inelastic scattering (SIDIS) cross sections. In the case of SIDIS, we illustrate how to separate the small transverse momentum region, where transverse momentum dependent (TMD) pdfs are needed, from a purely collinear large transverse momentum region, and we examine the influence of subleading power corrections. We also review the steps for formulating TMD factorization in transverse coordinate space, and we study the effect of transforming to the well-known $\bstarsc$-scheme.  Within the Yukawa theory, we investigate the consequences of switching to a generalized parton model (GPM) approach, and compare with a fully factorized approach. Our results highlight the need to address similar or analogous issues in QCD.  
\end{abstract}

\maketitle

\section{Introduction}
\label{s.intro}

It is often the goal of hadronic scattering experiments to gain an increased understanding of the intrinsic properties of the scattered hadrons. To this end, the parton model~\cite{Feynman:1972}, wherein hadrons are viewed as collections of nearly free 
point-like constituents, is indispensable as a framework for constructing models of intrinsic structure and relating them to high-energy scattering observables. 
However, in theories that require renormalization, the operators that count the number of elementary particles in a target are beset by divergences, and the steps for dealing with them sometimes require modifications of the intuitive expectations that arise from a purely parton model framework. A recently discussed example concerns the question of whether parton density functions must be strictly positive definite~\cite{Candido:2020yat}; a literal probability density interpretation would imply strict positivity, but it turns out that the pdfs in a typical renormalizable quantum field theory can violate positivity, depending on the choice of renormalization scheme and the scales~\cite{Collins:2021vke}. While issues such as these are naturally relevant when interpreting  measurements in terms of partonic constituents, they also have important practical and phenomenological consequences. For example, with regard to the positivity question, it is important to know if strict positivity should be imposed directly on fit parametrizations. This is relevant to, among other things, recent debates about the evidence for an intrinsic charm component in the proton pdf~\cite{Ball:2022qks,Courtoy:2022ocu,Guzzi:2022rca}. Indeed, charm quark pdf extractions do appear to require that negative pdfs be allowed (see, for example, Figure~1 of \cite{Ball:2022qks}).  
Other examples of the role of positivity-related constraints can be found in Refs.~\cite{DAlesio:2020vtw,Aamodt:2010pp,Gamberg:2022kdb}.

In quantum field theory (QFT), the parton model gets placed on firmer footing through the factorization theorems~\cite{Collins:1987pm,Collins:1989gx,Collins:2011qcdbook}. 
Nowadays, however, generalizations of a basic partonic picture play a role in scenarios far beyond the original leading power descriptions of inclusive processes like inclusive deep inelastic scattering~\cite{Liu:2019ntk,AbdulKhalek:2021gbh}. They appear, for example, in the factorization theorems for semi-inclusive processes where TMD pdfs and fragmentation functions (ffs) are important. The parton model also accounts for spin dependent effects, and there are partonic descriptions of higher-twist behavior~\cite{Efremov:1981sh,Qiu:1991pp,Jaffe:1991ra,Ji:1993ey,Tangerman:1994bb,Mulders:1995dh,Bacchetta:2006tn,Qiu:2020oqr}. In all such cases, it is important to stress-test the limits of any assumptions that are rooted in a parton model picture. Confinement, non-Abelian gauge invariance, and the general complexity of strongly coupled nonperturbative quantum field theory makes this difficult in real QCD. However, many of the steps in standard factorization derivations are not specific to QCD, but instead apply rather generally to most of the simpler relativistic renormalizable quantum field theories found in the introductory chapters of textbooks. By retracing the steps of factorization in those theories, where contributions from all spacetime scales can be handled perturbatively, it becomes straightforward to confirm the most basic consequences of factorization while also probing their limits. Indeed, it is possible to find departures from parton model expectations arising from the need for renormalization alone. 
A recently discussed example is the use of a scalar Yukawa theory~\cite{Collins:2021vke}, already mentioned above, to illustrate the possibility of positivity violations in pdfs defined with the $\msbar$ renormalization scheme. Simple field theories are useful more generally for stress-testing other aspects of factorization and related assertions regarding pdfs and notions of intrinsicness in the presence of renormalization

This motivates us in the present paper, to further explore the factorization of simple QFTs in DIS. We will present calculations of the cross sections for both DIS and SIDIS in a scalar Yukawa model theory with non-zero masses for all fields first \emph{without} factorization. By keeping the coupling small, we ensure that all parts of these calculation can be handled simply with low order Feynman diagrams. Next, we will retrace the basic steps involved in factorizing the graphs in the large-$Q$, fixed Bjorken-$x$ deep inelastic limit. It then becomes possible to compare the unfactorized, unapproximated results with standard collinear and TMD factorization treatments. Sensitivity to the mass scales in the Lagrangian serves as a measure of sensitivity to intrinsic large distance dynamics, analogous to the sensitivity to confinement scale physics in QCD.  We will compare factorized and unfactorized versions of the same calculations and note how the sizes of the differences between them can provide guidance on questions relevant to implementations of both TMD and collinear factorization at moderate hard scales. 

Some of the questions to be addressed are: 
\begin{itemize}
\item 
What are typical sizes of subleading powers (or higher twist) at moderate $Q$ and how important are they for maintaining reasonable agreement with the unfactorized cross section?
\item What are typical consequences of switching between different precise definitions for objects like pdfs? For example, what is the effect of using a collinear pdf defined as the cutoff integral of a TMD pdf as opposed to the usual renormalized definition for the pdf?
\item What are the relative sizes of contributions from large and small transverse momentum in TMD parton densities, and how important are the large-$\Tsc{q}{}$ corrections to TMD factorization in transverse momentum dependent cross sections? 
\item In TMD factorization, it is standard to transform from transverse momentum space to coordinate $\T{b}{}$  space. Then, one identifies small-$\T{b}{}$ contributions with collinear factorization contributions and the large-$\T{b}{}$ contributions with the intrinsic or nonperturbative properties of hadrons.  In standard approaches to TMD factorization, the large-$\T{b}{}$ is sequestered in the exponential of functions labeled $g$. What is the typical impact of implementing this separation, and how sensitive are results to the choice of large-$\T{b}{}$ modeling? 
\end{itemize}

We will begin in~\sref{crosssecionts} by reviewing basic DIS and SIDIS kinematics for the general case and by describing our notation and conventions. In \sref{yukawa}, we will define the specific version of the scalar Yukawa theory that we will use throughout the rest of this paper. In~\sref{pdfs}, we will review the operator definitions of pdfs and TMD pdfs, discuss their basic properties, and show how they are calculated in the Yukawa theory.  In \sref{collinearsteps}, we will step through the basic procedure for factorizing the inclusive DIS structure functions in collinear factorization into a hard part and a pdf, and in \sref{TMDandSIDIS} we do the same for SIDIS with TMD factorization. In both cases, we will compare factorized and unfactorized calculations of the same quantities. We will also discuss the steps for recovering collinear factorization by integrating TMD factorized expressions over all transvere momentum, and we will compare with an approach that only uses TMD pdfs (the GPM). In \sref{input}, we will comment on the lower boundary in $Q$ where agreement between factorized and unfactorized expressions begins to break down. In \sref{coord}, we will convert the TMD factorization treatment into transverse coordinate space, and consider the effect of switching to the $\bstarsc$ method for isolating large and small transverse coordinate contributions. We will summarize our results in~\sref{conclusion}, where we will also comment on the limitations and risks of comparing them with real QCD. 

\section{Cross Sections and Structure Functions}
\label{s.crosssecionts}

Before turning to DIS for the specific case of the Yukawa theory, we review the notation and conventions of DIS cross sections and structure functions in the general case in this section. In later sections, we will work with both DIS and its extension to SIDIS. In both cases, a proton (or, more generally, any hadron) moves in the $+z$ direction with four momentum $p$. Except where specified, it is to be assumed that we are working in the Breit frame, where the photon four-momentum is $q = \parz{0,0,0,-Q}$ (see \fref{dissidis} (a)). In the SIDIS case, the observed final state hadron carries momentum $P_B$ (see \fref{dissidis} (b)).
The kinematical variables are mostly standard:
\begin{align}
q& = (l - l')  , &q^2& = -Q^2  , &\xbj& = \frac{Q^2}{2 p \cdot q} , \no
y& = \frac{p \cdot q}{p \cdot l} \, , &\zh& = \frac{P_B \cdot p}{p \cdot q} , &\xn& = -\frac{q^+}{p^+}  ,  \no
\zn& =\frac{P_B^-}{q^-}\, ,
\end{align}
where $l$ and $l'$ are the initial and recoil leptons. The light cone ratios $\xn$ (Nachtmann-$x$) and $\zn$ are expressed in terms of $q^\pm$, $p^+$, and $P_B^-$; for fragmentation, the light-cone ratio $\zn$ is the analogue of $\xn$, and 
in the massless limit they equal $z_h$ and Bjorken $\xbj$, respectively.  
 Our conventions for the light cone variables for a four vector $V$ are defined by 
\begin{equation}
V^{\mu} = \parz{V^+, V^-, \T{V}{}} \, , 
\end{equation}
where 
\begin{equation}
V^+ = \frac{V^0 + V^z}{\sqrt{2}}, \qquad V^- 
	= \frac{V^0 - V^z}{\sqrt{2}}, \qquad  \T{V}{} = \parz{V^x, V^y}\, .
\end{equation}
\begin{figure}
\centering
  \begin{tabular}{c@{\hspace*{.008mm}}c@{\hspace*{.008mm}}}
    \includegraphics[scale=0.3]{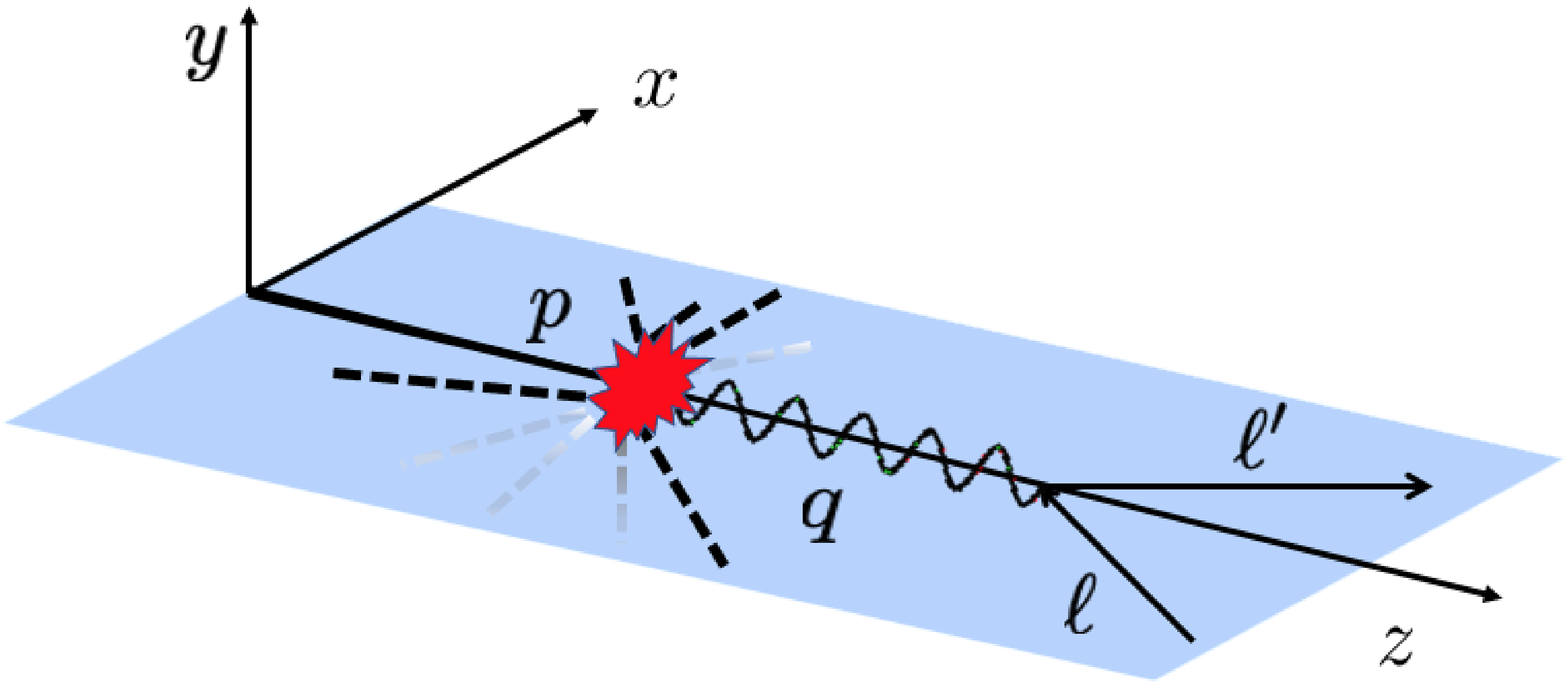}
    &
    \includegraphics[scale=0.3]{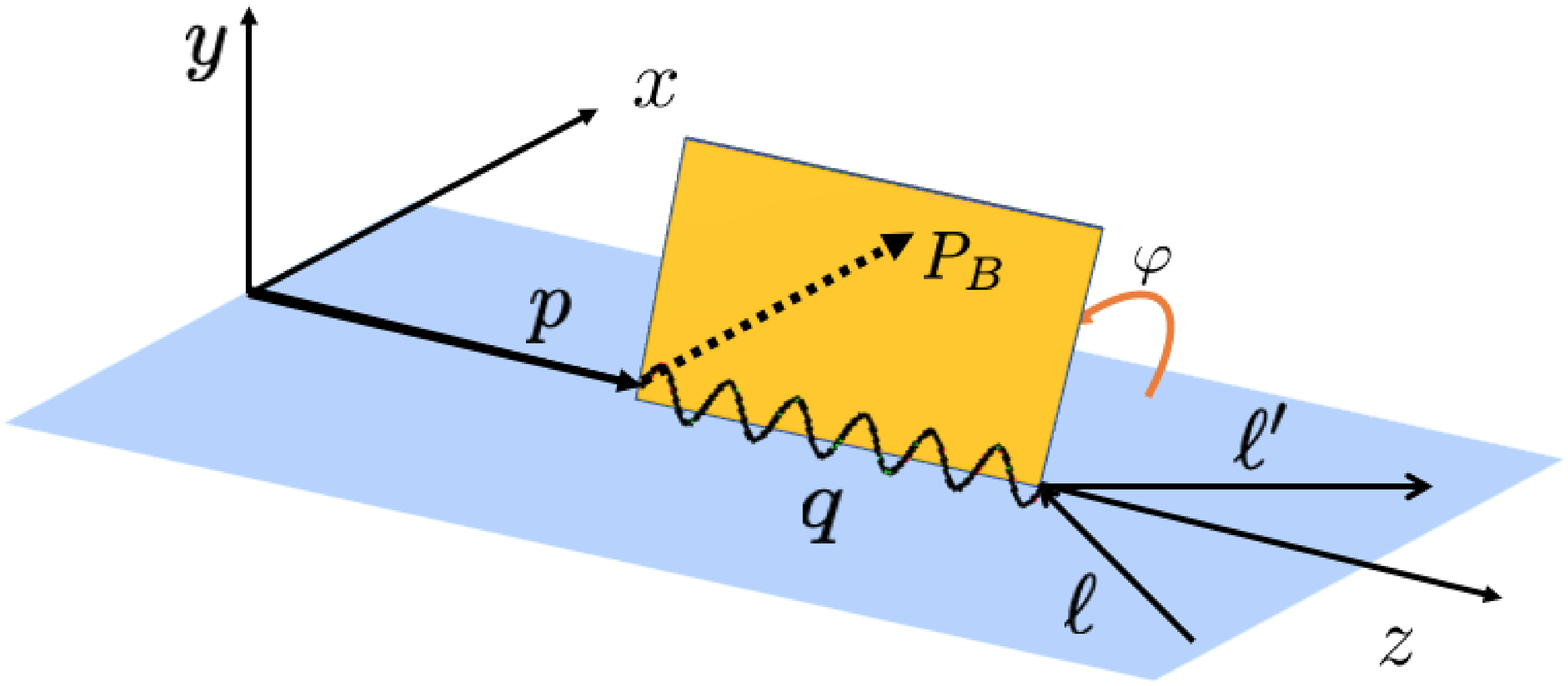}
  \\
  (a) & (b) 
  \end{tabular}
  \caption{An illustration of the kinematic configuration of  DIS (a) and SIDIS (b) events in  the Breit frame (photon frame). The incoming ($\ell$) and outgoing ($\ell^\prime$) lepton momenta form the lepton plane (shown in blue) (a) The dashed lines represent the unobserved DIS particles. (b) $P_B$ is the momentum of the produced hadron (B). The azimuthal angle of the  hadron plane (shown in yellow) is measured counterclockwise with respect to the lepton plane. 
 }
\label{f.dissidis}
\end{figure}
Our conventions for separating cross sections into structure functions match those of Ref.~\cite{Boglione:2019nwk}, which mostly follow typical DIS and SIDIS notation (see also Ref.~\cite{Bacchetta:2006tn}). The differential cross section for DIS is
\begin{equation}
    \frac{\diff{\sigma_{\rm DIS}}{}}{\diff \xbj \diff{ y} \diff{\psi}} = \frac{\alpha_{\rm em}^2y}{Q^4}L_{\mu\nu}W^{\mu\nu}_{\rm DIS},
\end{equation}
where $\psi$ is the azimuthal angle of the scattered lepton. For SIDIS it is
\begin{equation}
    \frac{\diff{\sigma_{\rm SIDIS}}{}}{\diff \xbj \diff{ y} \diff{\psi}\diff{\zn}\diff{^2\T{P}{B}}} = \frac{\alpha_{\rm em}^2y}{4Q^4 \zn}L_{\mu\nu}W^{\mu\nu}_{\rm SIDIS}\, .
\end{equation}
The hadronic tensors are
\begin{align}
\label{e.hadronictensor}
&{}W_{\rm DIS}^{\mu \nu}  \equiv 4 \pi^3 \sum_X \delta^{(4)}(p + q - P_X) \langle p, S | j^{\mu}(0) | X \rangle \langle X | j^{\nu}(0) | p, S \rangle \, 
\end{align}
for standard DIS, and 
\begin{equation}
    \begin{split}
    &{}W^{\mu\nu}_{\rm SIDIS} \equiv \sum_X \delta^{(4)}(p+q-P_{\rm B}-p_X) \bra{p,S}j^{\mu}(0)\ket{P_{\rm B},X}\bra{P_{\rm B},X}j^{\nu}(0)\ket{p,S}\, 
    \label{eq:SIDIS_hadronic_tensor}
\end{split}
\end{equation}
for SIDIS. The usual decomposition into structure functions is
\begin{equation}
    \begin{split}
        W^{\mu\nu} &= \Big(-g^{\mu\nu} +\frac{q^{\mu}q^{\nu}}{q^2}\Big)F_1
        + \Big(p^{\mu} - \frac{p\cdot q}{q^2}q^{\mu}\Big)\Big(p^{\nu} - \frac{p\cdot q}{q^2}q^{\nu}\Big)\frac{F_2}{p\cdot q}\\
        &+i\epsilon^{\mu\nu\alpha\beta} q^{\alpha}S^{\beta}\frac{g_1}{p\cdot q}
        +i\epsilon^{\mu\nu\alpha\beta} q^{\alpha}[(p\cdot q)S^{\beta} - (S\cdot q)p^{\beta}]\frac{g_2}{(p\cdot q)^2} + \cdots\, ,
    \end{split}
    \label{e.hadronic_tensor_decompositionp}
\end{equation}
and we will use this for both the DIS and SIDIS cases. 
In the DIS case, the arguments of the structure functions are $\xbj$ and $Q^2$, while in the SIDIS case there are additional $\zh$ and $\T{P}{B}$ arguments. The ``$\cdots$'' indicate that there are structure functions in SIDIS beyond what are 
shown explicitly in \eref{hadronic_tensor_decompositionp} (see Ref.~\cite{Bacchetta:2006tn}). These vanish after an integration over azimuthal angle, and we will not consider them further in this paper. The leptonic tensor is
\begin{equation}
\label{eq:lepttensor}
L_{\mu \nu} = 2 (l_\mu l^\prime_\nu + l^\prime_\mu l_\nu - g_{\mu \nu} \; l \cdot l^\prime) \, .
\end{equation}

The SIDIS and DIS structure functions are related though integrals over all $\zn$ and transverse momentum. Specifically,
\begin{equation}
    \sum_{\rm B}\int \frac{\diff{^2\textbf{P}_{\rm B,T}}{}\diff{\zn}{}}{4 \zn} F_{1,2}(\xbj ,Q^2,\zh,\T{P}{B})\\
 = \langle N\rangle F_{1,2}(\xbj ,Q^2)\, .
\end{equation}
Note the normalization $1/(4 \zn)$ on the left-hand side and the multiplicity $\langle N \rangle$ on the right-hand side (see 
Section 6 in Ref.~\cite{Boglione:2019nwk} for details). Also,  by convention, the structure function arguments are $\zh$ and $\xbj$ rather than $\zn$ and $\xn$. The sum is over all types $B$ of final state particles. The unpolarized structure functions are projected from the hadronic tensor as follows,
\begin{align}
F_{1,2}\parz{\xbj,Q} = {\rm P}_{1,2}^{\mu\nu}\, W_{\mu\nu}(\pp,q),
\end{align}
where the projectors for the structure functions are given by
\begin{subequations}
\label{e.F12proj}
\begin{align}
{\rm P}_1^{\mu\nu}
&=- \frac{1}{2} g^{\mu\nu}
+ \frac{2 Q^2 \xn^2}{(Q^2 + \tarmass^2 \xn^2)^2} p^\mu p^\nu\  \approx - \frac{1}{2}
   \left[ g^{\mu\nu}
	- 4 \xbj^2 \frac{\pp^\mu \pp^\nu}{Q^2}
   \right] \, ,
\label{e.P1}\\
{\rm P}_2^{\mu\nu}
&= \frac{12 Q^4 \xn^3 \left(Q^2-\tarmass^2 \xn^2\right)}
       {\left(Q^2 + \tarmass^2 \xn^2\right)^4}
	\bigg( p^\mu p^\nu
	     - \frac{\left(Q^2 + \tarmass^2 \xn^2\right)^2}{12 Q^2 \xn^2}
		g^{\mu\nu}
	\bigg) \approx - \xbj
   \left[ g^{\mu\nu}
	- 12 \xbj^2
	  \frac{\pp^\mu \pp^\nu}{Q^2}
   \right]\, .
\label{e.P2}
\end{align}
\end{subequations}%
The SIDIS versions of the structure functions are obtained by projecting on the integrand of the hadronic tensor $W^{\mu \nu}_\text{SIDIS}$,
\begin{equation}
F_{1,2}(\xbj,Q,\zh,\T{P}{B}) =  \parz{\contractortot{1,2}}_{\mu \nu} W^{\mu \nu}_\text{SIDIS} \, .
\end{equation}

\section{The Theory}
\label{s.yukawa}

Our test case for factorization in DIS is the real scalar Yukawa field theory with the following interaction term,
\begin{align}
\mathcal{L}_{\rm int}
  = -\lambda\, \overline{\Psi}_N\, \psi_q\, \phi\ +\ {\rm H. C.}\, .
\label{e.lagrangian}
\end{align}
A  $\Psi_N$ particle is taken to be the spin-$1/2$ target, and we will refer to it as a 
``nucleon'' with mass $\tarmass$. In addition, there is a spin-1/2 ``quark'' field $\psi_q$ 
with mass $\mquark$, and a chargeless scalar ``diquark'' or ``scalar gluon" field $\phi$ with 
a mass $\mgluon$.  See Chapter 6 of \cite{Collins:2011qcdbook} for similar illustrations of principle using a Yukawa theory. The numerical value of $\lambda$ fixes the strength of this 
interaction.  It is useful to use the notation
\begin{equation}
a_\lambda \equiv \frac{\lambda^2}{16 \pi^2} \, ,
\end{equation}
by analogy with similar notation $a_s = g_s^2/(16 \pi^2)$,  common in QCD. We will choose $\lambda$ to be so small that the fixed order perturbative calculation of the graphs in \fref{basicmodel} give an arbitrarily good approximation to the inelastic ($\xbj < 1$) single photon structure tensors $W^{\mu \nu}_\text{DIS}$ and $W^{\mu \nu}_\text{SIDIS}$. Like QCD, the theory is renormalizable, though it is not asymptotically free. Also like QCD, it is a finite range interaction, characterized by time and distance scales less then order $\sim 1/(\text{intrinsic mass scales})$. The ``intrinsic'' scales analogous to nonperturbative effects in QCD are the masses, $\tarmass$, $\mquark$, and $\mgluon$ that appear in the Lagrangian density, 
and these correspond to any ``$m$" in the error terms:
\begin{equation}
\label{e.masses}
m \in \left\{\tarmass, \mquark, \mgluon \right\} \, .
\end{equation}
\begin{figure}[t]
\centering
  \begin{tabular}{c@{\hspace*{.01mm}}c@{\hspace*{.01mm}}c}
    \includegraphics[scale=0.15]{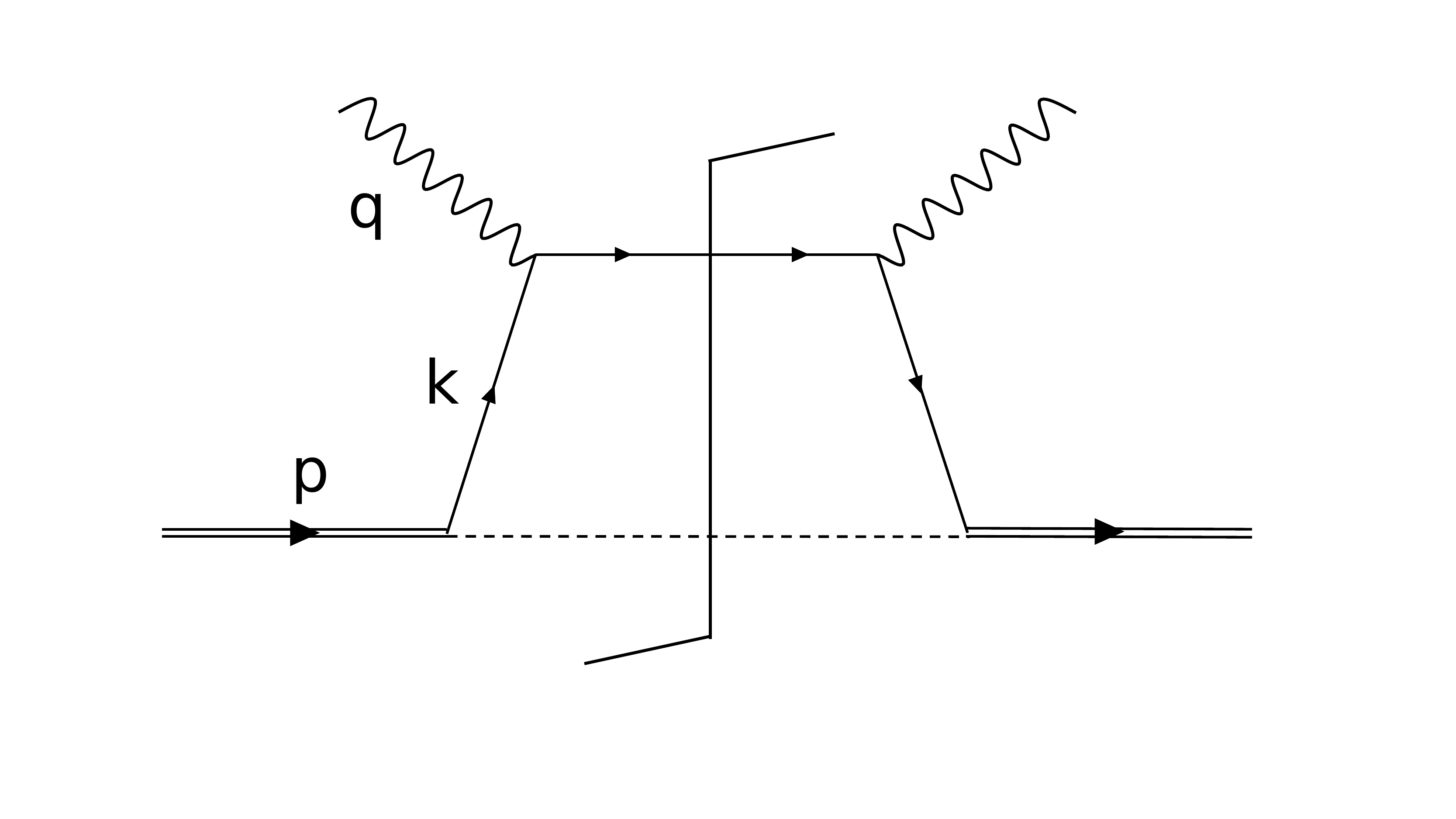}
    \hspace{0.1cm}
    &
    \hspace{0.1cm}
    \includegraphics[scale=0.15]{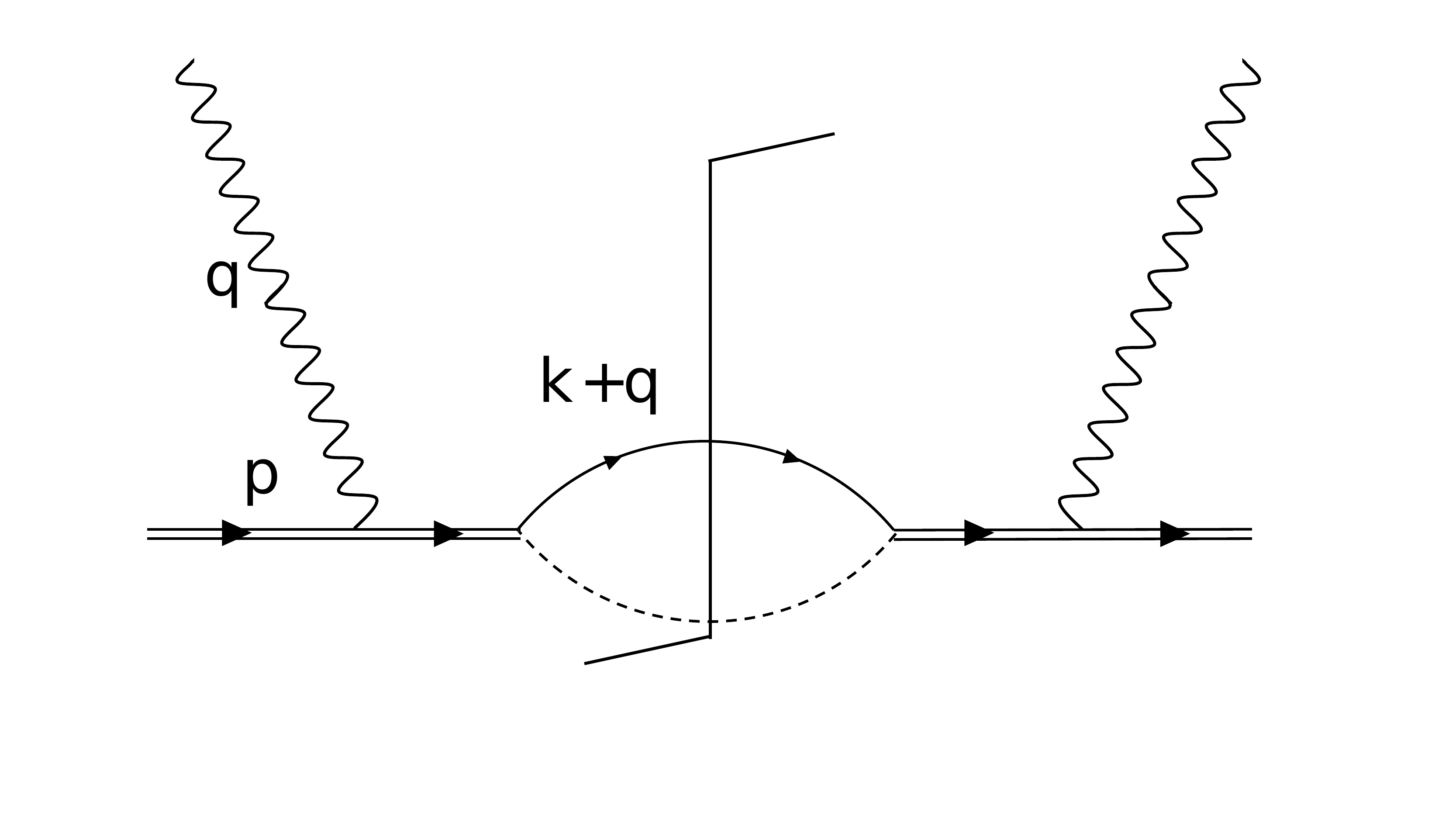}
    &
    \hspace{0.1cm}
    \includegraphics[scale=0.15]{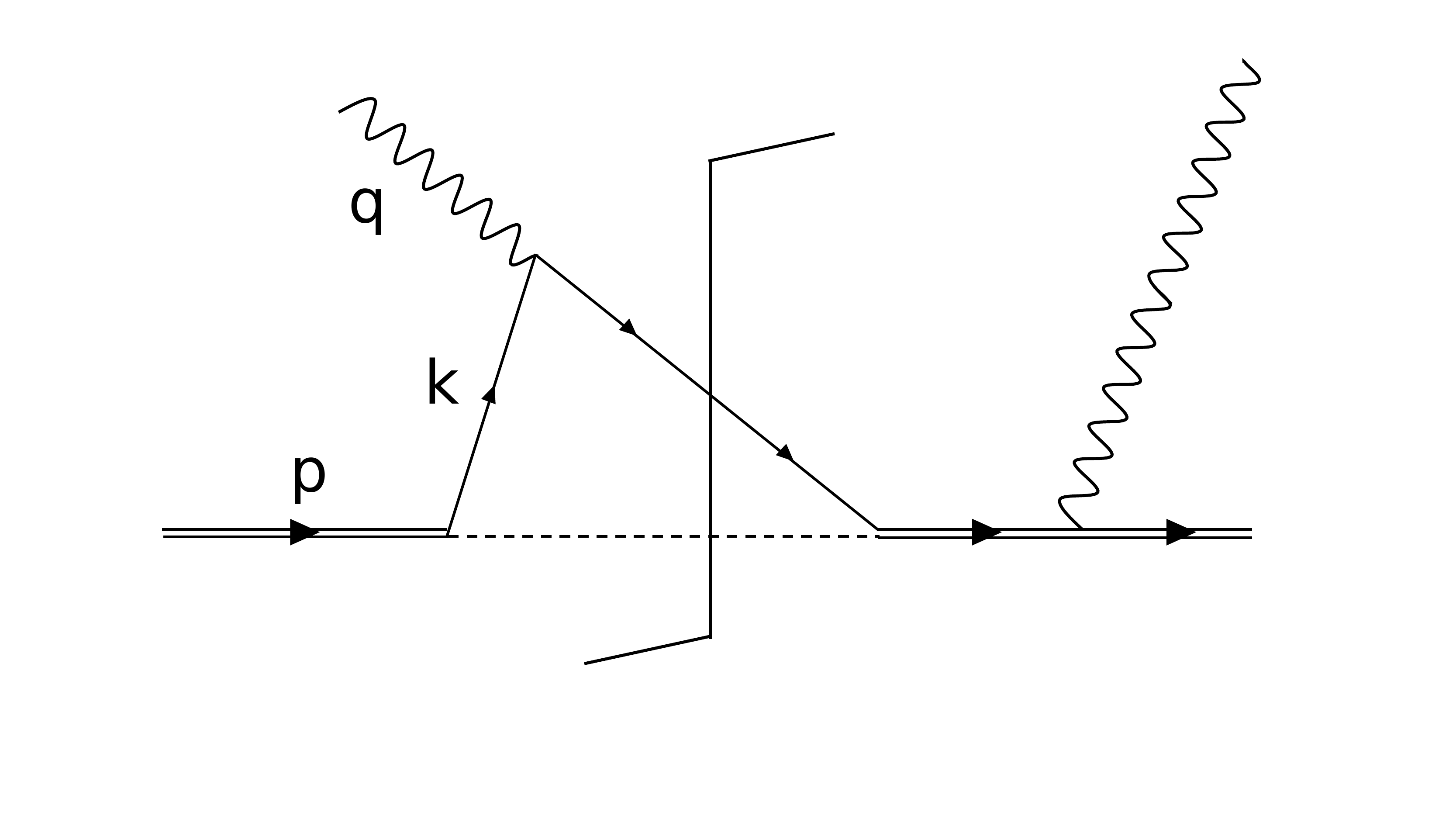}
  \\
  (a) & (b) & (c)
  \end{tabular}
\caption{Contributions to DIS from \eref{lagrangian} at $\order{a_\lambda}$.
  Graph~(a) is the handbag diagram that contributes at leading power and small 
  transverse momentum. $k$ labels the struck quark momentum. Graphs~(b) and (c) contribute at leading power to large 
  $\Tsc{k}{}$ (the Hermitian conjugate for (c) is not shown). The momenta of the virtual 
  photon is ($q$) and the target nucleon is ($p$).
}
\label{f.basicmodel}
\end{figure}

We will handle all ultraviolet divergences with dimensional regularization. In $d$ spacetime dimensions,  we take $\epsilon$ to be defined through $d \equiv 4 - 2 \epsilon$. We also define the factor
\begin{equation}
S_{\epsilon} \equiv \frac{(4 \pi)^\epsilon}{\Gamma(1 - \epsilon)} \, ,
\end{equation}
which multiplies $1/\epsilon$ poles in implementations of $\msbar$ renormalization. Renormalization of the Lagrangian should be understood to have been performed in the $\msbar$ scheme with a dimensional regularization scale $\mu$. Keeping the lowest nonvanishing order beta function and neglecting scalar self-interactions gives for the running coupling, 
\begin{equation}
a_\lambda(\mu) \approx \frac{a_\lambda(\mu_0)}{1 - 10 a_\lambda(\mu_0) \ln \frac{\mu}{\mu_0} }
\end{equation}
relative to a reference scale $\mu_0$. For this paper, we will assume $a_\lambda(\mu_0)$ is small enough that the running can be entirely neglected.

As just mentioned, the graphs that contribute to $W^{\mu \nu}_\text{DIS}$ away from the $\xbj = 1$ elastic limit at the first non-vanishing order in $a_\lambda$ are all shown in \fref{basicmodel}. These graphs also give the SIDIS structure tensor $W^{\mu \nu}_\text{SIDIS}$ if we identify the ``observed final state particle'' $P_B$ with the final state ``quark'' $q$. Calculating the structure functions in \eref{hadronic_tensor_decompositionp}, with no approximations on the graphs in \fref{basicmodel}, is straightforward in the Yukawa theory, though the steps are somewhat tedious when all the masses are allowed to be general. To avoid breaking the flow of our discussion we have provided these steps and other useful results in \aref{exact}. 

The factorization theorem for DIS states that, in the asymptotic $m/Q \to 0$ limit with fixed $0 < \xbj < 1$, the cross section separates into a process-specific short distance (or high virtuality) factor and one or more universal large distance factors.  The short distance factors are insensitive to the dynamics that govern large distance dynamics, so we should find that they are insensitive to the $m$ scales from \eref{masses}. The large distance factors do depend on $m$, but they are universal in the sense that they are defined by explicit operator matrix elements (see \sref{pdfs}) for objects like pdfs and do not reference DIS or any other specific physical process. The well-known steps for deriving factorization in a gauge theory carry over straightforwardly to the scalar Yukawa theory, and indeed are much simpler due to the absence of large coupling, confinement, gauge degrees of freedom and the need for Wilson lines. Moreover, since there are no large coupling parts involved in the calculation of \fref{basicmodel}, both the long and short distance parts are calculable in perturbation theory if we just choose $a_\lambda$ to be very small. By comparing calculations for various values of $Q/m$,   before and after factorization, we hope to obtain a sense of the size of the errors induced by factorization. We will do this for both the DIS and SIDIS cases in the next few sections. 

Notice that the graphs in \fref{basicmodel} have no divergences at all, neither in the ultraviolet (UV) nor the infrared (IR)/collinear regions. The divergences that do appear in our calculations of those graphs at intermediate stages are, therefore, artifacts of factorization approximations. Thus, our calculations will help clarify the nature of those divergences. 

As a prelude to the later discussion of factorization approximations, we may anticipate the result by examining the structure functions as they appear before there are any approximations. We show examples in \fref{exactplots} for a selection of values for $Q$ and with $\tarmass = \mgluon = 1.0$~Gev and $\mquark = 0.3$~GeV chosen to mimic typical small mass scales in QCD.  Vertical dashed lines show the kinematical maximum (see \eref{xmax}) of $\xbj$ for each $Q$, given the specific values of the intrinsic mass scales we have chosen.  As $Q$ increases, the curves for $F_1$ and $F_2$ become relatively smooth over the full range of $0 < \xbj < 1$. We should expect that, once we obtain the factorized approximations in later sections, the corresponding curves will match those of \fref{exactplots} with high accuracy for the larger $Q$ cases. 
\begin{figure}[t]
\centering
  \begin{tabular}{c@{\hspace*{.01mm}}c}
    \includegraphics[scale=0.30]{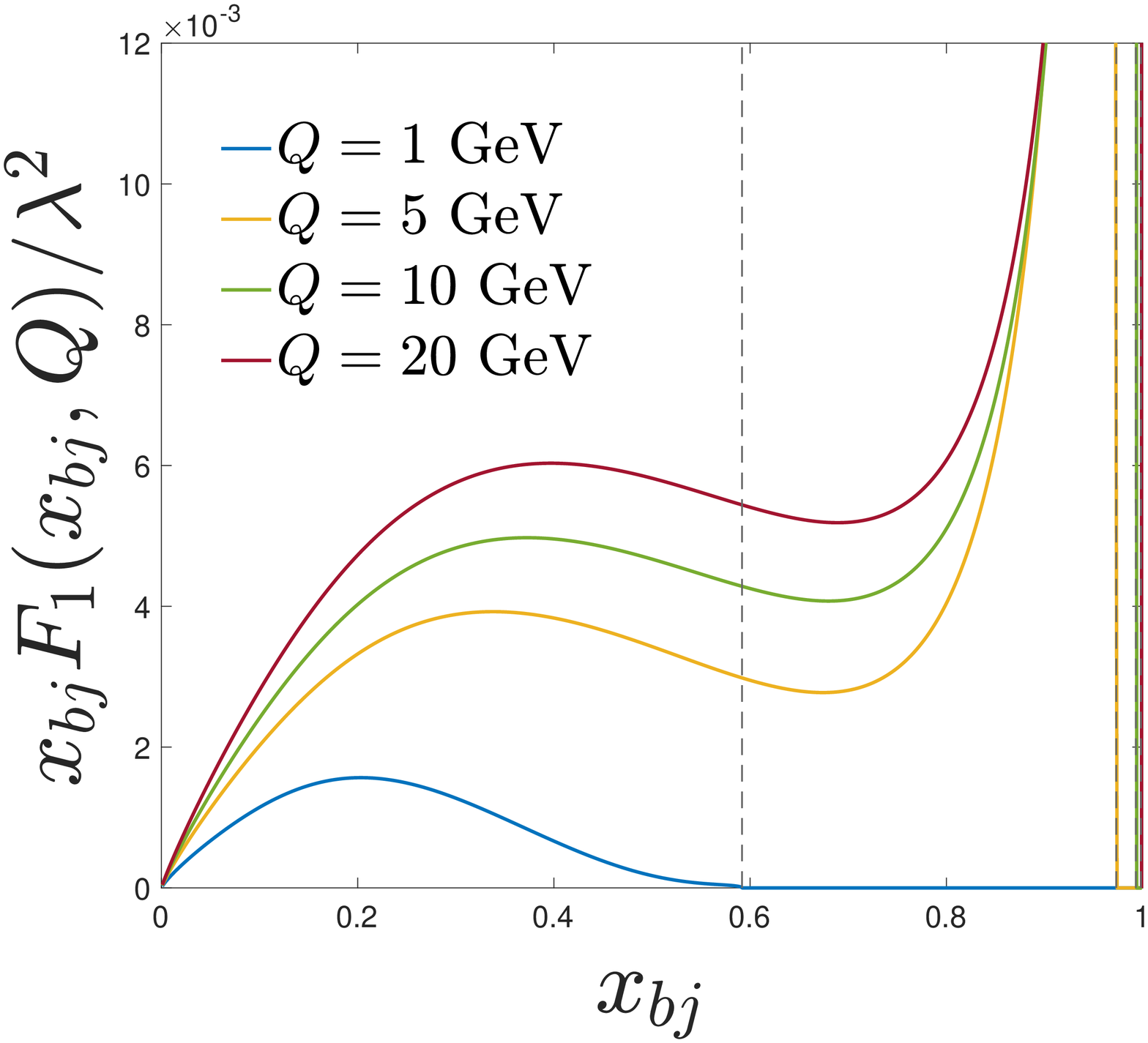}
    &
    \hspace{0.1cm}
    \includegraphics[scale=0.30]{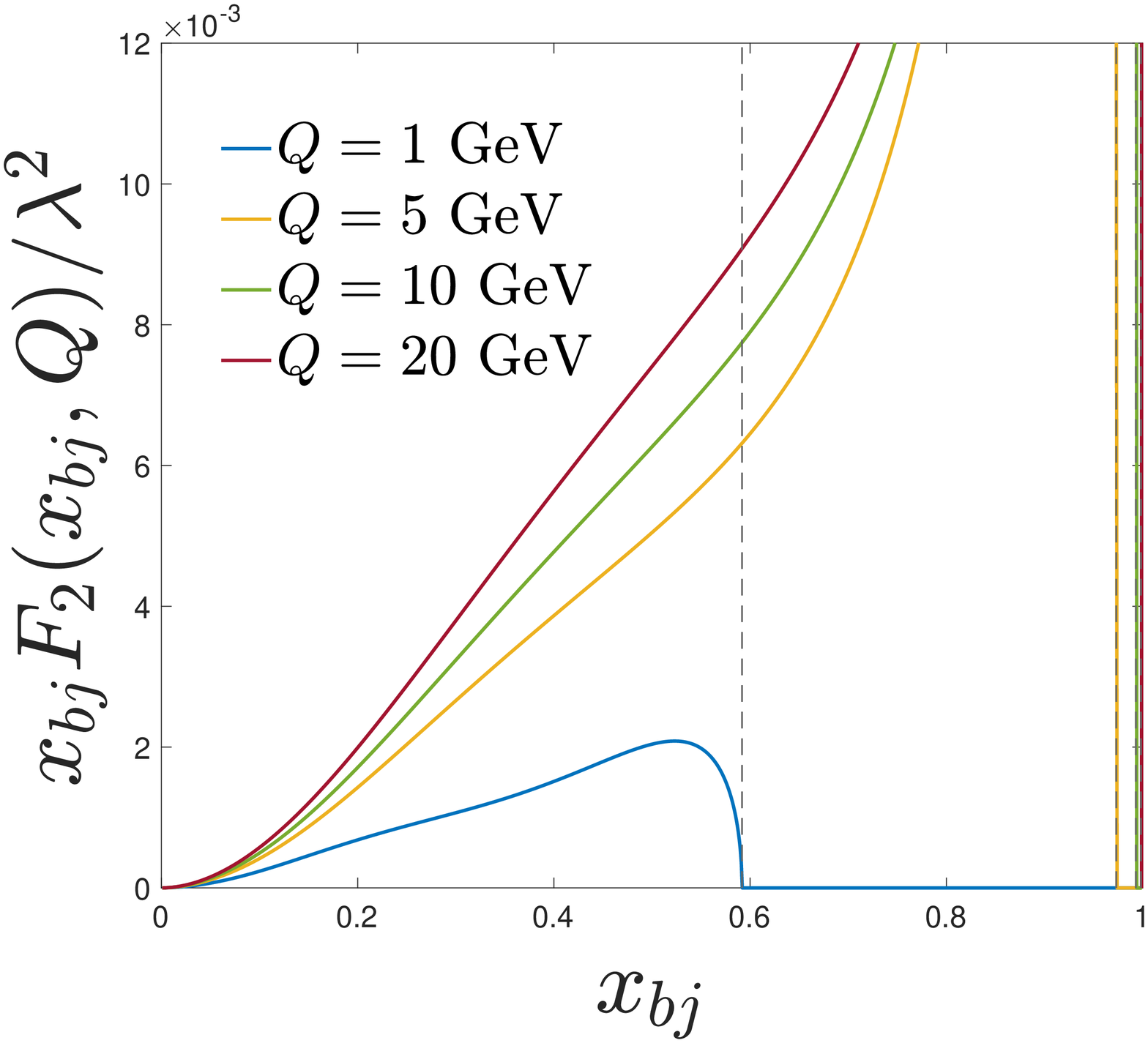}
  \\
  (a) & (b) 
  \end{tabular}
\caption{The unfactorized (a) $F_1(\xbj,Q)$ and (b) $F_2(\xbj,Q)$ corresponding to \fref{basicmodel} with $\tarmass = \mgluon = 1.0$~GeV and $\mquark = 0.3$~GeV. The vertical dashed lines indicate the kinematical upper limits on $\xbj$ for each value of $Q$ (see \eref{xmax}).
}
\label{f.exactplots}
\end{figure}

\section{Parton Densities}
\label{s.pdfs}

At various stages in the discussion it will be necessary to identify contributions 
to pdfs. Therefore, we postpone the treatment of factorization and focus momentarily on reviewing the properties of the operator definitions for pdfs, now specializing to the Yukawa theory from~\sref{yukawa}. 

The bare parton density for a fermion of flavor $i$ inside a fermion $p$ is 
defined by the usual matrix element of bare number density operators:
\begin{equation}
f_{0,i/p}(\xi) = \int \frac{\diff{w^-}{}}{2 \pi} \, e^{-i \xi p^+ w^-} 
\; \langle p | \, \bar{\psi}_{0,i}(0,w^-,\T{0}{}) {\frac{\gamma^+}{2}} 
\psi_{0,i}(0,0,\T{0}{}) \, | p \rangle \, . \label{e.pdfdef}
\end{equation}
Without a UV regulator, the bare pdf is divergent. Ultimately, we work with a renormalized collinear parton density 
\begin{align}
{f_{i/p}(\xi;\mu)} = Z_{i/i'} \otimes {f_{0,i'/p}}
\equiv{}& \sum_{i'} \int \frac{\diff{z}}{z} 
\,  Z(z,a_\lambda(\mu))_{i/i'} \, f_{0,i'/p}(\xi/z)  \, ,  \label{e.pdf}
\end{align}
where $Z_{i/i'}$ is a renormalization factor. Expanding the DIS cross section in a factorized form through order $a_\lambda$ will require $f^{(0)}_{p/p}$ and $f^{(1)}_{q/p}$, with the superscripts indicating the order in perturbation theory.
The last line uses the standard convolution integral notation,
\begin{equation}
\left[ A \otimes B \right](x) \equiv \int_{x}^1 \frac{\diff{\xi}{}}{\xi}  A(x/\xi) B(\xi) \, . \label{e.convdef}
\end{equation}
Note carefully that our use of the term ``bare'' for the pdf is in the track A sense of Ref.~\cite{Collins:2021vke}.
Notice also the absence of a Wilson line operator in \eref{pdfdef} as compared to what we would need in a gauge theory like QCD.

Implementing dimensional regularization, expanding \eref{pdf} through order $a_\lambda$, and applying $\overline{{\rm MS}}$ renormalization by subtracting the $S_\epsilon/\epsilon$ pole gives
\begin{align}
f^{(0)}_{p/p}(\xi;\mu) 
&{}=  \int \frac{\diff{k^-}{}\diff{^{2 - 2 \epsilon }\T{k}{}}}{(2 \pi)^{4 - 2 \epsilon}} 
{\rm Tr} \left[  \frac{\gamma^+}{2} \picineq{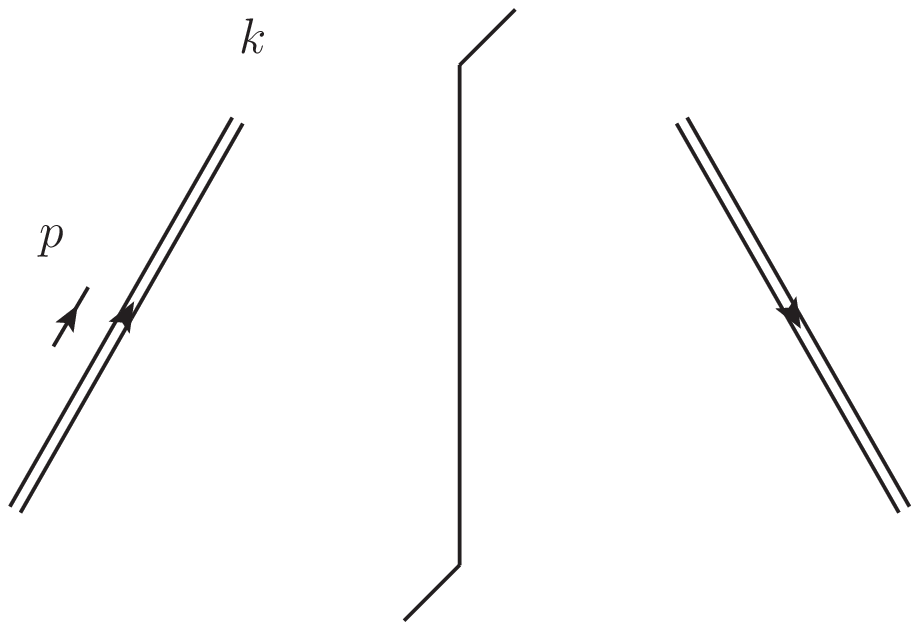} \right] = \delta(1 - \xi) \, , \label{e.PinP} \\
                         \nonumber \\
f^{(1)}_{q/p}(\xi;\mu) &{}\stackrel{\xi \neq 1}{=} 
\int \frac{\diff{k^-}{}\diff{^{2 - 2 \epsilon }\T{k}{}}}{(2 \pi)^{4 - 2 \epsilon}} {\rm Tr} 
\left[  \frac{\gamma^+}{2} \picineq{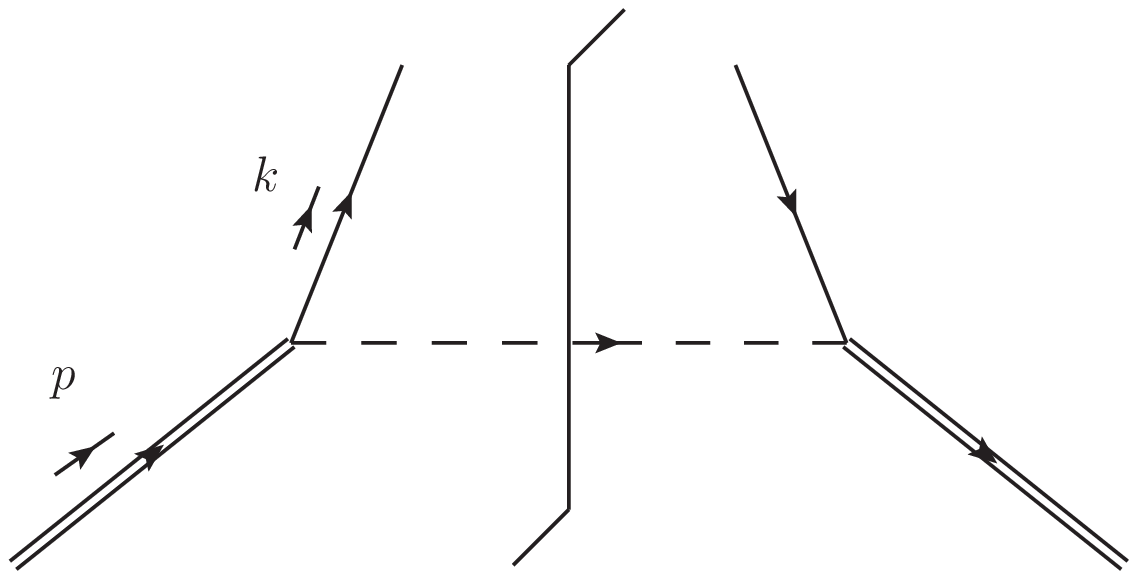} \right] + \overline{{\rm MS}} \;\; 
\text{C.T.} 
\nonumber \\
                         &{}=  a_\lambda(\mu) (1 - \xi) 
 \parz{\frac{\chi(\xi)^2}{\Delta(\xi)^2} 
 + \ln \left[ \frac{\mu^2}{\Delta(\xi)^2} \right] - 1} \, , \label{e.finP}
\end{align}
where in the last line we have used the abbreviations
\begin{equation}
\Delta(\xi)^2 \equiv \xi \mgluon^2 + (1 - \xi) \mquark^2 - \xi (1-\xi) \tarmass^2 \, , \qquad \chi(\xi)^2 \equiv (\mquark + \xi \tarmass)^2 \, ,
\end{equation}
and where the $\overline{{\rm MS}}$ counterterm is 
\begin{equation}
\overline{{\rm MS}} \;\; \text{C.T.} 
= -a_\lambda(\mu) (1 - \xi) \frac{S_\epsilon}{\epsilon} \, . 
\label{e.msct}
\end{equation}
Eq.~\eqref{e.finP} is obtained in dimensional 
regularization after we calculated the integral
\begin{equation}
\frac{a_\lambda(\mu)}{\pi} (2 \pi \mu)^{2 \epsilon} (1-\xi)
\int \diff{^{2 - 2 \epsilon} \T{k}{}}{}  \frac{ \Tscsq{k}{} 
+ \chi(\xi)^2 }
	  {\big[\Tscsq{k}{} + \Delta(\xi)^2
	   \big]^2} + \overline{{\rm MS}} \;\; \text{C.T.} \, , \label{e.unint}
\end{equation}
with the counterterm added,  and where we set $\epsilon = 0$.

The bare quark TMD pdf for a flavor $i$ in hadron $p$ is similarly defined as
\begin{equation}
f_{0,i/p}(\xi,\T{k}{}) = \int \frac{\diff{w^-}{} \diff{^{2} \T{w}{} }{} }{(2 \pi)^3} \, e^{-i \xi p^+ w^- + i \T{k}{} \cdot \T{w}{} } 
\; \langle p | \, \bar{\psi}_{0,i}(0,w^-,\T{w}{}) {\frac{\gamma^+}{2}} 
\psi_{0,i}(0,0,\T{0}{}) \, | p \rangle \, . \label{e.tmdpdfdef}
\end{equation}
To get a renormalized TMD pdf in the Yukawa theory, we only need to switch to the renormalized field
\begin{equation}
f_{0,i/p}(\xi,\T{k}{}) = Z_2 \int \frac{\diff{w^-}{} \diff{^{2} \T{w}{} }{} }{(2 \pi)^3} \, e^{-i \xi p^+ w^- + i \T{k}{} \cdot \T{w}{} } 
\; \langle p | \, \bar{\psi}_{i}(0,w^-,\T{w}{}) {\frac{\gamma^+}{2}} 
\psi_{i}(0,0,\T{0}{}) \, | p \rangle \, = Z_2 f_{i/p}(\xi,\T{k}{};\mu) . \label{e.tmdpdfdefren}
\end{equation}
Since the wavefunction renormalization has the form $Z_2 = 1 + \order{a_\lambda}$, there is no $\order{a_\lambda}$ self-energy contribution in the $\order{a_\lambda}$ graphs in \fref{basicmodel}, so we will have no explicit $Z_2$ contribution to our $\order{a_\lambda}$ quark-in-hadron TMD pdf. Therefore, the expression for the TMD pdf $f_{q/p}(\xi,\T{k}{};\mu)$ is obtained by simply dropping the counterterm in \eref{unint}, keeping the integrand of the first term, and taking the limit to 4 dimensions:
\begin{align}
f^{(1)}_{q/p}(\xi,\T{k}{};\mu)
&{}=\frac{a_\lambda(\mu)}{\pi} (1-\xi) \frac{ \Tscsq{k}{} + \chi(\xi)^2 }
	  {\big[ \Tscsq{k}{} + \Delta(\xi)^2
	   \big]^2} \, . \label{e.TMDpdfresult} 
\end{align}
We will need these expressions in later sections. 

Dealing with divergences and evolution in the Yukawa theory is far simpler than in a gauge theory due to the absence of Wilson lines or light-cone divergences.  
In the Yukawa theory above, TMD evolution equation for the quark-in-hadron TMD pdf is also very simple because it only involves the wavefunction normalization $Z_2$ in \eref{tmdpdfdefren}. The TMD evolution equation~\cite[Sec.~8.71]{Collins:2011qcdbook} is just 
\begin{equation}
\label{e.tmd_evol}
\frac{\diff{}}{\diff{\ln \mu}{}} \ln f_{q/p}(\xi,\T{k}{};\mu) = -2 \gamma_2(a_\lambda(\mu)) \, ,  
\end{equation}
where 
\begin{equation}
\gamma_2(a_\lambda(\mu)) \equiv \frac{1}{2} \frac{\diff{\ln Z_2}{}}{\diff{\ln \mu}{}} \, .
\end{equation}
At lowest order, 
\begin{equation}
\gamma_2^{(1)}(a_\lambda(\mu)) = - \frac{a_\lambda(\mu)}{2} \, .
\end{equation}
The general solution to the TMD evolution equation is
\begin{equation}
f_{q/p}(\xi,\T{k}{};\mu) = f_{q/p}(\xi,\T{k}{};\mu_0) \exp\left\{-2 \int_{\mu_0}^\mu \frac{\diff{\mu}{}}{\mu} \gamma_2(a_\lambda(\mu)) \right\} \, , \label{e.evolved}
\end{equation}
where evolution is from a reference scale $\mu_0$ up to a generic large scale $\mu$. 

There are alternative ways to provide an exact definition to a collinear pdf free of UV divergences. One way that very closely coincides with parton model intuition is to define it as the TMD pdf integrated up to a cutoff $k_c$,
\begin{equation}
\label{e.cuttoff_def}
f^c_{q/p}(\xi;\mu;k_c) \equiv \pi \int_0^{k_c^2} \diff{\Tscsq{k}{}}{} f_{q/p}(\xi,\T{k}{};\mu) \, .
\end{equation}
Normally, $k_c$ is set equal to $\mu$, but this need not be the case.
This approach is preferred in some areas of small-$x$ physics, e.g. ~\cite{Kimber:2001sc,Watt:2003mx,Guiot:2019vsm}, where the relation is taken as a definition for the TMD pdf, and is usually called an ``unintegrated'' pdf. With the TMD pdf calculated in \eref{TMDpdfresult}, the cutoff definition for the collinear pdf is
\begin{equation}
\label{e.cuttoff_express}
f^c_{q/p}(\xi;\mu;k_c) =  a_\lambda(\mu) (1 -\xi) \left[ \ln \parz{\frac{\Delta(\xi)^2 + k_c^2}{\Delta(\xi)^2}} - \frac{k_c^2}{k_c^2 + \Delta(\xi)^2} + \frac{k_c^2 \chi(\xi)^2}{\Delta(\xi)^2 \left[ \Delta(\xi)^2 + k_c^2 \right]} \right] \, .
\end{equation}
This definition of the collinear pdf only equals the standard $\msbar$ definition in \eref{finP} if $k_c = \mu$ and $\order{m^2/\mu^2}$ corrections are neglected. Beyond lowest order, the connection between the cutoff and the renormalized definitions can also involve non-power-suppressed terms, and in gauge theories there are complications with the Wilson line in relations like \eref{cuttoff_def} that we will not address here (see, however, the detailed discussion in \cite{Collins:2003fm}).  

\section{collinear factorization}
\label{s.collinearsteps}

Now that we have identified the pdf contributions, we may build up the factorized expressions for structure functions starting from the exact results for \fref{basicmodel} and applying the approximations appropriate to the deeply inelastic regime. We seek the form of the standard collinear factorization theorem for inclusive DIS,
\begin{align}
W^{\mu \nu}(\pp,q) 
&{}= \sum_{i,i'} \int_{\xbj}^1 \frac{\diff{\xi}}{\xi} 
\widehat{W}^{\mu \nu}_{i/i'}(\xbj/\xi,q;\mu) f_{i'/p}(\xi;\mu) 
+ \order{\frac{m^2}{Q^2}} \,  \no
&{}= \sum_{i,i'} \widehat{W}^{\mu \nu}_{i/i'} \otimes f_{i'/p} + \order{\frac{m^2}{Q^2}}  \, ,
\end{align}
where $\widehat{W}^{\mu \nu}_{i/i'}(\xbj/\xi,q)$ is a partonic structure tensor (with suitable subtractions, to be discussed below) for a 
massless, on-shell partonic target of flavor $i'$, $f_{i'/p}(\xi;\mu)$ is a pdf for a parton 
flavor $i'$ in target $p$, $\mu$ is the renormalization group scale, and $\sum_{i,i'}$ is a 
sum over all flavors. 
The analogous expressions for structure functions are
\begin{align}
F_1(\xbj,Q^2)  &{}= \sum_{ii'} \int_{\xbj}^{1} \frac{\diff{\xi}}{\xi} 
  \widehat{F}_{1,i/i'}(\xbj/\xi,\mu^2/Q^2;\mu) f_{i'/p}(\xi;\mu) + \order{\frac{m^2}{Q^2}} \,   
  \label{e.f1p} \\
  			&{}= \sum_{i,i'} \widehat{F}_{1,i/i'} \otimes f_{i'/p} + \order{\frac{m^2}{Q^2}} \, , \no
F_2(\xbj,Q^2)  &{}= \sum_{i,i'} \int_{\xbj}^{1} \diff{\xi} 
  \widehat{F}_{2,i/i'}(\xbj/\xi,\mu^2/Q^2;\mu) f_{i'/p}(\xi;\mu) + \order{\frac{m^2}{Q^2}} \,   
  \label{e.f2p} \\
  			&{}= \sum_{ii'} \xi \widehat{F}_{2,i/i'} \otimes f_{i'/p} + \order{\frac{m^2}{Q^2}} \, . 
			\nonumber
\end{align}
In the limit that the $\order{\frac{m^2}{Q^2}}$ terms are negligible, the structure functions have 
process-specific hard parts, $\widehat{F}_{1,2}$, that are insensitive to large spacetime distances. But the collinear pdfs $f_{i'/p}$ account for the intrinsic properties of the target, so we should expect them to retain sensitivity to $m$. 

In this section, we will systematically step through the approximations necessary to factorize the graphs in \fref{basicmodel} as in \erefs{f1p}{f2p} for the Yukawa theory.  
The specific task is to expand in small 
$m^2/Q^2$ and confirm that factorization is satisfied order-by-order in $a_\lambda$. For example, a structure function (say $F_1$) becomes
\begin{align}
F_1(\xbj,Q)  
	&{}=\sum_{ii'}  (\widehat{F}_1^{(0)} + \widehat{F}_1^{(1)} + \cdots )_{i/i'} 
	\otimes (f^{(0)} + f^{(1)} + \cdots)_{i'/p} + \order{\frac{m^2}{Q^2}}  \nonumber \\
	&{}=\sum_{ii'} \widehat{F}_{1,i/i'} \otimes f_{i'/p} + \order{\frac{m^2}{Q^2}} \, , 
	\label{e.fact1}
 \end{align}
where the superscript ``$(n)$'' refers to the order in $a_\lambda$, and the ``$\cdots$" refers to higher orders in $a_\lambda$. 

\subsection{Outline Of Steps}

Building up all the terms in \eref{fact1} entails a  mixture of different 
approximations, each corresponding to a different region of momentum. Together, they are a simple example of the method of matched asymptotic expansions~\cite[Ch. 4]{nayfeh}. There is a nested chain of increasingly larger kinematical regions, and different approximations apply in each one. We will step through the procedure below for the case of the $F_1$ structure function.

First, consider the zeroth order term in \eref{fact1}, which corresponds to elastic scattering and is just the 
convolution of a zeroth order pdf (\eref{PinP}) with a zeroth order partonic $\widehat{F}_1$ 
in which $p$ is the target,
\begin{align}
\widehat{F}_{1,p/p}^{(0)}(\xi/\xbj,Q) &{}= \frac{1}{2} 
\delta(1 - \xbj/\xi) = \picineq{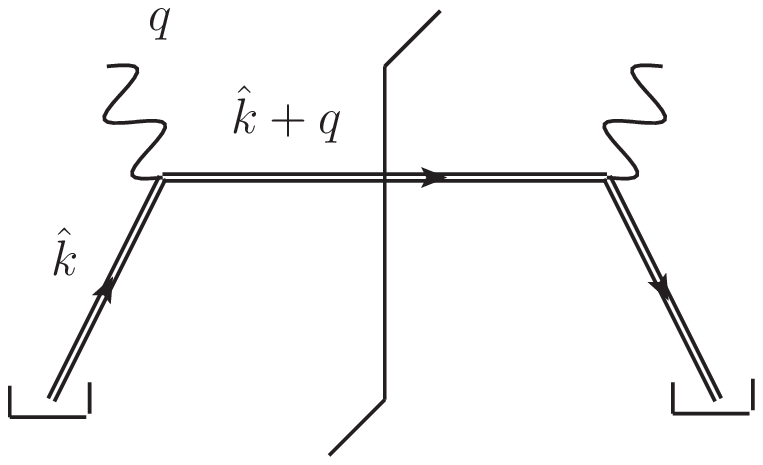} \label{e.hard0} \, .
\end{align}
The hooks at the bottom of the graph notate the approximation 
that all lines above them are to be treated as massless and on-shell. Thus, 
\begin{align}
F_1^{(0)}(\xbj,Q) 
= \widehat{F}_{1,p/p}^{(0)} \otimes f^{(0)}_{p/p} + \order{\frac{m^2}{Q^2}} 
			   &{}= \picineq{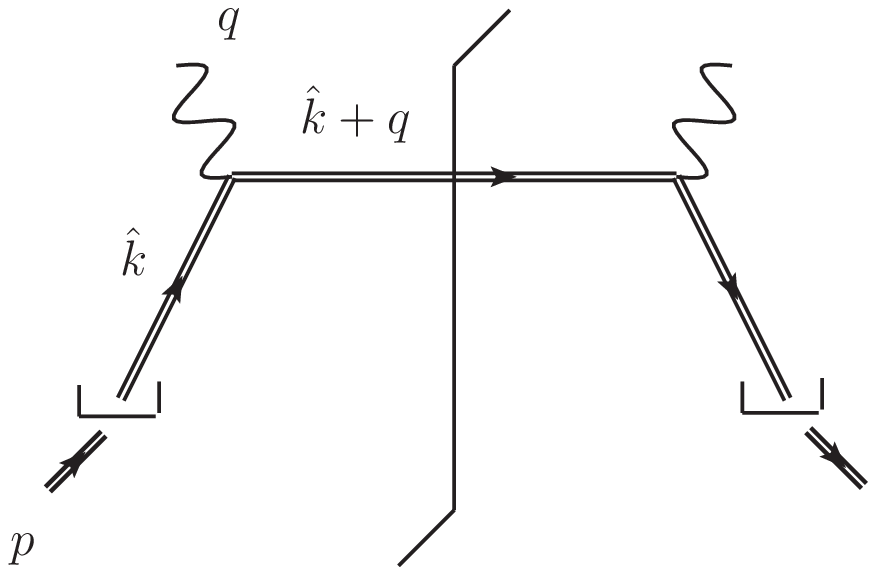} + \order{\frac{m^2}{Q^2}}  \no
			   &{}= \frac{1}{2} \delta(1 - \xbj) + \order{\frac{m^2}{Q^2}} \, . 
			   \label{e.f10}
\end{align}
The second line above introduces additional graphical conventions that we will use throughout this paper. The hard part $\widehat{F}_{1,p/p}^{(0)}(\xi/\xbj,Q)$ from \eref{hard0} is placed above the graph in \eref{PinP} for the integrand of the pdf.  This symbolizes the 
convolution integral, \eref{convdef}. The $\widehat{F}_{2,i/i'}^{(0)}(\xi/\xbj,Q)$ contribution is 
similarly 
\begin{equation}
\widehat{F}_{2,p/p}^{(0)}(\xi/\xbj,Q) =  \delta(1 - \xbj/\xi)  \, . \label{e.hard0f2}
\end{equation}
The form of \eref{hard0} is the same for the zeroth order structure function of any fermionic parton target, including a quark,
\begin{align}
\widehat{F}_{1,q/q}^{(0)}(\xi/\xbj,Q) &{}= \frac{1}{2} 
\delta(1 - \xbj/\xi)  =  \picineq{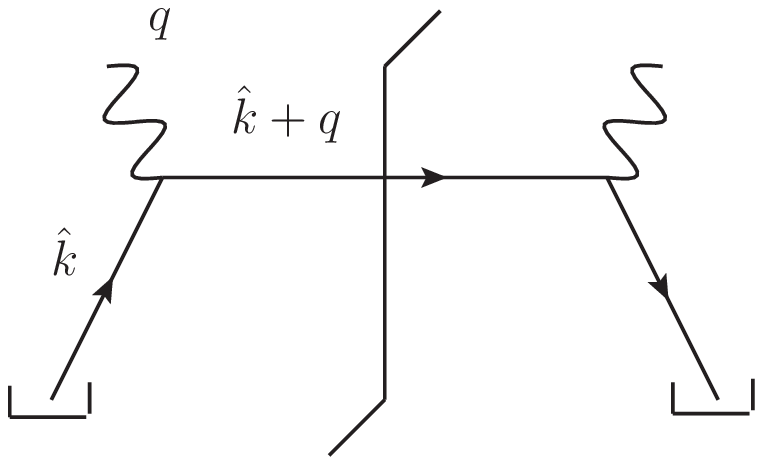} \, . \label{e.hard02} 
\end{align}


Turning now to $\xbj < 1$ and the graphs in \fref{basicmodel}, the smallest region that we need to approximate corresponds to a neighborhood of $\Tsc{k}{} = 0$ extending to not much larger than $\order{m}$. In that region, graphs (b) and (c) of \fref{basicmodel} are subleading in ${\rm max}(m^2,\Tscsq{k}{})/Q^2$, so part of the approximation is to keep only graph (a) and neglect graphs (b) and (c). In the upper part of graph (a), it is only the $k^+$ component of $k$ that is important in the small transverse momentum region, so the small $k^-$ and $\Tsc{k}{}$ components can be neglected there. The details of the small $\Tsc{k}{}$ approximations are reviewed extensively 
in many other places~\cite{Collins:2011qcdbook} so we will not repeat
them here. The important point is that they separate 
graph (a) into a zeroth order hard part and an $\order{a_\lambda}$ quark-in-hadron pdf, up to $\order{{\rm max}(m^2,\Tscsq{k}{})/Q^2}$ corrections. 

To symbolize the small-$\Tsc{k}{}$ approximation, we will use the ``approximator'' notation $\TI$. This is an instruction to replace the internal $k$ line of the object on its left by an approximate version and to drop power suppressed errors. Applied to \fref{basicmodel}(a),
\begin{equation}
F_{1}(\xbj,Q,\T{k}{}) = \TI F_{1}(\xbj,Q,\T{k}{}) 
+ \order{\frac{{\rm max}(m^2,\Tscsq{k}{})}{Q^2}} \, . \label{e.T1app}
\end{equation}
The approximation with the error term made explicit is
\begin{equation}
F_{1}(\xbj,Q,\T{k}{})  = \TI F_{1}(\xbj,Q,\T{k}{}) 
+ \left[ F_{1}(\xbj,Q,\T{k}{})  - \TI F_{1}(\xbj,Q,\T{k}{}) \right] \, . \label{e.sub1}
\end{equation}
The error in braces is the $\order{{\rm max}(m^2,\Tscsq{k}{})/Q^2}$-suppressed contribution from~\eref{T1app}, so the $\TI$ approximation is no longer accurate once $\Tsc{k}{}$ is comparable to $Q$. 
The $\TI$ approximator does not yield the expansion for the full integral over
 $\Tsc{k}{}$ necessary for fully inclusive scattering. 

To fix this, we next consider the larger $\Tsc{k}{} \approx Q$ region. Specifically, we consider the region where the ratio $\Tsc{k}{}/Q$ is fixed and expand the graphs in \fref{basicmodel} in powers of $m/\Tsc{k}{}$. The approximation is accurate in a neighborhood of $m < \Tsc{k}{} \lesssim Q$, and it exploits the smallness of the mass scales ($\mgluon$, $\mquark$, 
$\tarmass$) relative to the large transverse momentum $\Tsc{k}{}$. We call the corresponding approximator $\TII$ and write
\begin{equation}
F_{1}(\xbj,Q,\T{k}{})  
= \TII F_{1}(\xbj,Q,\T{k}{}) + \order{\frac{m^2}{\Tscsq{k}{}}} \, .
\end{equation}
In the Yukawa theory model calculations of \fref{basicmodel}, the large $\Tsc{k}{}$ approximation amounts to simply setting 
all small masses to zero.

The error term in \eref{sub1} is only significant when $\Tsc{k}{}$ is large relative to $m$, so  
the final step is to apply $\TII$ to the entire contribution in the braces in \eref{sub1}. Since that term is 
already $\order{\Tscsq{k}{}/Q^2}$, the resulting overall error is now 
$\order{m^2/Q^2}$ point-by-point in $\Tsc{k}{}$:\footnote{Note that the $\TII$ 
approximation sets all masses to zero in $F_1$, so the $\TI$ approximation 
ultimately contributes an error of size $\order{\Tscsq{k}{}/Q^2}$ instead of 
$\order{{\rm max}(m^2,\Tscsq{k}{})/Q^2}$.}
\begin{align}
F_{1}(\xbj,Q,\T{k}{}) 
&{}=  \TI F_{1}(\xbj,Q,\T{k}{}) 
+ \TII \left[ F_{1}(\xbj,Q,\T{k}{})  - \TI F_{1}(\xbj,Q,\T{k}{}) \right] 
+ \order{\frac{\Tscsq{k}{}}{Q^2} \times \frac{m^2}{\Tscsq{k}{}} } \, . \no
&{}= \text{W-term} + \text{Y-term} + \order{\frac{m^2}{Q^2}} \, .\label{e.sub2}
\end{align}
In a common jargon, the first term on the second line is labeled the ``W-term'' and the second term is the ``Y-term,'' as indicated on the last line. 
Now we may integrate \eref{sub2} over the whole kinematically accessible region of $\Tsc{k}{}$ and be assured 
that the overall error in our calculation of the inclusive $F_{1}(\xbj,Q)$ vanishes like 
$m^2/Q^2$ in the large $Q$ limit.

Integrating the $W$-term over transverse momentum gives 
the contribution to $F_{1}(\xbj,Q)$ with an $\order{a_\lambda^0}$ hard part and an $\order{a_\lambda^1}$ pdf, while integrating 
the $Y$-term produces the contribution to $F_{1}(\xbj,Q)$ with an $\order{a_\lambda^1}$ hard part and an $\order{a_\lambda^0}$ pdf. So, to order $\order{a_\lambda^1}$ the fully factorized approximation is
\begin{align}
  F^{(1)}_{1}(\xbj,Q) &{}\stackrel{\xbj \neq 1}{=}  
  \sum_i \widehat{F}_{1,q/i}^{(0)} \otimes f_{i/p}^{(1)} 
  +  \sum_i \widehat{F}_{1,q/i}^{(1)} \otimes f_{i/p}^{(0)} 
  + \order{a_\lambda^2,\frac{m^2}{Q^2}} \, . 
  \label{e.finalcol}
\end{align}
We will illustrate the above with explicit expressions in the next few subsections. 

\subsection{Small Transverse Momentum}
\label{s.smalltrans}

Retracing the steps of the last subsection, the first is to apply the $\TI$ directly to graph \fref{basicmodel}(a). The result is
\begin{align}
\TI F_{1}(\xbj,Q,\T{k}{}) 
&{}= \frac{a_\lambda(\mu)}{2 \pi} (1-\xbj) \frac{\Tscsq{k}{} + \chi(\xbj)^2 }
	  {\big[\Tscsq{k}{} + \Delta(\xbj)^2    \big]^2} \, . \label{e.TMDfact}
\end{align}
Up to a factor of $1/2$ from the hard coefficient, this is just the TMD pdf in \eref{TMDpdfresult}. 
Integrating it over all transverse momentum 
gives  
\begin{align}
\widehat{F}_{1,q/q}^{(0)} \otimes f_{q/p}^{(1)} &{}\stackrel{\xbj < 1}{=} \picineq{gallery/lowktfact} 
=\frac{a_\lambda(\mu)}{2} (1 - \xbj) 
\parz{\frac{\chi(\xbj)^2}{\Delta(\xbj)^2} 
+ \ln \left[ \frac{\mu^2}{\Delta(\xbj)^2} \right] - 1} \, . \label{e.F0f1}
\end{align}
A side effect of the $\TI$ approximation is that there is a UV divergence in the integral 
over $\Tsc{k}{}$ that did not exist in the original unapproximated graph. To deal with it, we have taken $\epsilon \neq 0$ and applied $\msbar$ renormalization to the pdf, as in \eref{unint}, before returning to $4$ dimensions. Note that we could have chosen to instead use the cutoff definition in \erefs{cuttoff_def}{cuttoff_express} for the pdf, but the difference between the two choices amounts only to power-suppressed errors in the cross section calculation. 

The graphical notation in \eref{F0f1} is analogous to that of  
\eref{f10}. As before, hooks on the target quark lines symbolize the 
approximations on the $k$-momentum that flows into the top of the graph.  As the diagrammatic notation in \eref{F0f1} emphasizes, this result is 
constructed from pieces that we already know from earlier sections, namely~\eref{hard02} and \eref{finP}.
Placing the 
\eref{hard02} graph over the \eref{finP} integrand represents the convolution of 
the hard part at the top with the pdf at the bottom. 

\subsection{Large Transverse Momentum}

The $\TII$ approximation sets all 
masses to zero in the unapproximated \fref{basicmodel}, and convolutes the result 
with the trivial zeroth order proton-in-proton pdf, \eref{PinP}. At large $\Tsc{k}{}$, all the graphs in \fref{basicmodel} are leading power, and none can be neglected. The $\order{a_\lambda}$ hard part is
\begin{align}
\widehat{F}^{(1)}_{1,q/p}(\xi,Q;\mu)_{\rm unsub} &{}= \picineq{gallery/hard2} =  \frac{a_\lambda(\mu)}{2 \pi} 
\int_{k_{\rm cut}^2}^{\ktms^2(\xi)} \diff{^2 \T{k}{}}{} 
\parz{ \frac{1 - \xi}{ \kappa(\xi) \Tscsq{k}{}} 
- \frac{\xi (2 \xi^2 - 6 \xi + 3) }{Q^2 (1 - \xi)^2 \kappa(\xi)} } \, . \label{e.FO}
\end{align}
The blobs on the left and right represents the sum of \emph{all} graphs in \fref{basicmodel}. We have 
used 
\begin{equation}
\kappa(\xi) \equiv \sqrt{1 - \frac{\Tscsq{k}{}}{\ktms^2(\xi)} }, 
\end{equation}
where
\begin{equation}
\ktms^2(\xi) \equiv \frac{(1 - \xi) Q^2}{4 \xi} \, 
\end{equation}
is the kinematical upper bound on transverse momentum in the massless 
approximation. (The ``hat'' is to distinguish this from the exact kinematical upper bound in \eref{ktmaxval}.) The integral over $\Tsc{k}{}$ in \eref{FO} diverges at $\Tsc{k}{} = 0$, 
so we have temporarily introduced a lower cutoff, $k_{\rm cut}$.  (We will find it 
unnecessary once we combine all terms.) The ``${\rm unsub}$'' 
subscript means we have yet to apply the $\TII \TI$ subtraction in \eref{sub2}. The 
hooks in \eref{FO} symbolize the $\TII$ approximation that sets all masses to zero in the lines above them. 

Taking the convolution of \eref{FO} with \eref{PinP} gives
\begin{align}
\TII F^{(1)}_1(\xbj,Q) &{}= \picineq{gallery/largektfact} 
=  \; \widehat{F}^{(1)}_{1,q/p,{\rm unsub}} \otimes f^{(0)}_{p/p}  \, \nonumber \\
      			       &{}                                      
			        =  \; \frac{a_\lambda(\mu)}{2} \int_{k_{\rm cut}^2}^{\ktms^2(\xbj)} 
			        \diff{\Tscsq{k}{}}{} 
			       \parz{ \frac{1 - \xbj}{ \kappa(\xbj) \Tscsq{k}{}} 
			       - \frac{\xbj (2 \xbj^2 - 6 \xbj + 3) }{Q^2 (1 - \xbj)^2 \kappa(\xbj)} } \, . 
			       \label{e.FO2}
\end{align}
The hooks in \eref{FO2} now denote the separation between the hard subgraph, where 
all masses are negligible relative to the hard scale, from the pdf of \eref{PinP}.\footnote{Compared with textbook calculations of partonic scattering, these expressions may look somewhat odd since there are no ``$+$''-distributions. In fact, we could combine these results with self energy graphs and reexpress them in terms of ``$+$''-distributions, but this is unnecessary for the $\xbj < 1$ region of the graphs in \fref{basicmodel}.  A nice feature of this toy model is that we are able to avoid using distributions.}

To complete the large-$\Tsc{k}{}$ approximation we need to subtract from \eref{FO2} the term that 
corresponds to the combined $\TII \TI F_{1}(\xbj,Q,\T{k}{})$ 
approximation in 
\eref{sub2}. This amounts to the same low $\Tsc{k}{}$ factorization approximation we made in \sref{smalltrans}, but now
with all masses set to zero in accordance with the $\TII$ approximation. Since we already have 
\eref{FO2}, we simply need to apply to it  $\TI$. It amounts to the handbag approximation 
again, but now with all particle masses set to zero, including in the pdf itself,
\begin{equation}
\TII \TI F_{1}(\xbj,Q,\T{k}{}) = \picineq{gallery/asymptot} 
= \frac{a_\lambda(\mu)}{2 \pi } \frac{(1 - \xbj)}{\Tscsq{k}{}} \, . \label{e.asymp}
\end{equation}
The two sets of hooks indicate graphically that the two combined approximations are being applied simultaneously.
When integrating \eref{asymp} over transverse momentum, the $\TI$ is to be  read as an instruction to apply $\msbar$ 
renormalization in the integration over $\Tscsq{k}{}$, just as we did when $\TI$ was applied to the unapproximated graph (a). Thus,
\begin{equation}
\TII \TI F^{(1)}_1(\xbj,Q) =\frac{a_\lambda(\mu)}{2} \mu^{2 \epsilon} S_\epsilon 
\int_{k_{\rm cut}^2}^\infty \diff{\Tscsq{k}{}}{} (\Tscsq{k}{})^{-\epsilon} \frac{(1 - \xbj)}{\Tscsq{k}{}} 
- \frac{a_\lambda(\mu)}{2} (1 - \xbj) \frac{S_\epsilon}{\epsilon} \, . \label{e.sub4}
\end{equation}
The same lower $k_{\rm cut}^2$ cutoff in \eref{FO2} needs to be imposed also in \eref{sub4}. It 
is simple to verify by direct calculation
that (for this low order graph) the $\msbar$ subtraction in \eref{sub4} is equivalent to 
applying an upper cutoff of $\mu^2$ on the $\Tscsq{k}{}$ integral:
\begin{align}
\TII \TI F^{(1)}_1(\xbj,Q) &{}= \frac{a_\lambda(\mu)}{2} \int_{k_{\rm cut}^2}^{\mu^2} 
\diff{\Tscsq{k}{}}{} \frac{(1 - \xbj)}{\Tscsq{k}{}} \nonumber \\
&{}= \frac{a_\lambda(\mu)}{2} \int_{k_{\rm cut}^2}^{\ktms^2(\xbj)} 
\diff{\Tscsq{k}{}}{} \frac{(1 - \xbj)}{\Tscsq{k}{}} 
+ \frac{a_\lambda(\mu)}{2} (1 - \xbj) \ln \frac{\mu^2}{\ktms^2(\xbj)} \, . \label{e.subtractterm}
\end{align}
Subtracting \eref{subtractterm} from \eref{FO2} gives the full collinear $\widehat{F}^{(1)}_1$. Including the trivial 
convolution with the zeroth order pdf in \eref{PinP} produces the second term of \eref{finalcol},
\begin{align}
& \widehat{F}^{(1)}_{1,q/p} \otimes f^{(0)}_{p/p}  \nonumber \\
&=\frac{a_\lambda(\mu)}{2} \int_{0}^{\ktms^2(\xbj)} 
\diff{\Tscsq{k}{}}{} \parz{ \frac{(1 - \xbj)(1 - \kappa(\xbj))}{ \kappa(\xbj) \Tscsq{k}{}} 
- \frac{\xbj (2 \xbj^2 - 6 \xbj + 3) }{Q^2 (1 - \xbj)^2 \kappa(\xbj)} } 
- \frac{a_\lambda(\mu)}{2} (1 - \xbj) \ln \frac{\mu^2}{\ktms^2(\xbj)} \, . \label{e.F1f0}
\end{align}
Now the integral over $\Tsc{k}{}$ is well-behaved in both the UV and IR limits, confirming that 
no lower cutoff was needed, so we have removed the $k_{\rm cut}$ from \eref{F1f0}. 
Equation~\eqref{e.F0f1} accounts for the zeroth order contribution to the hard partonic structure function while 
\eref{F1f0} accounts for the first $\order{a_\lambda}$ contribution. Both must 
be present in order to have complete factorization with only power suppressed and $\order{a_\lambda^2}$ errors.   

\subsection{Fully factorized result}
\label{s.fullresult}

Combining \eref{F0f1} and \eref{F1f0} gives all of the leading terms in \eref{finalcol}, and evaluating the integrals explicitly gives
\begin{align}
F_1(\xbj,Q)  &{}= \underbrace{\frac{a_\lambda(\mu)}{2} (1 - \xbj) 
                         \parz{\frac{(\mquark + \xbj \tarmass)^2}{\Delta(\xbj)^2} 
                         + \ln \left[ \frac{\mu^2}{\Delta(\xbj)^2} \right]-1}}_{ \hat{F}_{1,q/q}^{(0)}
                           \otimes f^{(1)}_{q/p} } + \nonumber \\
                           &{}+  \underbrace{ \parz{\frac{a_\lambda(\mu)}{2} \left[ 2 (1 - \xbj) 
                           \ln \parz{2} 
                           - \frac{2 \xbj^2 - 6 \xbj + 3}{2 (1 - \xbj)} \right]
                           - \frac{a_\lambda(\mu)}{2} (1 - \xbj) \ln \frac{\mu^2}{\ktms^2(\xbj)}}}_{ \hat{F}_{1,q/p}^{(1)}
                           \otimes f^{(0)}_{p/p}} \, \no
                           &{}+ \order{\frac{m^2}{Q^2}, a_\lambda^2} \, . \label{e.finalfact} 
\end{align}
Equation~\eqref{e.finalfact} can be written in a more explicitly factorized way, with
\begin{align}
&{}F_{1}(\xbj,Q)  = \sum_i \int_{\xbj}^1 \frac{\diff{\xi}{}}{\xi} \times \no
 {}&\times \underbrace{\frac{1}{2}  \left\{\delta\parz{1 - \frac{\xbj}{\xi}} \delta_{qi} 
+ a_\lambda(\mu) \parz{1 - \frac{\xbj}{\xi}} \left[ \ln \parz{4} 
- \frac{\parz{\frac{\xbj}{\xi}}^2 - 3 \frac{\xbj}{\xi} + \frac{3}{2}}{\parz{1 - \frac{\xbj}{\xi}}^2} 
-  \ln \frac{4 \xbj \mu^2}{Q^2 (\xi - \xbj)} \right] \delta_{pi} \right\}}_{\hat{F}_{1,q/i}(\xbj/\xi,\mu/Q;a_\lambda(\mu))} \times \no
\no
&{} \times \underbrace{\left\{ \delta\parz{1 - \xi} \delta_{ip}
+ a_\lambda(\mu) (1 - \xi) \left[  \frac{\chi(\xi)^2}{\Delta(\xi)^2} 
+ \ln \parz{\frac{\mu^2}{\Delta(\xi)^2}} - 1 \right] \delta_{iq} \right\}}_{f_{i/p}(\xi;\mu)} \no
&{}+ \order{\frac{m^2}{Q^2}, a_\lambda^2} \, .
 \label{e.finalfactorized}
\end{align}
The second line is now the (subtracted) hard partonic structure function through $\order{a_\lambda}$, and the third line is the pdf through $\order{a_\lambda}$.

Equation~\eqref{e.finalfactorized} is the factorization of \fref{basicmodel} in the form of \eref{fact1} that we sought. It is an approximation whose accuracy increases as $m/Q \to 0$. More general treatments of factorization show that the pattern continues to all orders in $a_\lambda$.  

Several well-known features of factorization are recognizable in \eref{finalfactorized}.  First, the hard factor $\hat{F}_{1,q/i}$ on the second line is a partonic DIS 
structure function, it is process-specific, and it depends on the process-specific kinematical variable $Q$. However, it is independent of any of the small mass scales like $\mquark$, $\mgluon$ or $\tarmass$ that govern intrinsic structure over large spacetime scales. 
Conversely, the pdf $f_{i/p}$ on the third line does depend on intrinsic scales, but it is universal in that it follows directly (\eref{PinP} and \eref{finP}) from the operator definition in \eref{pdfdef}.  
Second, the logarithmic dependence on $\mu$ cancels between the second and third lines through order $a_\lambda(\mu)$, demonstrating the renormalization group independence. Any 
residual $\mu$ dependence is in the $\order{a_\lambda(\mu)^2}$ 
running of parameters like $a_\lambda(\mu)$, but this too would vanish with higher 
orders.  

In an asymptotically free theory like QCD, the goal would be to ensure that higher order terms in the perturbative expansion of the hard part remain small or finite as $Q \to \infty$. Thus, logarithms like those in $\hat{F}_{1,q/i}(\xbj/\xi,\mu/Q;a_\lambda(\mu))$ need to be kept under control by choosing to set $\mu \propto Q$.
With such a choice, all $Q$-dependence \emph{in the hard part} resides in the running of the coupling, which vanishes in the DIS limit. In the non-asymptotically free Yukawa toy theory that we are using here, there are fewer advantages to doing this, but the steps nevertheless very clearly illustrate the procedure. Thus, we are generally interested in the pdf defined with its scale of order $Q$, $f_{q/p}(\xi;Q)$. 

The steps above apply in the same way to the $F_2$ structure function, giving
\begin{align}
&{}F_{2}(\xbj,Q)  = \sum_i \int_{\xbj}^1 \diff{\xi}{} \times \no
 {}&\times \underbrace{ \left\{\delta\parz{1 - \frac{\xbj}{\xi}} \delta_{qi} 
+ a_\lambda(\mu)\frac{\xbj}{\xi} \parz{1 - \frac{\xbj}{\xi}} \left[ \ln \parz{4} 
- \frac{3\parz{\frac{\xbj}{\xi}}^2 - 5 \frac{\xbj}{\xi} + \frac{3}{2}}{\parz{1 - \frac{\xbj}{\xi}}^2} 
-  \ln \frac{4 \xbj \mu^2}{Q^2 (\xi - \xbj)} \right] \delta_{pi} \right\}}_{\hat{F}_{2,q/i}(\xbj/\xi,\mu/Q;a_\lambda(\mu))} \times \no
\no
&{} \times \underbrace{\left\{ \delta\parz{1 - \xi} \delta_{ip}
+ a_\lambda(\mu) (1 - \xi) \left[  \frac{\chi(\xi)^2}{\Delta(\xi)^2} 
+ \ln \parz{\frac{\mu^2}{\Delta(\xi)^2}} - 1 \right] \delta_{iq} \right\}}_{f_{i/p}(\xi;\mu)} \no
&{}+ \order{\frac{m^2}{Q^2}, a_\lambda^2} \, .
 \label{e.finalfactorizedf2}
\end{align}
It is worth verifying graphically and numerically that the factorized expressions for $F_1$ and $F_2$ match the unapproximated calculations of the graphs in \fref{basicmodel} when $m/Q$ approaches zero. This is illustrated in~\fref{factplots}. The solid curves are the same as those in \fref{exactplots}, but now overlaid on top are the calculations with factorization, obtained from \eref{finalfactorized} and \eref{finalfactorizedf2} and shown as dot-dashed lines. As expected, the unapproximated and factorized calculations agree as $Q$ increases above $\approx 1$~GeV.
\begin{figure}[t]
\centering
  \begin{tabular}{c@{\hspace*{.01mm}}c}
    \includegraphics[scale=0.30]{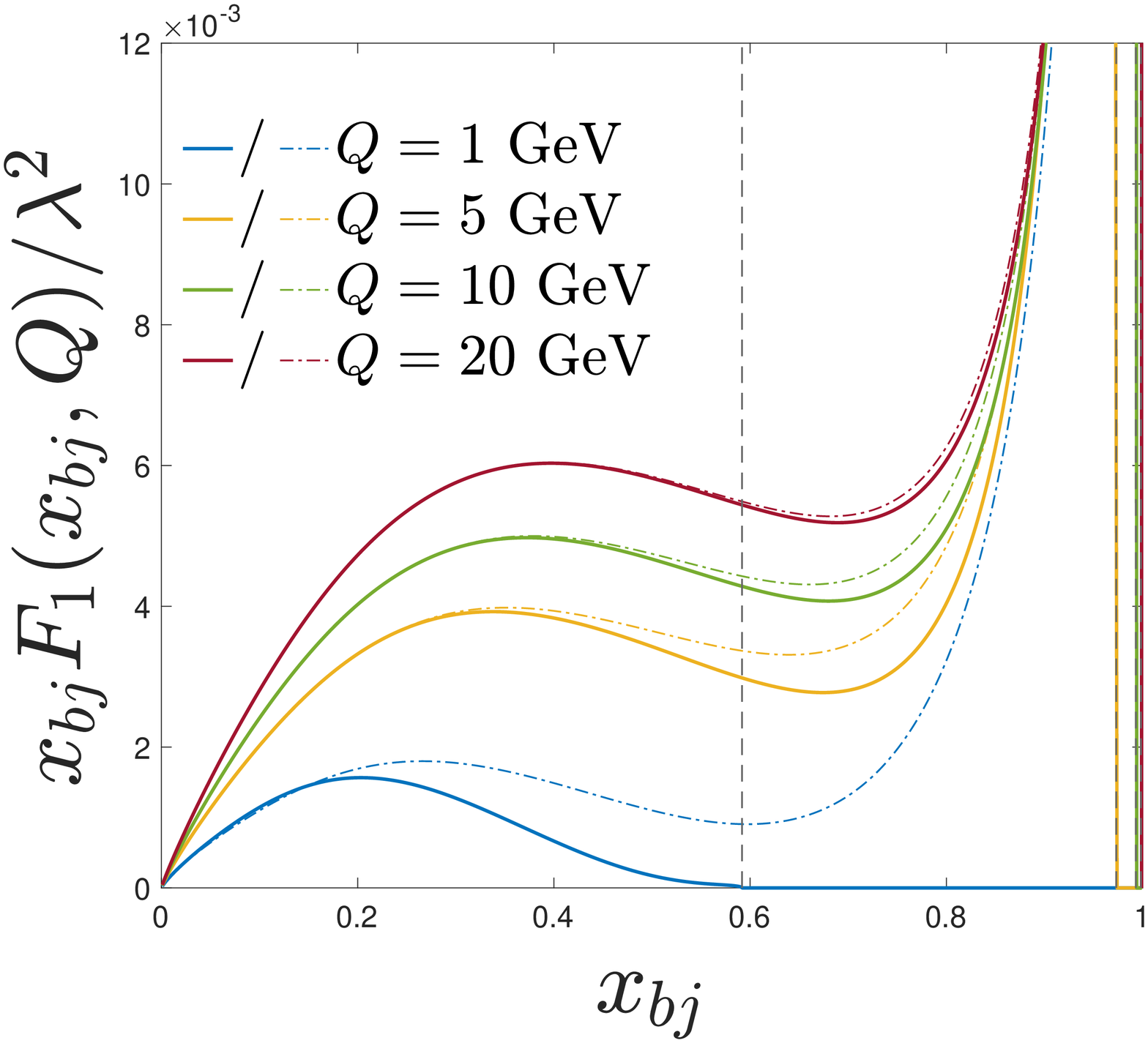}
    &
    \hspace{0.1cm}
    \includegraphics[scale=0.30]{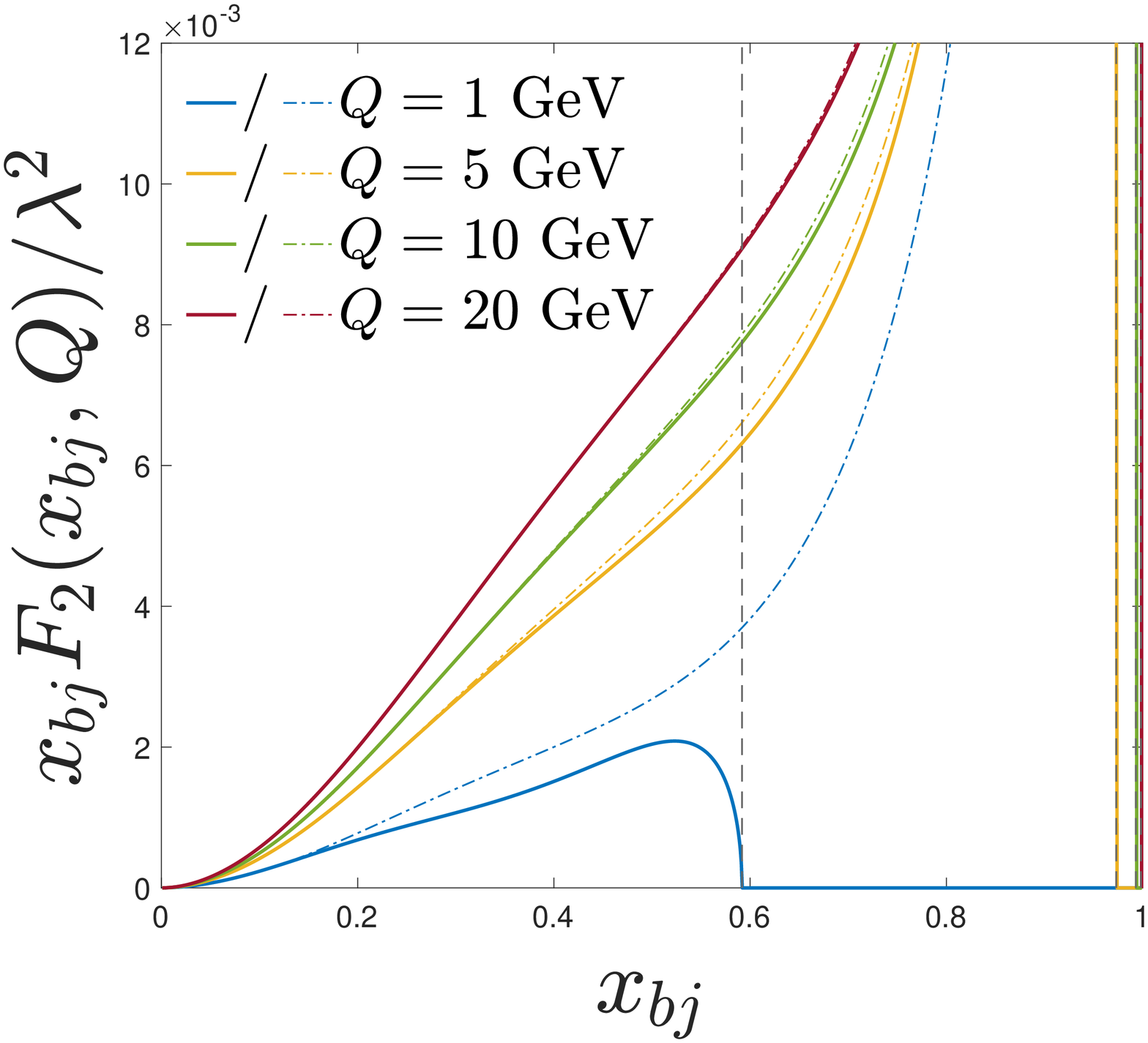}
  \\
  (a) & (b) 
  \end{tabular}
\caption{The same curves as in \fref{exactplots}, but now with the factorized expressions for $F_1$ and $F_2$ from \eref{finalfactorized} and \eref{finalfactorizedf2} also shown as the dot-dashed curves. The vertical dashed curves are the kinematical maximum (\eref{xmax}) corresponding to each value of $Q$.}
\label{f.factplots}
\end{figure}
If we ignore the running of $a_\lambda(\mu)$, as we will in all plots here, then $F_1$ and $F_2$ are exactly independent of the numerical value used for $\mu$, though the \emph{relative} contribution from each factor in \eref{finalfactorized} and \eref{finalfactorizedf2} changes with $\mu$. This is illustrated in \fref{F1mudep}. Note that while the value of $\mu$ is arbitrary, certain choices minimize or maximize the contribution from $\widehat{F}_{1,2}$. For instance, when $\mu$ is chosen to be equal to the hard scale $Q = 20$ GeV (dashed green curves) we recover the naive parton model prediction in the low $\xbj$ region. On the other hand, the non trivial partonic contribution is dominant when the renormalization scale is chosen to be of the order of the nonperturbative mass scales of the model (dotted yellow curves). In fact, from \eref{finP} we see that there exists a functional form of $\mu\sim \mathcal{O}(m)$ that makes the non trivial pdf contribution vanish, namely
\begin{equation}
    \mu(\xi)^2 = \Delta(\xi)^2e^{1 - \frac{\chi(\xi)^2}{\Delta(\xi)^2}} \, , 
\end{equation}
although in general one lacks knowledge of ``nonperturbative'' quantities like $\chi(\xi)$ and $\Delta(\xi)$. 
\begin{figure}[!h]
	\centering
	\begin{subfigure}[b]{0.48\textwidth}
		\includegraphics[width=\textwidth]{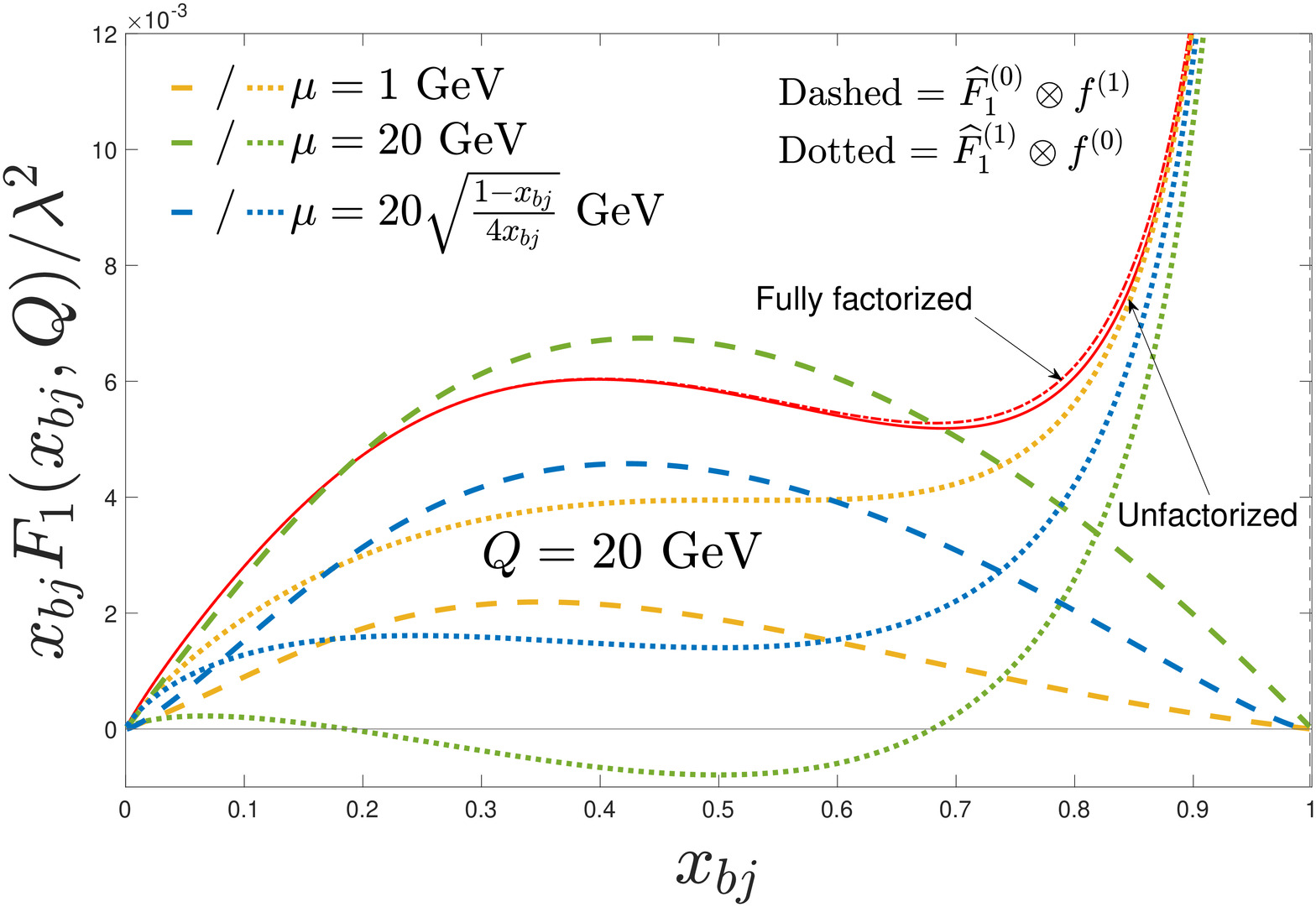}
		\caption{}
	\end{subfigure}
	\begin{subfigure}[b]{0.48\textwidth}
		\includegraphics[width=\textwidth]{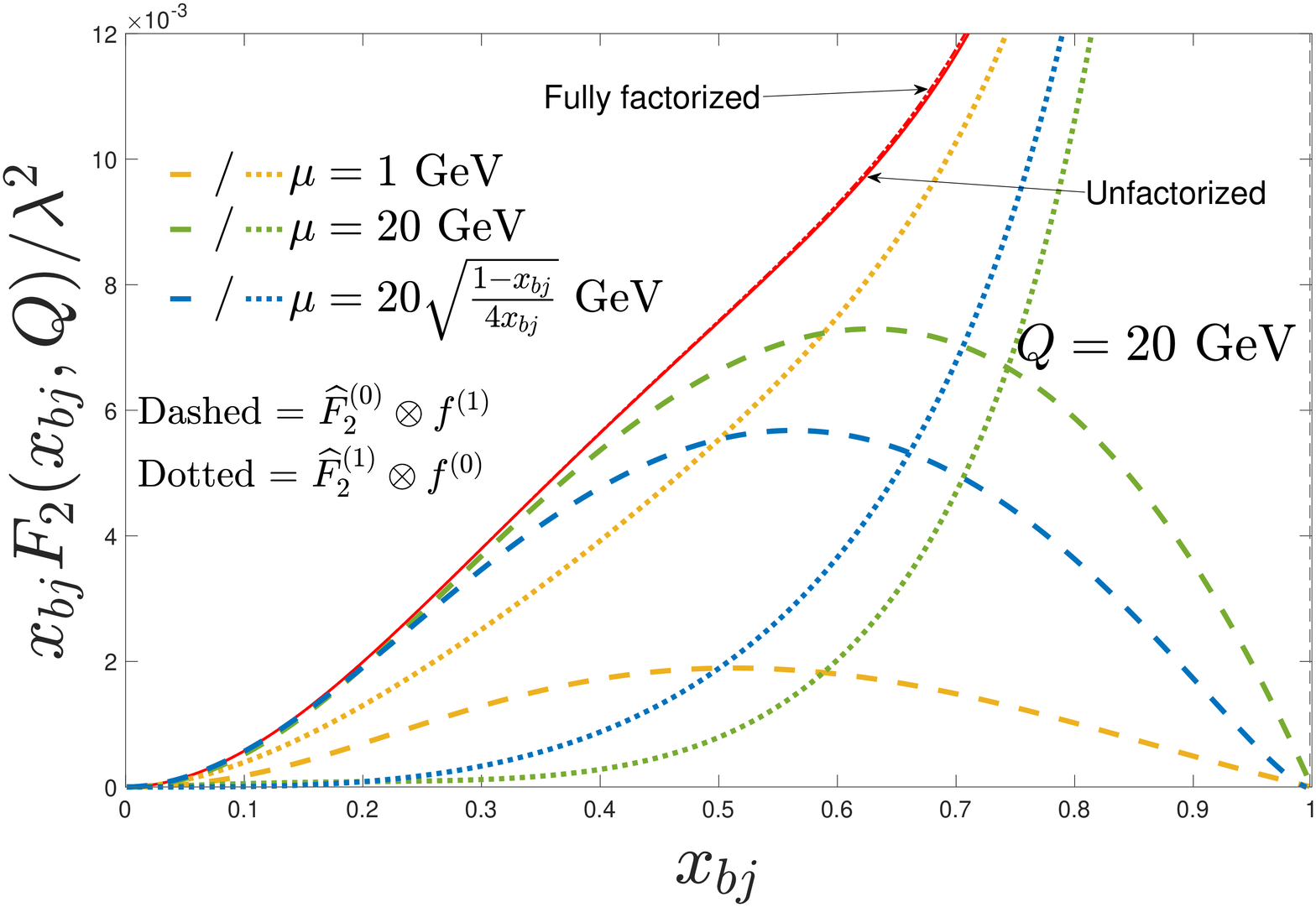}
		\caption{}
	\end{subfigure}
	\caption{The red solid and dot-dashed curves are the same $Q = 20$~GeV curves as those appearing in \fref{factplots} (a) and (b). Now we also show the terms $\hat{F}^{(0)}\otimes f^{(1)}$ (dashed) and $\hat{F}^{(1)}\otimes f^{(0)}$ (dotted), as in \eref{finalfact} (and similarly for $F_2$) for $\mu = 1$~GeV (yellow), $\mu = 20$~GeV (green) and $\mu = (20~\text{GeV}) \sqrt{(1-\xbj)/4\xbj}$ (blue). }
	\label{f.F1mudep}
\end{figure}
We will comment further on these plots in \sref{input}.

\section{TMD Factorization \& SIDIS}
\label{s.TMDandSIDIS}

So far, we have focused on collinear factorization and DIS, but we may regard the same set of graphs in \fref{basicmodel} as contributions to SIDIS and use TMD factorization. For this, we will take the observed final state particle $B$ to be the ``quark'' of the Yukawa theory in \sref{yukawa}. The TMD factorization formula for the hadronic tensor is then
\begin{align}
&{} W^{\mu \nu}(\xbj,Q,\zh,\T{P}{B}) \no
&{}= \sum_{ij} H_{ij}(\mu/Q;\mu)^{\mu \nu} \int \diff{^2 \T{k}{1}}{} \diff{^2 \T{k}{2}}{} f_{i/p}(\xbj,\T{k}{1};\mu) D_{B/j}(\zh, \zh\T{k}{2};\mu) \delta^{(2)} (\T{q}{} + \T{k}{1} - \T{k}{2})  + Y^{\mu \nu} + \order{\frac{m^2}{Q^2}} \no
&{}=\sum_{ij} H_{ij}(\mu/Q;\mu)^{\mu \nu} \int \frac{\diff[2]{\T{b}{}}}{(2 \pi)^2}
    ~ e^{-i\T{q}{}\cdot \T{b}{} }
    ~ \tilde{f}_{i/p}(\xbj,\T{b}{};\mu) 
    ~ \tilde{D}_{B/j}(\zh,\T{b}{};\mu) + Y^{\mu \nu} + \order{\frac{m^2}{Q^2}} \no
&{}=\sum_{ij} H_{ij}(\mu/Q;\mu)^{\mu \nu} \left[ f_{i/p}, D_{B/j} \right] + Y^{\mu \nu} + \order{\frac{m^2}{Q^2}} \, .
\label{e.hadrotens}
\end{align}
The form of the first term on the first line is well-known from TMD parton model treatments. The transverse momentum $\T{P}{B}$ is with respect to the Breit frame, while the momenta in the integrand are in the hadron frame. On the second line, the transverse momentum convolution has been re-expressed in transverse coordinate space, as is very common in treatments that implement TMD evolution in real QCD. We have used the transverse momentum variable
\begin{equation}
\T{q}{} \equiv -\frac{\T{P}{B}}{\zn} \, ,
\end{equation}
which is the hadron frame transverse momentum of the virtual photon. 
On the last line we have used a very standard bracket notation (e.g.~\cite{Tangerman:1994eh,Mulders:1995dh}) for transverse convolution integrals, especially in the context of hadron structure. The $Y^{\mu \nu}$ is the large-$\Tsc{q}{}$ correction term to account for $\Tsc{q}{} \approx Q$. The coordinate space correlation functions are 
\begin{align}
\label{e.fbspace}
\tilde{f}_{i/p}(\xi,\T{b}{};\mu) &{}= \int \diff{^2 \T{k}{}}{} 
e^{-i \T{k}{} \cdot \T{b}{} } f_{i/p}(\xi,\T{k}{};\mu) \\
\tilde{D}_{q/j}(\zeta,\T{b}{};\mu) &{}= \int \diff{^2 \T{k}{2}}{} 
e^{i \T{k}{} \cdot \T{b}{} } D_{q/i}( \zeta ,\zeta \T{k}{};\mu) \, .
\end{align}
Note the factor of $\zeta$ multiplying $\T{k}{}$ in the argument of the TMD FF. The hard factor is
\begin{equation}
H_{ij}(\mu/Q;\mu)^{\mu \nu} = \frac{\zh}{2} {\rm Tr} [\gamma^\nu \gamma^+ \gamma^\mu \gamma^-] |H(\mu/Q;\mu)|^2_{ij} \, , 
\end{equation}
where $|H(\mu/Q;\mu)|^2$ is a hard vertex factor that in perturbation theory takes the form
\begin{equation}
|H(\mu/Q;\mu)|^2_{ij} = 1 + \order{a_\lambda(\mu)} \, .
\end{equation}
The unpolarized quark structure functions follow from \eref{F12proj}, 
\begin{align}
&{}{\rm P}_1^{\mu\nu} H_{ij}(\mu/Q;\mu)_{\mu \nu} = H_1(\mu/Q;\mu) \delta_{iq} \delta_{jq} = 2\, \zh |H(\mu/Q;\mu)|^2_{ij} \, , \no
&{}{\rm P}_2^{\mu\nu} H_{ij}(\mu/Q;\mu)_{\mu \nu} = H_2(\mu/Q;\mu) \delta_{iq} \delta_{jq} = 4\, \zh \xbj |H(\mu/Q;\mu)|^2_{ij} \, .
\end{align}

The above is general, and applies equally to QCD and to the Yukawa theory. However, the expressions simplify considerably when we specialize to the low order Yukawa theory graphs of \fref{basicmodel}. Then, there is only one flavor of struck parton, so we may drop the sums over flavor indices. Also, there is only one particle flavor that can appear in the final state, namely the quark. So $B = q$ and we may drop the sum over $B$. The TMD FF has the trivial form in the current region of the $W$-term, 
\begin{equation}
\label{e.simpleFF}
D(\zh,\zh \T{k}{};\mu) = \delta(1-\zh) \delta^{(2)}(\zh \T{k}{}) \, .
\end{equation}
Therefore, we may integrate the cross section over $\zn$ to evaluate the $\delta$-function at $\zn = 1$. The cross section that we will consider, therefore, is actually
\begin{equation}
\label{e.zint}
\int \diff{\zn}{} \frac{\diff{\sigma_{\rm SIDIS}}{}}{\diff \xbj \diff{ y} \diff{\psi}\diff{\zn}\diff{^2 \T{P}{B}}} \, ,
\end{equation}
with $\zn$ approximated by $\zh$ in \eref{simpleFF}, as usual in a leading power approximation.
(The $Y$ term comes with an analogous $\delta$-function in the collinear FF that fixes the value of $\zn$.) Once \eref{simpleFF} is substituted into the second line of \eref{hadrotens}, two transverse momentum $\delta$-functions remain. Therefore, we may evaluate both the $\T{k}{1}$ and $\T{k}{2}$ integrals and the delta functions fix $\T{k}{2} = 0$ and $\T{k}{1} = -\T{q}{} = \T{P}{B}$. Finally, for the low order graphs considered here, 
\begin{align}
&{}{\rm P}_1^{\mu\nu} H_{ij}(\mu/Q;\mu)_{\mu \nu} \to {\rm P}_1^{\mu\nu} H_{\mu \nu} = H_1 = \frac{1}{2} \, , \no
&{}{\rm P}_2^{\mu\nu} H_{ij}(\mu/Q;\mu)_{\mu \nu} \to {\rm P}_2^{\mu\nu} H_{\mu \nu} = H_2 = \xbj \, .
\end{align}
Therefore, \eref{hadrotens} is
\begin{align}
W^{\mu \nu}(\xbj,Q,\T{k}{}) &{}= H^{\mu \nu} f_{q/p}(\xbj,\T{k}{};\mu) + Y^{\mu \nu} + \order{\frac{m^2}{Q^2}} \no
&{}= H^{\mu \nu} \int \frac{\diff[2]{\T{b}{}}}{(2 \pi)^2}
    ~ e^{i\T{k}{}\cdot \T{b}{} }
    ~ \tilde{f}_{q/p}(\xbj,\T{b}{};\mu) 
     + Y^{\mu \nu} + \order{\frac{m^2}{Q^2}} \, .
\label{e.hadrotensapp}
\end{align}
Or, 
\begin{align}
F_1(\xbj,Q,\T{k}{}) &= \frac{1}{2} f_{q/p}(\xbj,\T{k}{};\mu) + Y_1 + \order{\frac{m^2}{Q^2}} = \frac{1}{2} \int \frac{\diff[2]{\T{b}{}}}{(2 \pi)^2}
    ~ e^{i\T{k}{}\cdot \T{b}{} }
    ~ \tilde{f}_{q/p}(\xbj,\T{b}{};\mu) 
     + Y_{1} + \order{\frac{m^2}{Q^2}} \, , \label{e.hadrotensappF1} \\
F_2(\xbj,Q,\T{k}{}) &= \xbj f_{q/p}(\xbj,\T{k}{};\mu) + Y_2 + \order{\frac{m^2}{Q^2}} = \xbj \int \frac{\diff[2]{\T{b}{}}}{(2 \pi)^2}
    ~ e^{i\T{k}{}\cdot \T{b}{} }
    ~ \tilde{f}_{q/p}(\xbj,\T{b}{};\mu) 
     + Y_{2} + \order{\frac{m^2}{Q^2}} \, .
\label{e.hadrotensappF2}
\end{align}
We have used the shorthand $\T{k}{} = \T{P}{B} = \T{k}{1}$ to simplify notation. The absence of a $\zn$ argument on the left sides of \erefs{hadrotensapp}{hadrotensappF1} indicates that these are the TMD observables \emph{after} the $\zn$-integral in \eref{zint}. That is, it is the integral $\int \diff{\zn}{}/(4 \zn)$ of the SIDIS hadronic tensor. 
 
Is also useful to work other standard linear combinations of the two structure functions $F_1$ and $F_2$ like the longitudinal unpolarized structure function $F_L$ defined below
\begin{equation}
    F_L(\xbj,\T{k}{},Q)\equiv \left(1 + \frac{4\tarmass^2\xbj^2}{Q^2}\right)F_2(\xbj,\T{k}{},Q) - 2\xbj F_1(\xbj,\T{k}{},Q).
    \label{e.exactTMD_FL}
\end{equation}
which vanishes in accordance with the Callan-Gross relation.

Since we have already obtained the TMD pdfs in \eref{pdfdef} when we set up collinear factorization, explicit expressions for the W-term structure functions follow automatically from \erefs{hadrotensappF1}{hadrotensappF2}. Indeed, we already have \eref{TMDfact} for the first term of \eref{hadrotensappF1}. For $F_2$, the same expression applies but multiplied by $2 \xbj$. 

For $Y_1$, we need the second term of \eref{sub2}, which is the integrand of \eref{FO2} minus that of 
\eref{subtractterm},
\begin{equation}
Y_1 = \frac{a_\lambda(\mu)}{2 \pi} \parz{\frac{(1 - \xbj)(1-\kappa(\xbj))}{ \kappa(\xbj) \Tscsq{k}{}} 
			       - \frac{\xbj (2 \xbj^2 - 6 \xbj + 3) }{Q^2 (1 - \xbj)^2 \kappa(\xbj)}} \, ,
\label{e.F1_Yterm}
\end{equation}
and for the $F_2$ $Y$-term, 
\begin{equation}
Y_2 = \frac{a_\lambda(\mu)}{\pi} \xbj\parz{\frac{(1 - \xbj)(1-\kappa(\xbj))}{ \kappa(\xbj) \Tscsq{k}{}} 
			       - \frac{\xbj (6 \xbj^2 - 10 \xbj + 3) }{Q^2 (1 - \xbj)^2 \kappa(\xbj)}} \, .
			       \label{e.F2_Yterm}
\end{equation}

The plots in \fref{TMD_WY_vs_unfact} show the interplay between the W and Y term in all three structure functions. 
They show that the role of the Y term is necessary for describing the large $\Tsc{k}{}$ region independently of $Q$ and especially for regions of higher $\xbj$. 

The term in \eref{asymp} that is used to form the $Y$-term is an important ingredient in treatments of transverse momentum dependence. It is itself expressible as a version of collinear factorization,
\begin{equation}
\TII \TI F_{1}(\xbj,Q,\T{k}{}) = 
F^{\text{ASY}}_{1}(\xbj,Q,\T{k}{}) = \frac{a_\lambda(\mu)}{2 \pi} \frac{(1 - \xbj)}{\Tscsq{k}{}}  = \sum_j C^{F_1}_{q/j}(k_T) \otimes f_{j/p} \, ,
\end{equation}
where $C^{F_1}_{q/j}(k_T)$ is a hard coefficient that depends on $\Tsc{k}{}$. In \eref{asymp}, it is only the zeroth order $p$-in-$p$ pdf from \eref{PinP} that enters, so the factorization is rather trivial. In the literature on transverse momentum in QCD, it is often called the ``asymptotic term'' because it describes the limit where $\Tsc{q}{}/Q$ is a small but fixed ratio and $Q \to \infty$, so we label it with an $\text{ASY}$ superscript. Of course, the TMD pdf has its own asymptotic term, which we can read off from \erefs{hadrotensappF1}{hadrotensappF2},
\begin{equation}
f^{\text{ASY}}_{q/p}(\xbj,\T{k}{};\mu) = \sum_j C^{f_{q/p}}_{q/j}(k_T) \otimes f_{j/p}
 = \frac{a_\lambda(\mu)}{\pi} \frac{(1 - \xbj)}{\Tscsq{k}{}}  \, . \label{e.asym_pdf}
\end{equation}

The longitudinal TMD structure function $F_L(\xbj,\Tsc{k}{},Q)$ is subleading at small $\Tsc{k}{}$ relative to $F_2$ and $F_1$, so its dominant contribution is from the large $\Tsc{k}{}$ region. Thus, it is mostly described by its Y term contribution. In the TMD factorized version of the hadronic structure tensor, a non-zero W term contribution only arises from the subleading mass term in the projector in 
\eref{exactTMD_FL}, 
\begin{equation}
    F_L^{W}\left(\xbj,\Tsc{k}{},Q\right) = 4\frac{\tarmass^2\xbj^2}{Q^2}F_2^{W}\left(\xbj,\Tsc{k}{},Q\right),
\end{equation}
while
\begin{equation}
    F_2^W\left(\xbj,\Tsc{k}{},Q\right) - 2\xbj F_1^W\left(\xbj,\Tsc{k}{},Q\right) = 0,
\end{equation}
in line with the Callan-Gross relation in the naive parton model picture. 
\begin{figure}[!h]
	\centering
	\begin{subfigure}[b]{0.38\textwidth}
		\includegraphics[width=\textwidth]{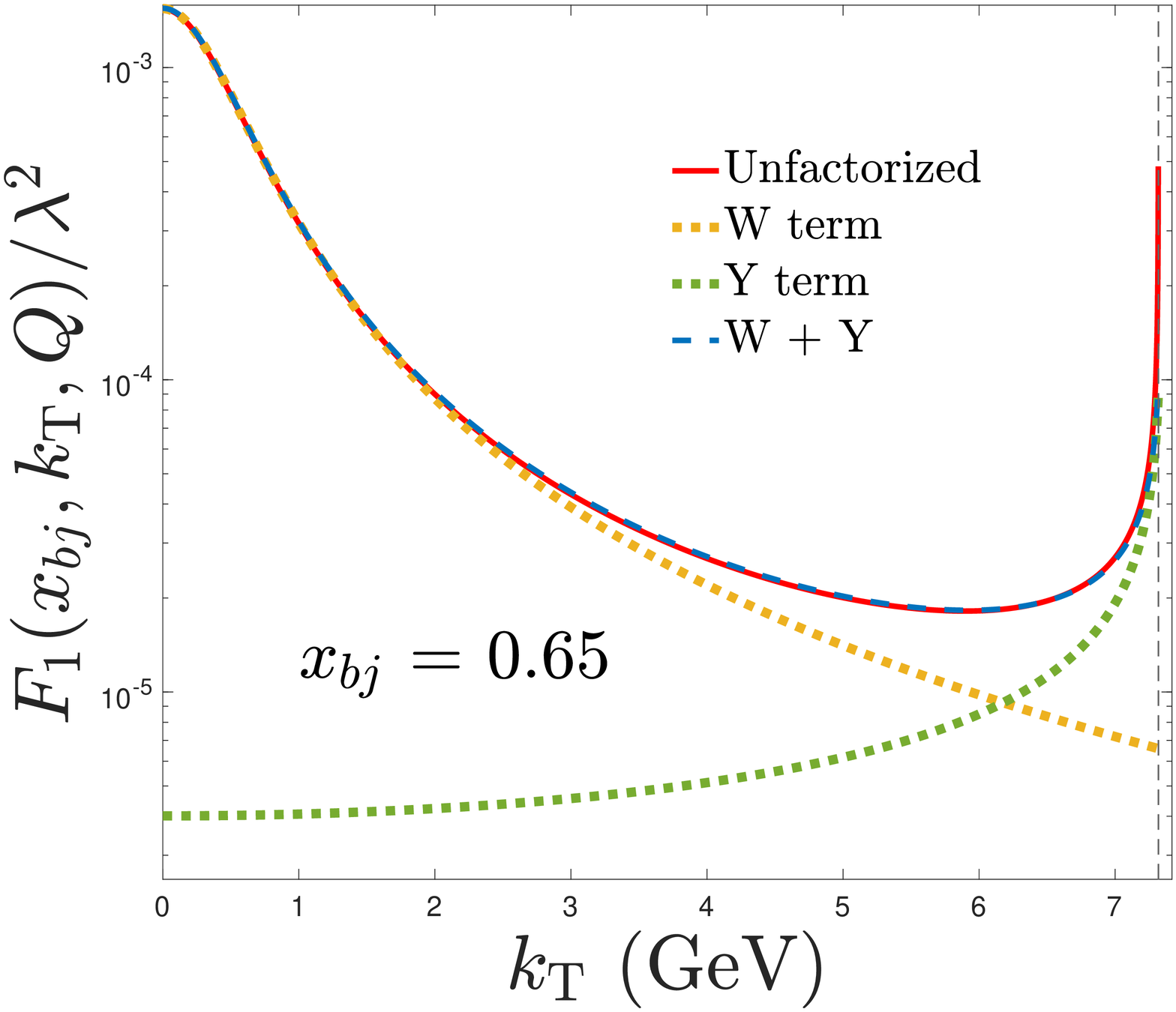}
		\caption{}
		\label{f.TMDF1_Q20_x065}
	\end{subfigure}
	\hspace{1mm}
	\begin{subfigure}[b]{0.38\textwidth}
		\includegraphics[width=\textwidth]{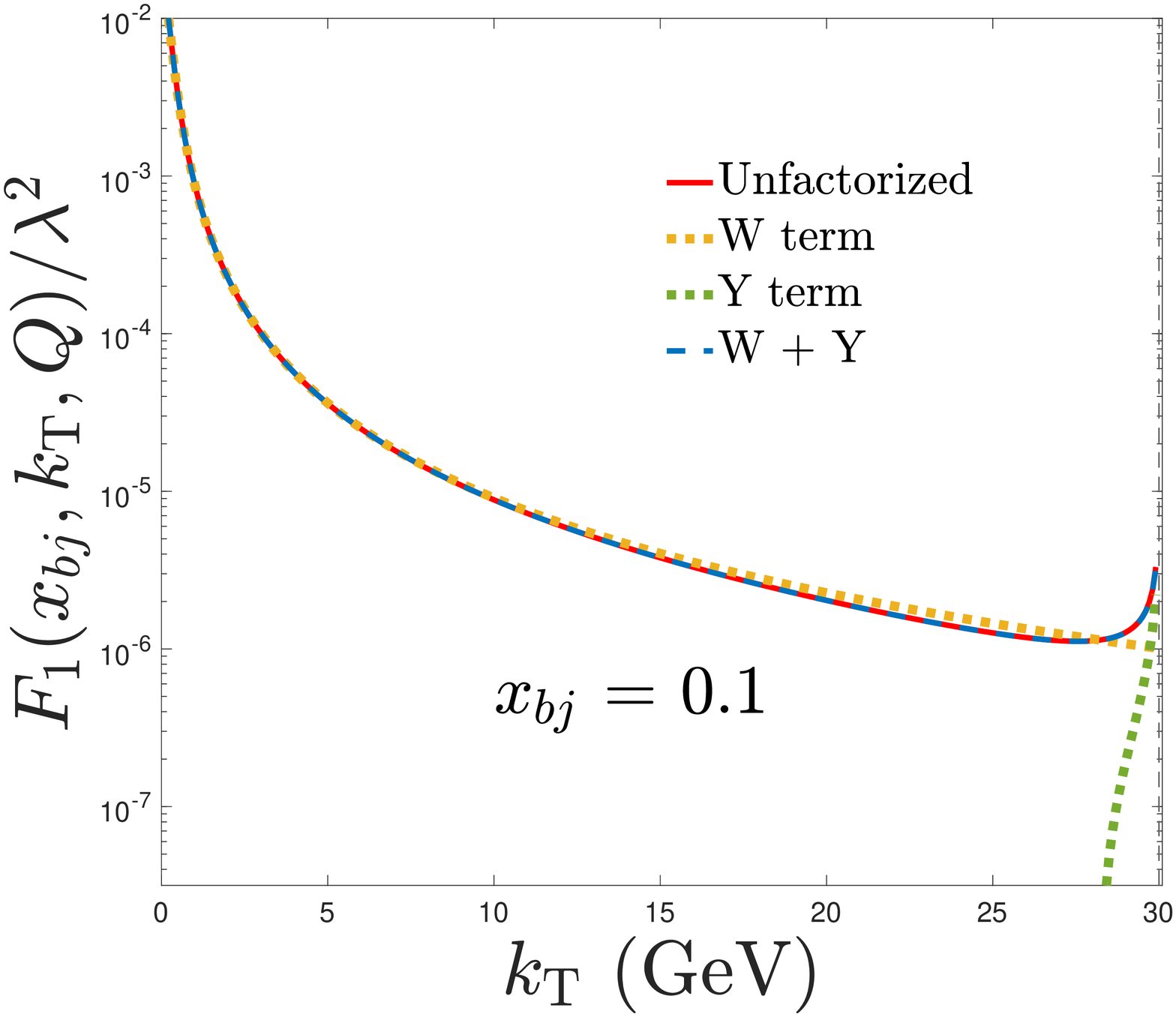}
		\caption{}
		\label{f.TMDF1_Q20_x0_10}
	\end{subfigure}
	\hspace{1mm}
	\begin{subfigure}[b]{0.38\textwidth}
		\includegraphics[width=\textwidth]{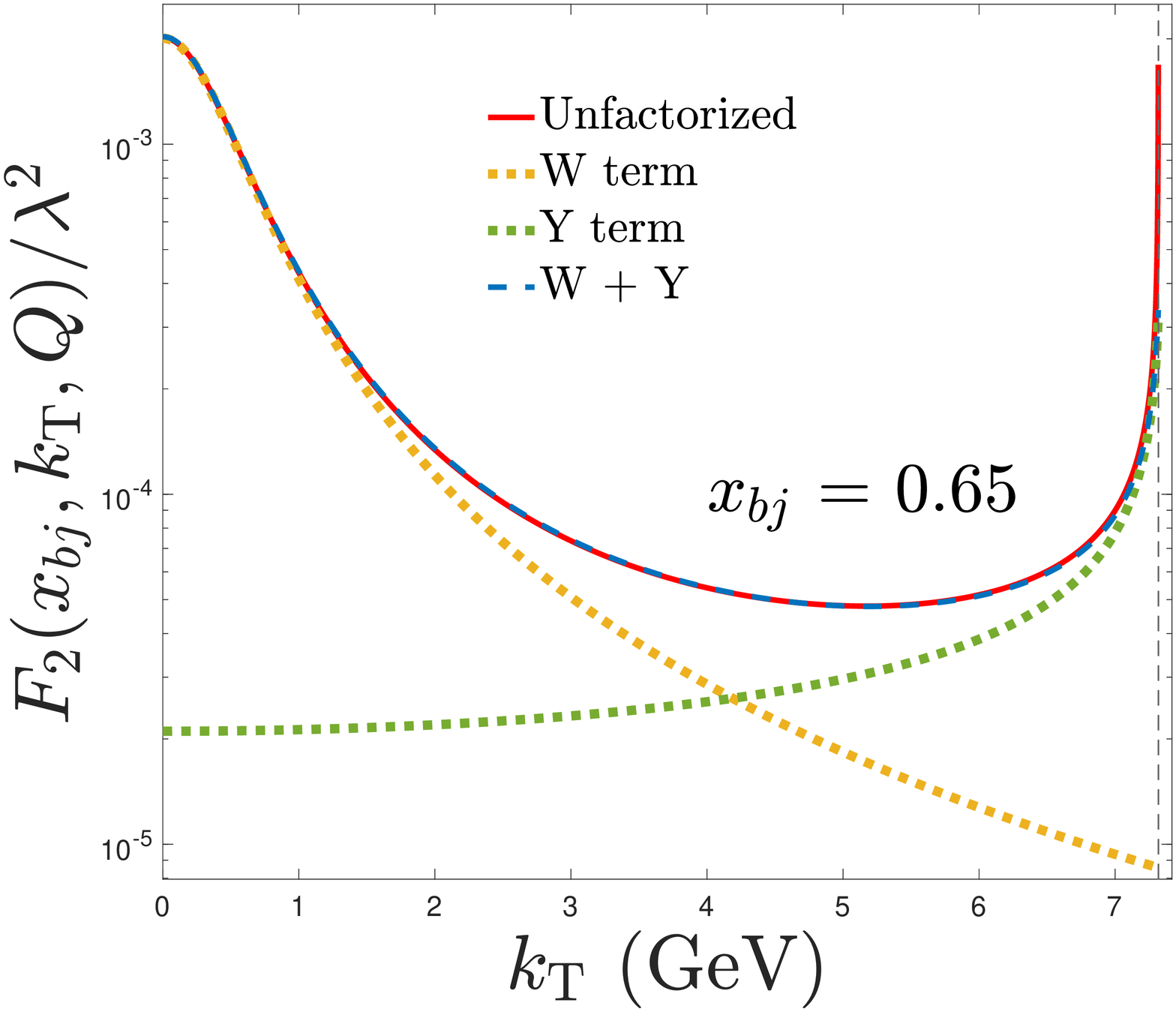}
		\caption{}
		\label{f.TMDF2_Q20_x065}
	\end{subfigure}
	\hspace{1mm}
	\begin{subfigure}[b]{0.38\textwidth}
		\includegraphics[width=\textwidth]{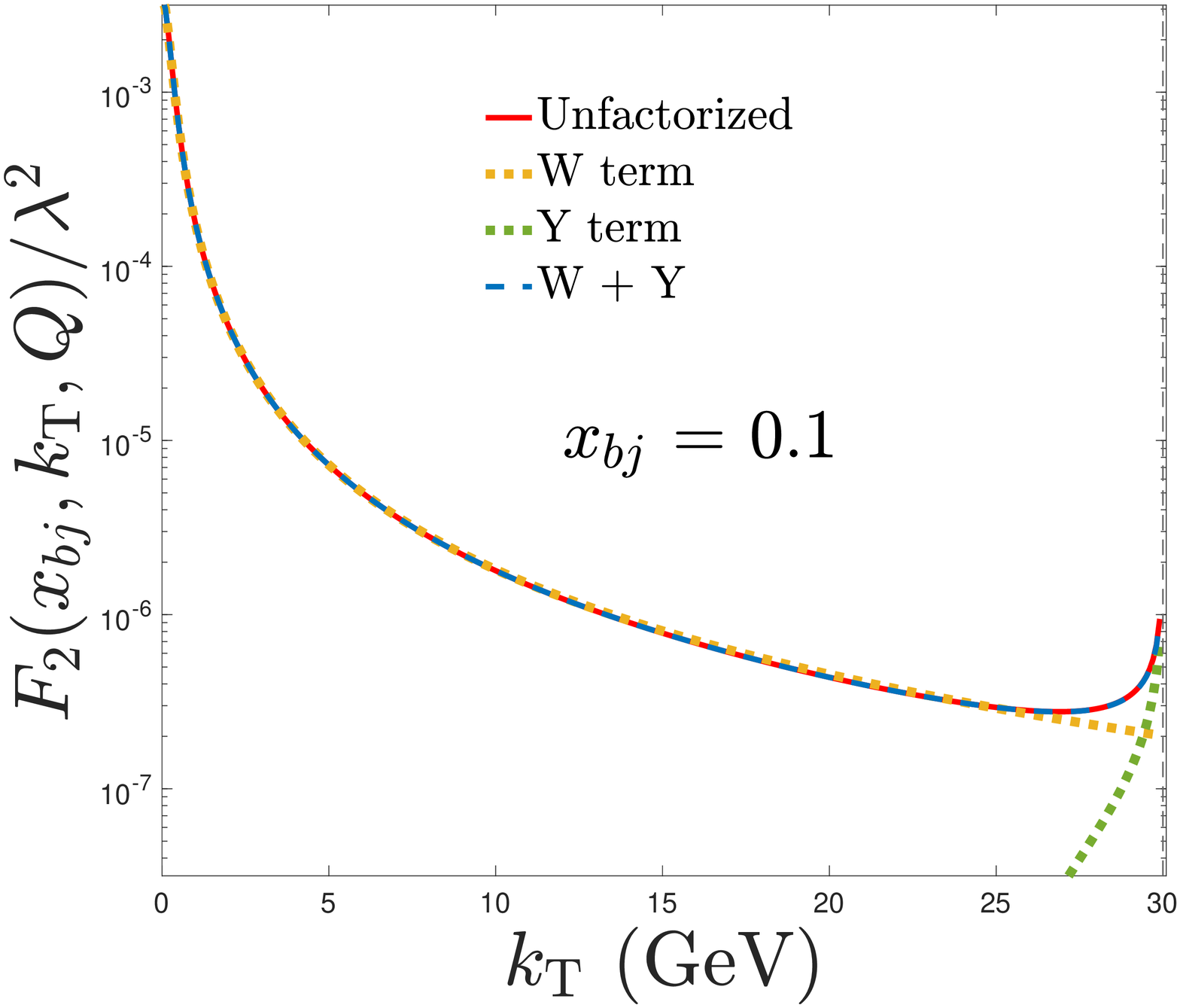}
		\caption{}
		\label{f.TMDF2_Q20_x0_10}
	\end{subfigure}
	\hspace{1mm}
	\begin{subfigure}[b]{0.38\textwidth}
		\includegraphics[width=\textwidth]{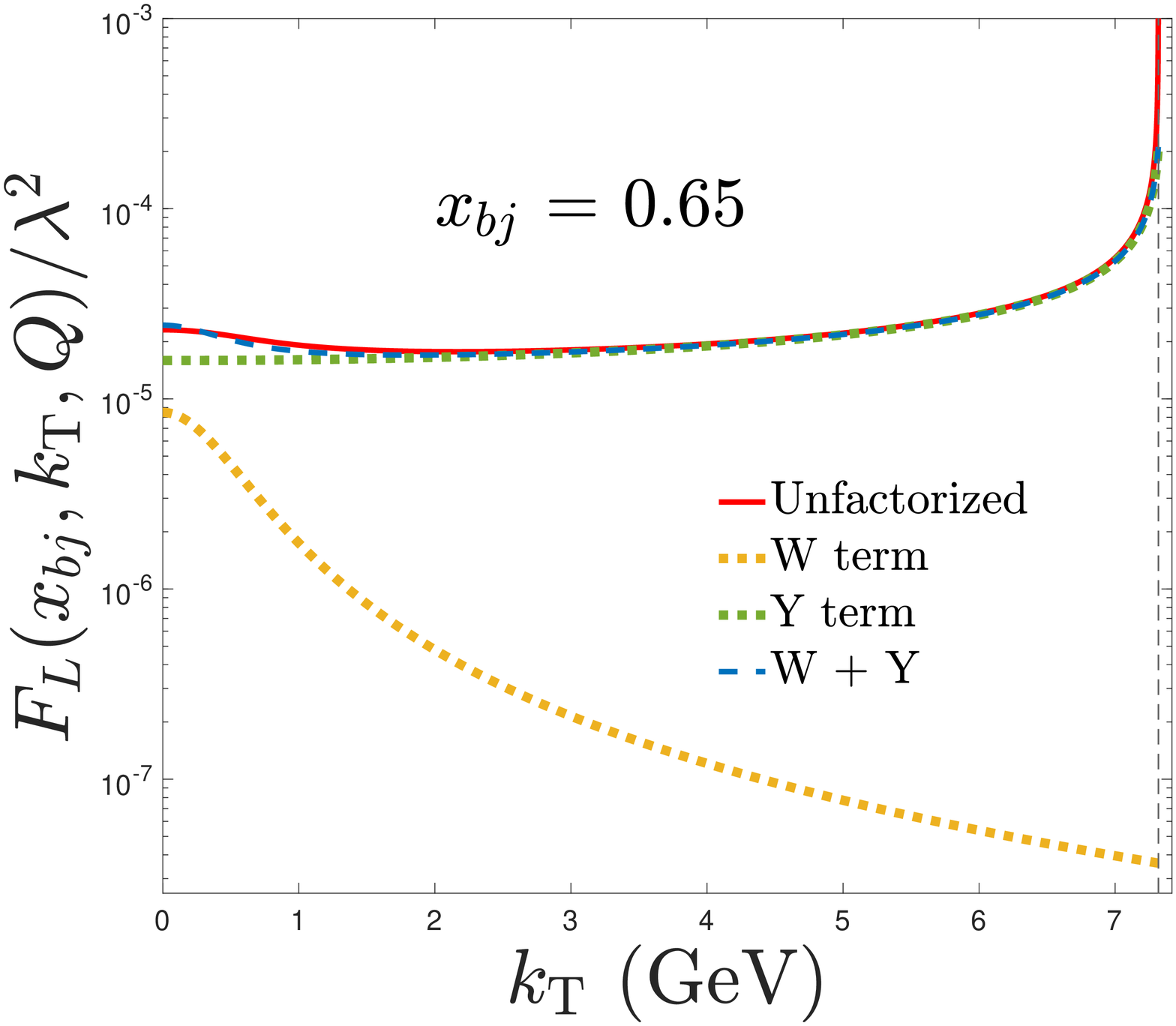}
		\caption{}
		\label{f.TMDFL_Q20_x065}
	\end{subfigure}
	\hspace{1mm}
	\begin{subfigure}[b]{0.38\textwidth}
		\includegraphics[width=\textwidth]{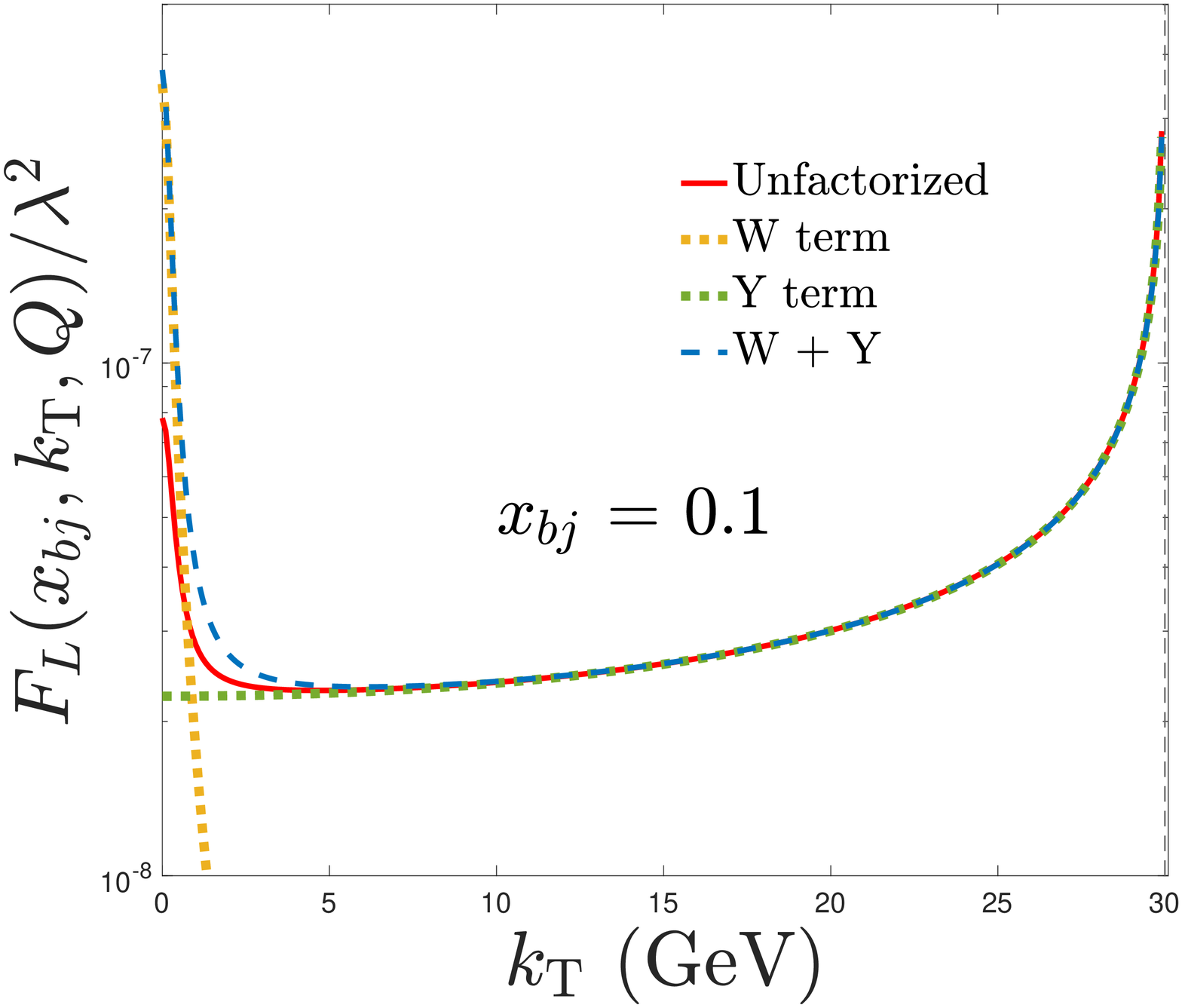}
		\caption{}
		\label{f.TMDFL_Q20_x0_10}
	\end{subfigure}
	\caption{The TMD structure functions $F_1(\xbj,\Tsc{k}{},Q)$ (a-b), $F_2(\xbj,\Tsc{k}{},Q)$ (c-d) and their linear combination $F_L(\xbj,\Tsc{k}{},Q)$ (e-f) as defined in equation \eref{exactTMD_FL} are shown for a specific value of $\xbj = 0.65$ (left) and $\xbj = 0.1$ (right). The dotted yellow and green curves show the contributions of the W and Y terms, respectively, while their sum (dashed blue curve), defined in \eref{hadrotensappF1} and \eref{hadrotensappF2}, approximates the unfactorized (solid red) curve. The dashed grey line indicates the maximum $\Tsc{k}{}$ that is kinematically allowed in the exact theory. The choice of the masses is still $\tarmass = \mgluon = 1$ GeV and $\mquark = 0.3$ GeV with a hard scale of $Q = 20$ GeV.}
	\label{f.TMD_WY_vs_unfact}
\end{figure}

Another way to illustrate the importance of both the W and Y terms in factorization is to consider how each contributes in the reduction to standard collinear factorization when integrating over all $\Tsc{q}{}$ to get the standard integrated structure functions $F_1(\xbj,Q)$ and $F_2(\xbj,Q)$. 

Past phenomenological approaches to TMD factorization, particularly in the context of hadron structure studies, ignore the role of the $Y$-term. Thus, instead of 
\begin{equation}
\int \frac{\diff{\zn}{} \diff{^2\T{P}{B}}{}}{4 \zn} F_{1,2}(\xbj ,Q^2,\zh,\T{P}{B}) = F_{1,2}(\xbj ,Q^2) \, ,
\end{equation}
one uses 
\begin{equation}
\int \frac{\diff{\zn}{} \diff{^2\T{P}{B}}{}}{4 \zn} F^W_{1,2}(\xbj ,Q^2,\zh,\T{P}{B}) \stackrel{??}{=} F_{1,2}(\xbj ,Q^2) \, , \label{e.gpmapprox}
\end{equation}
where the ``$??$'' on the equal sign is to emphasize that this is a type of conjectured approximation. This makes manipulating integrals of TMD functions very simple. Equation~\eqref{e.hadrotens} for a specific structure function is
\begin{equation}
F_{1,2}(\xbj,Q,\zh,\T{P}{B}) = \sum_{ij} \widehat{F}^{ij}_{1,2} \int \diff{^2 \T{k}{1}}{} \diff{^2 \T{k}{2}}{} f_{i/p}(\xbj,\T{k}{1}) D_{B/j}(\zh, \zh\T{k}{2}) \delta^{(2)} (\T{q}{} + \T{k}{1} - \T{k}{2})  \, , 
\label{e.hadrotens2}
\end{equation}
where $\widehat{F}^{ij}_{1,2}$ is the result of projecting with 
\eref{F12proj} on $H^{\mu \nu}$. (Since \eref{hadrotens2} is a simplified parton model version of the factorization theorem, we have dropped dependence on auxiliary variables like $\mu$ in the pdf and ff.) Integrating as in \eref{gpmapprox} and evaluating the $\delta$-functions gives 
\begin{equation}
F_{1,2}(\xbj ,Q^2) \stackrel{??}{=} \frac{1}{4} \sum_{ij} \widehat{F}^{ij}_{1,2} \parz{\int \diff{^2 \T{k}{1}}{} f_{i/p}(\xbj,\T{k}{1}) } \parz{\int \diff{\zh} \int \diff{^2 \T{k}{2}}{} \zh D_{B/j}(\zh, \zh\T{k}{2})} \, , 
\end{equation}
where we have used $\zn \approx \zh$. Then, using the parton model relations 
\begin{align}
& \int \diff{^2 \T{k}{1}}{} f_{i/p}(\xbj,\T{k}{1}) \stackrel{??}{=} f_{i/p}(\xbj) \label{e.gpmpdf} \\
& \int \diff{\zh} \int \diff{^2 \T{k}{2}}{} \zh D_{B/j}(\zh, \zh\T{k}{2}) \stackrel{??}{=} 1 \, 
\label{e.gpmff}
\end{align}
gives the naive parton model expectation,
\begin{equation}
F_{1,2}(\xbj ,Q^2) \stackrel{??}{=} \frac{1}{4} \sum_{ij} \widehat{F}^{ij}_{1,2}\; f_{i/p}(\xbj) \, . \label{e.gpm}
\end{equation}
See Eqs.(59)-(60) 
of \cite{Boer:2003cm} for an example of an application of the approximations in \erefs{gpmapprox}{gpm}. The question marks over the equal signs in \erefs{gpmpdf}{gpmff} are a reminder that in theories that require renormalization, like QCD and the Yukawa theory, the integrals are UV divergent. The equalities are only strictly valid in a literal probability interpretation for the pdf and ff. The appearance of UV divergences is an artifact of integrating transverse momentum outside the region where the small transverse momentum approximations hold. In other words, they come from neglecting the Y-term. Because they are in line with expectations from a naive parton model, but extended to TMD functions~\cite{Gardiner:1970wy}, the set of approximations conjectured in \erefs{gpmapprox}{gpm} are sometimes called the generalized parton model~\cite{Bomhof:2007wd,Gamberg:2010tj} (GPM). It continues to be common for the GPM and its extensions to be used in applications to hadron structure phenomenology. See, for example,
Refs.~\cite{Anselmino:2013lza,Signori:2013mda,Anselmino:2016uie}
 and more recently 
Refs.~\cite{Cammarota:2020qcw,Aslan:2022kmd}.
One may examine the typical size of the effect of the GPM approximation in the special case of the Yukawa theory of \sref{yukawa} by using the above results of this section. Then, \eref{gpmapprox} is simply what is obtained from the W-term when integrating \erefs{hadrotensappF1}{hadrotensappF2} over all kinematically accessible transverse momentum to get the full integrated cross section. Doing this integral gives
\begin{equation}
F_{1,2}(\xbj,Q;\mu) = F_{1,2}^W(\xbj,Q;\mu) +F_{1,2}^Y(\xbj,Q;\mu) + \order{\frac{m^2}{Q^2}} \, . \label{e.wplusy}
\end{equation}
where 
\begin{align}
        F_{1,2}^W(\xbj,Q;\mu) & \equiv H_{1,2}f^c(\xbj;\mu;k_c = k_m) \, , \label{e.Wintegrated} \\
        F_{1,2}^Y(\xbj,Q;\mu) & \equiv \pi\int_0^{k_m^2}\diff{\Tsc{k^2}{1}}{}Y_{1,2} \, . \label{e.Yintegrated} 
\end{align}
Equation~\eqref{e.Wintegrated} is just \eref{gpm} specialized to the Yukawa theory example and using the cutoff definition of the collinear pdf from \eref{cuttoff_def}. The full collinear factorization result in \erefs{finalfactorized}{finalfactorizedf2} is the result of dropping the power-suppressed terms in \eref{wplusy}, 
\begin{equation}
F_{1,2}^{\text{Full Fact.}}(\xbj,Q;\mu) = F_{1,2}^W(\xbj,Q;\mu) +F_{1,2}^Y(\xbj,Q;\mu) \, ,
\end{equation}
while the GPM approximation of \eref{gpmapprox} is recovered if we drop both the $\order{\frac{m^2}{Q^2}}$ and the Y-term $F_{1,2}^Y(\xbj,Q;\mu)$,
\begin{equation}
F_{1,2}^{\text{GPM}}(\xbj,Q;\mu) = F_{1,2}^W(\xbj,Q;\mu) \, ,
\end{equation}
The validity of the GPM, as compared with with full factorization, can be tested by looking at the relative contributions from $F_{1,2}^W(\xbj,Q;\mu)$, $F_{1,2}^Y(\xbj,Q;\mu)$, and $F_{1,2}$. Examples, are shown 
in \fref{F12_WY_term_ratio} and \fref{F12WY_ratio_mp1_ms1_mq03_x}. 

The statement that the W term yields the most contribution would imply its ratio with the unfactorized expression to be in the neighborhood of 1 independently of $\xbj$ for sufficiently large $Q$. In our examples in \fref{F12_WY_term_ratio}, where the hard scale has already been fixed to a value much larger than the other nonperturbative mass scales, only the sum of both W and Y terms well approximates the unfactorized structure functions while the small transverse momentum contribution is very rapidly dominated by its large transverse momentum counterpart already at  $\xbj\sim 0.5$ for $F_1$ and even earlier for $F_2$. Similarly, in \fref{F12WY_ratio_mp1_ms1_mq03_x}, the relative contributions for fixed $\xbj$ over an extended range of $Q$ are shown to satisfy the factorization statement only when both of them are accounted for. In fact, for relatively small $\xbj$ the small transverse momentum contribution is still a relatively good approximation to the full unapproximated $F_1$, although less so for $F_2$, which already suffers from the neglected $Y$ term contribution, as it is evident in the example with $\xbj = 0.3$ even at extremely large hard scales. However, for increasingly larger $\xbj$ the situation degrades even more rapidly and only the correct prescription is able to approximate the unfactorized expressions.   
\begin{figure}[!h]
	\centering
	\begin{subfigure}[b]{0.47\textwidth}
		\includegraphics[width=\textwidth]{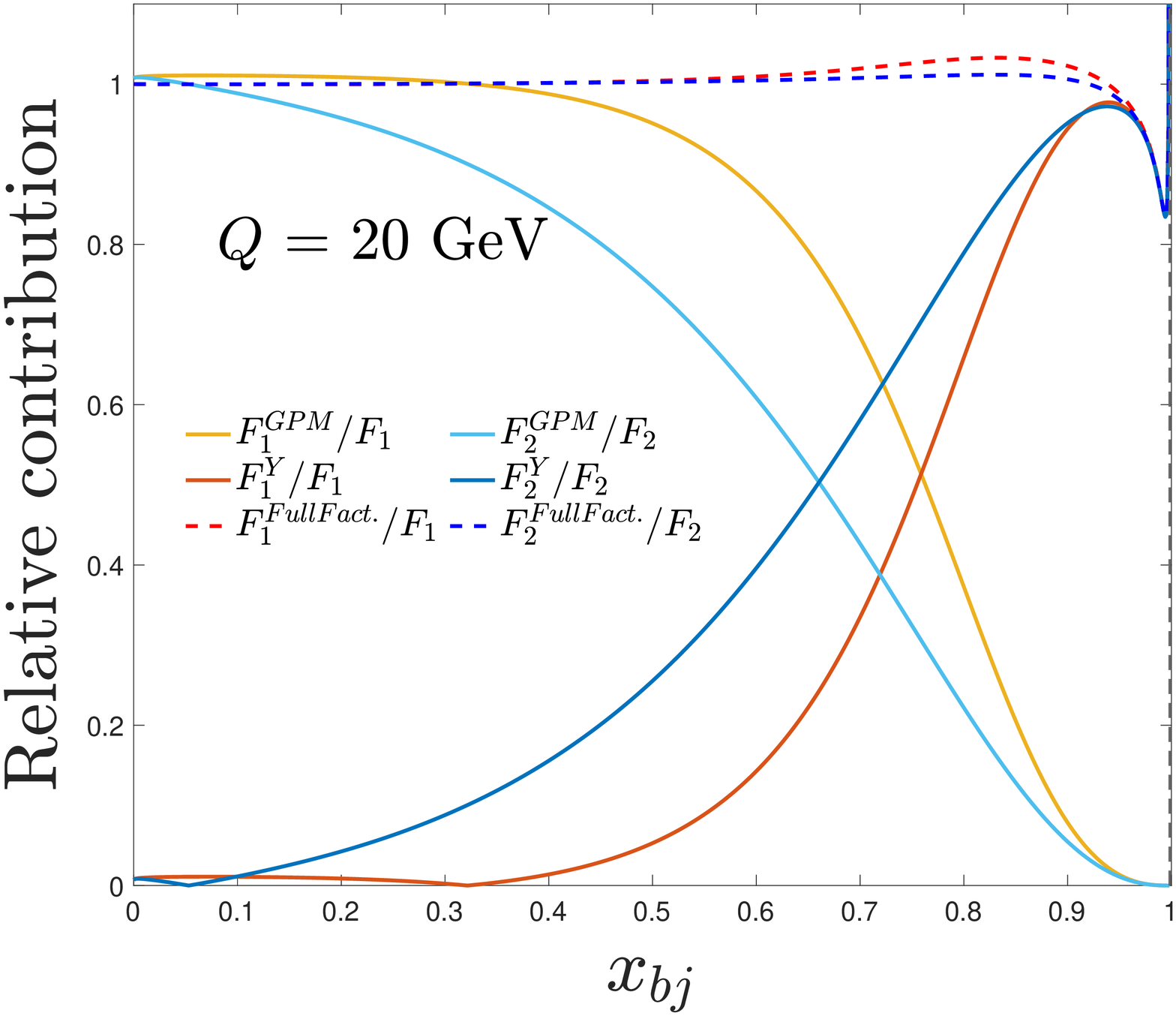}
		\caption{}
	\end{subfigure}
	\begin{subfigure}[b]{0.47\textwidth}
		\includegraphics[width=\textwidth]{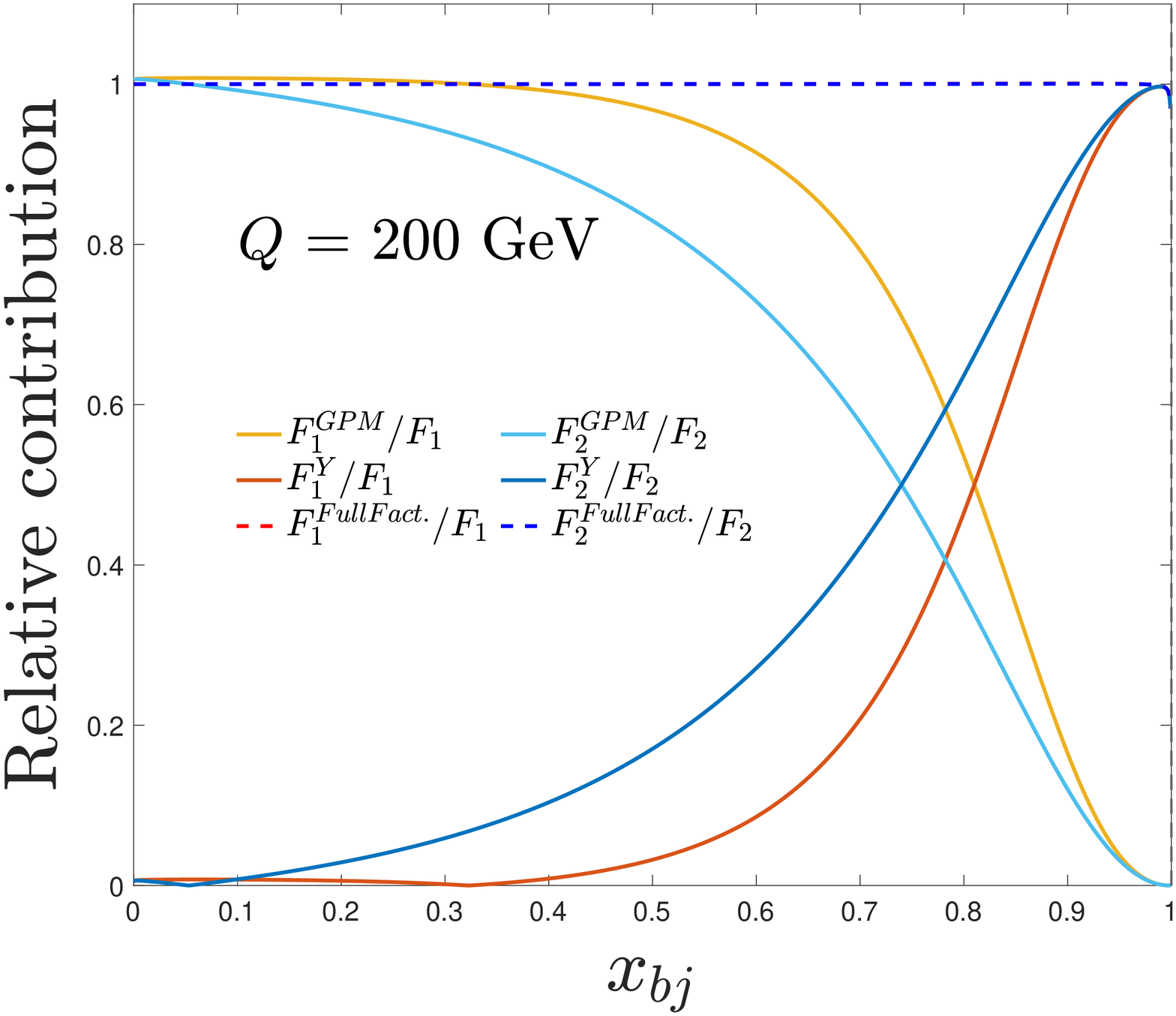}
		\caption{}
	\end{subfigure}
	\caption{Comparison between small and large transverse momentum contributions to the collinear structure functions $F_1$ and $F_2$. The ratios of the W and Y terms against the unfactorized structure functions $F_1$ and $F_2$  are shown for two values of the hard scale $Q = 20$ GeV (a) and $Q = 200$ GeV (b). The dashed lines show the better approximation given by their sum for both $F_1$ (dashed red) and $F_2$ (dashed blue). }
	\label{f.F12_WY_term_ratio}
\end{figure}

\begin{figure}[!h]
	\centering
	\begin{subfigure}[b]{0.45\textwidth}
		\includegraphics[width=\textwidth]{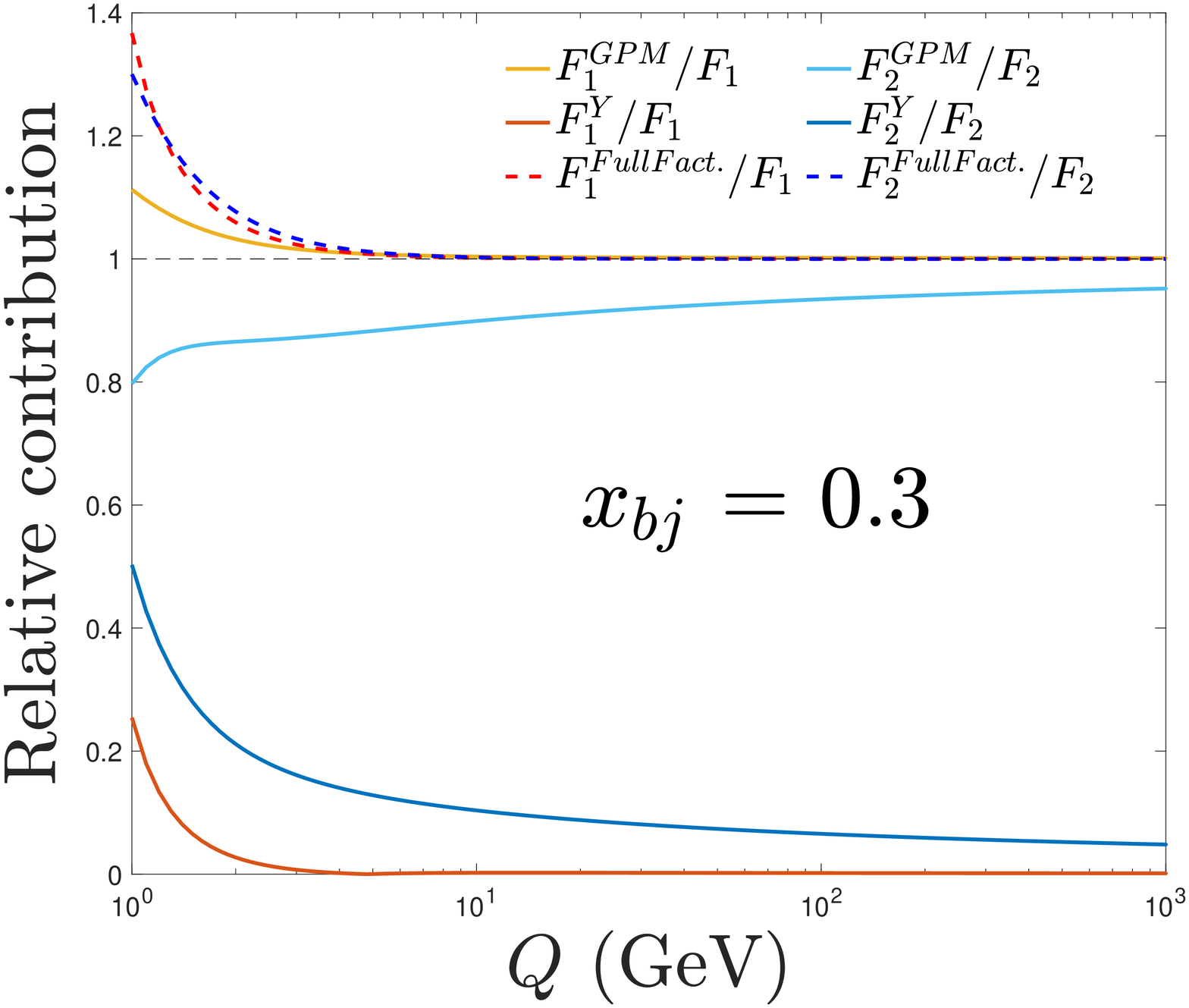}
		\caption{}
	\end{subfigure}
	\begin{subfigure}[b]{0.45\textwidth}
		\includegraphics[width=\textwidth]{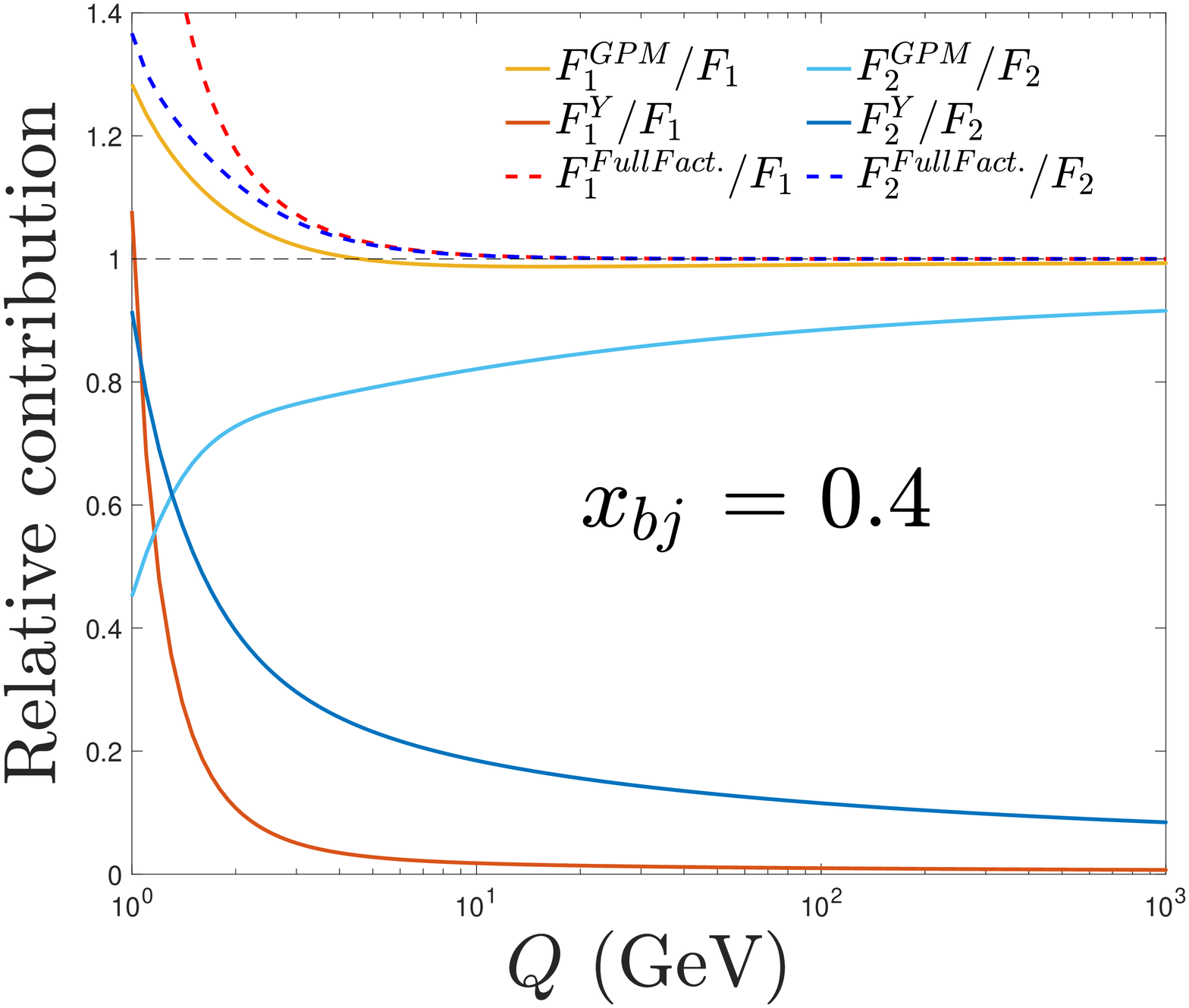}
		\caption{}
	\end{subfigure}
	\begin{subfigure}[b]{0.45\textwidth}
		\includegraphics[width=\textwidth]{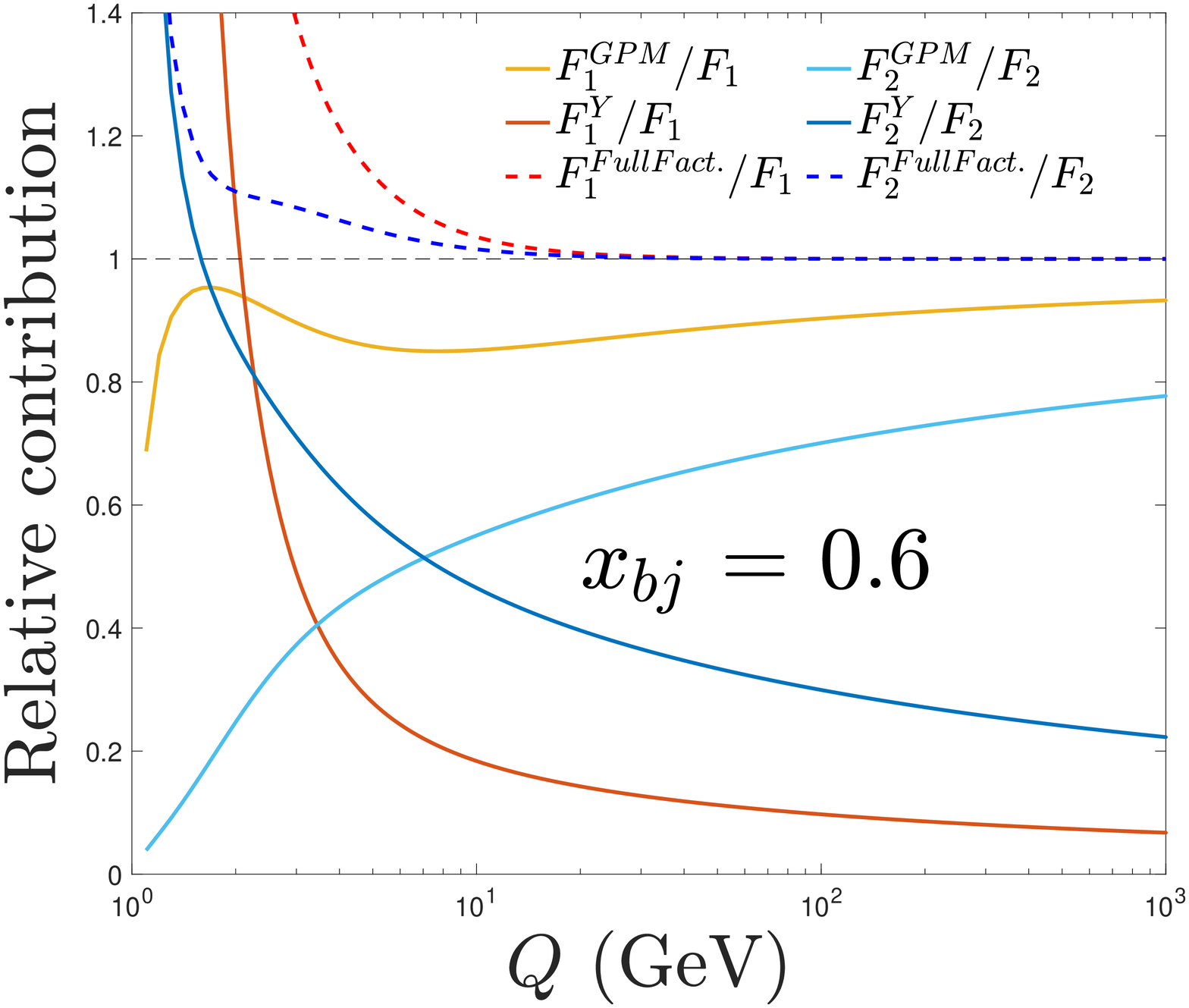}
		\caption{}
	\end{subfigure}
	\begin{subfigure}[b]{0.45\textwidth}
		\includegraphics[width=\textwidth]{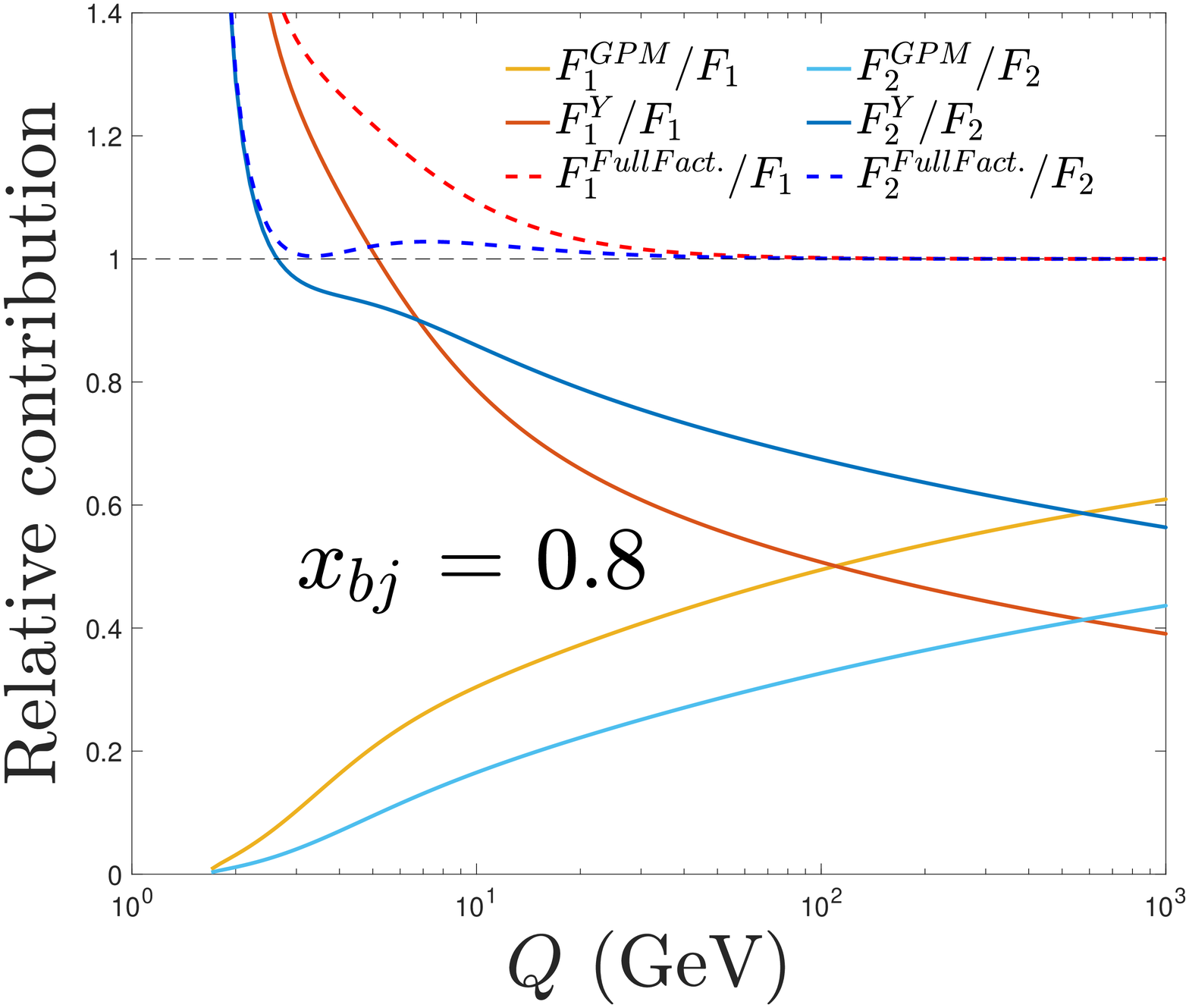}
		\caption{}
	\end{subfigure}
	\caption{The solid lines represent the ratios of the W and Y terms of the unpolarized structure functions $F_1$ and $F_2$ against their unfactorized expression for a wide range of the hard scale $Q$. The dashed lines show their sum which is compared with the dashed black line representing unity. The four figures share the same choice of the masses but each one of them represents the ratios taken at a different $\xbj$.}
	\label{f.F12WY_ratio_mp1_ms1_mq03_x}
\end{figure}
\section{The input scale $Q_0$}
\label{s.input}

The steps required to construct both the collinear and TMD factorization expressions in the previous two sections relied on expansions in $m^2/Q^2$, so the factorized expressions are useful approximations only when $Q$ is sufficiently large compared with intrinsic mass scales. Below some value, the justification for any truncation in powers of $m/Q$ fails. Therefore, applications of factorization generally require one to specify a (possibly $\xbj$-dependent) minimum $Q = Q_0$ below which the expansion is no longer trusted. Typically, one tries to choose $Q_0$ to be as small as can be reasonably justified so as to maximally exploit factorization techniques over the widest possible kinematical range. Sometimes, this is achieved by including parametrizations of subleading power behavior~\cite{Accardi:2009br,Accardi:2016qay}. In standard treatments of DIS, a typical $Q_0$ is usually between approximately $1$ and $4$~GeVs.

The sample curves from the Yukawa theory shown in \fref{factplots} demonstrate the limited validity of the factorization method as $Q$ decreases. With the values of $\mquark$, $\mgluon$, and $\tarmass$ chosen in those figures, the choice of $Q_0$ should be no smaller than around $1$~GeV, although for small $\xbj$ is appears to be possible to push $Q_0$ lower.\footnote{The reason is that the errors in the power expansion go like $\xbj m^2/Q^2$ at small $\xbj$, at least for these low order graphs.} 
 
Notice that it is the only size of the external kinematical variables $Q$ and $\xbj$ relative to the intrinsic that determines the level of agreement between the unfactorized and factorized expressions. If we neglect the running of the parameters, then there is no dependence at all upon the auxiliary parameter $\mu$ (see \eref{finalfact} and \fref{F1mudep}). In QCD, that arbitrariness in the choice of renormalization scale is exploited to minimize the size of higher order errors. 

One way to introduce an extra adjustable parameter in a way that might allow the factorized expression to be improved along the lines of \cite{Accardi:2009br,Accardi:2016qay}, for at least some regions of kinematics, is to 
switch out the $\msbar$ renormalized definition for the pdf with the cutoff definition in \eref{cuttoff_def}.  
The new parameter this introduces is the cutoff scale $k_c$, and one may attempt to adjust this to extend agreement between the factorized and unfactorized expressions to smaller $Q$. 

For a generic unfactorized structure function $F$, the percent errors introduced by factorization are 
\begin{equation}
    \begin{split}
        \delta F^{\msbar}\equiv \left|\frac{F-F^{\msbar}}{F}\right| \cdot 100 ,\hspace{1cm}
        \delta F^{c}\equiv \left|\frac{F-F^{c}}{F}\right| \cdot 100 \, ,
    \end{split}
\end{equation}
where the $c$ and $\msbar$ superscripts indicate if it is the $\msbar$ or the cutoff definitions of the collinear pdfs that are used in the calculation. 
While $k_c \approx \mu$ is the natural choice in the latter case, the size of $\delta F^c$ depends on the exact value of the cutoff. Interesting scenarios are those where one can find a $k_c$ such that $\delta F^c < \delta F^{\msbar}$ for at least some range of kinematics. Whether this is possible depends on the details of the long-range dynamics of the specific theory like the size of the intrinsic mass scales. However, it is interesting to investigate how difficult it is to construct examples. One example in the Yukawa theory is shown in \fref{F1MSbarvsCutoff}. The figure compares the result of using two definitions of the collinear pdf by considering the disagreement with the unfactorized $F_1$ at a hard scale $Q = 2.5$ GeV. The dotted yellow curve is with the standard $\msbar$ collinear pdf while the dotted green curve is with the cutoff definition from \eref{cuttoff_express}.  
Note that the largest mass is the ``spectator mass'' $\mgluon = 2$ GeV and it has been modified from the previous examples to introduce larger $\sim m/Q$ errors at large $\xbj$. Along with the unfactorized and factorized versions of $F_1$, we plot the percent error defined as
\begin{equation}
\%\,\text{err} \equiv \left|\frac{F-F^r}{F}\right| \cdot 100   ,
\end{equation}
where in the case of those examples it is $F=F_1$ and the subscript $r$ is a placeholder for the choices of the UV scheme. In this particular example, it appears that the use of a cutoff scheme reduces the error relative to the $\msbar$ scheme in the large $\xbj$ region. To relate the cutoff $k_c$ to external kinematics, we try the following form,
\begin{equation}
k_c(\xbj,Q) = \frac{Q}{Q_0} \left(a + b \xbj + c e^{(x - x_\text{max})^2/d^2} \right) \, ,
\end{equation}
and tune the parameters $a$, $b$, $c$ and $d$ to improve agreement with the exact $F_1$ at $Q \approx Q_0$. The plot in \fref{F1MSbarvsCutoff} is for the case $Q_0 = 2.5$~GeV. The form of the cutoff is quite ad hoc, but the exercise shows that it is in principle possible to use a pdf definition that extends the range of agreement between the factorized expressions and the unapproximated structure functions. 
\begin{figure}
    \centering
    \includegraphics[scale=.4]{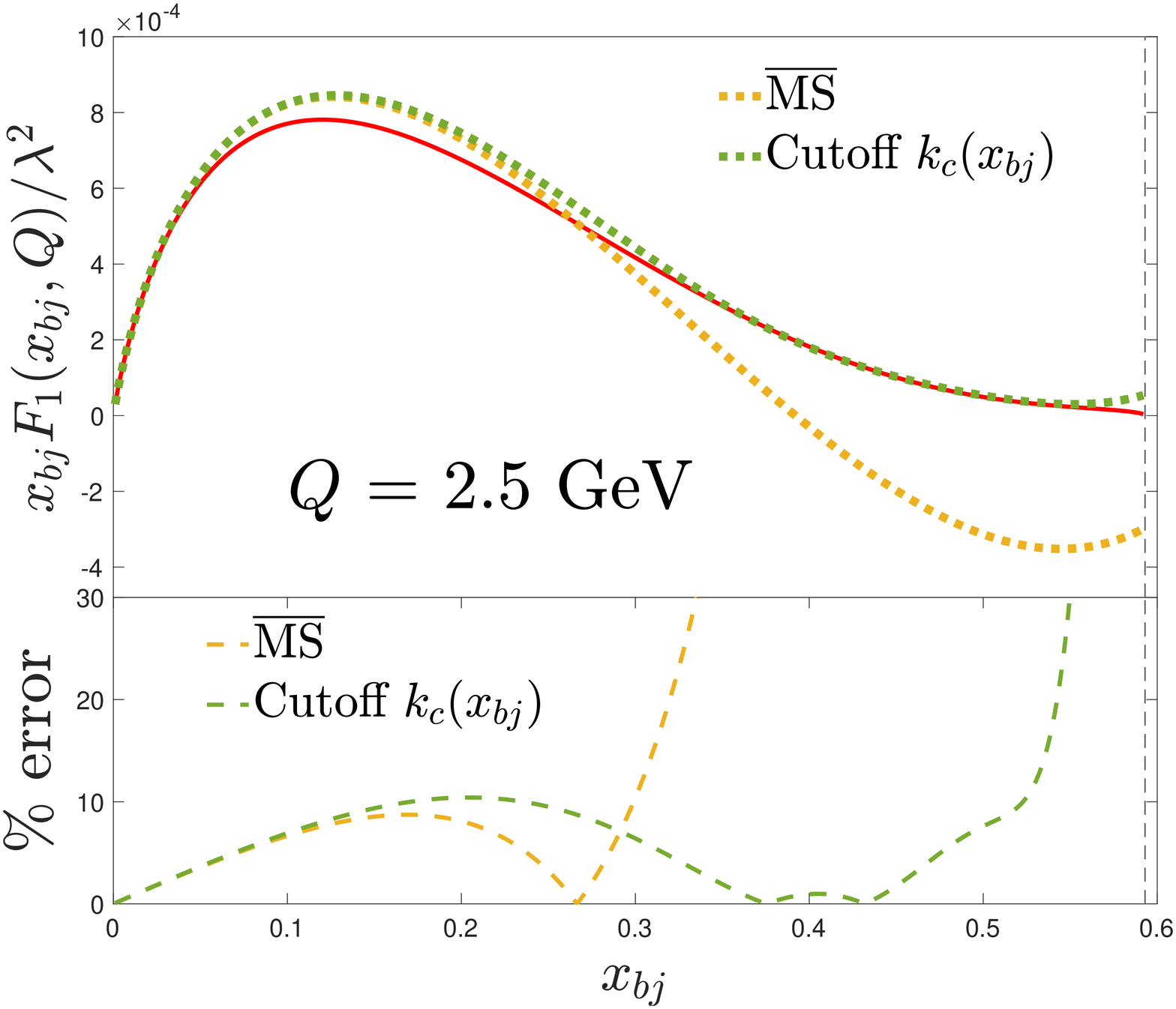}
    \caption{The factorized structure function $F_1$ is shown in its $\msbar$ (dotted yellow) and cutoff (dotted green) versions compared against its unfactorized (solid red) expression. The chosen functional form of the cutoff reads $k_c(\xbj) = -35.9678  + 5.68514 \,\xbj + 35.0979\, e^{(\xbj - x_{max})^2}$. Its explicit expression corresponds to a trial parametrization whose coefficients have been found by fitting the solutions to $\left|F_1(\xbj,Q) - F_1^c(\xbj,k_c,Q) \right|= 0$, i.e. the difference between the cutoff-factorized and unfactorized structure function. Here the masses have been chosen to be $\tarmass = 1$ GeV, $\mgluon = 2$ GeV and $\mquark = 0.3$ GeV with a hard scale $Q = 2.5$ GeV. }
    \label{f.F1MSbarvsCutoff}
\end{figure}
Of course, any such improvement is only possible when $Q$ is not especially large, i.e. near $Q_0$. For large $Q$, both $\delta F^{\msbar}$ and $\delta F^c$ vanish like a power of $m/Q$.  Using the cutoff definition for the pdf has the effect of including power corrections to the standards renormalized. It is worth noting that phenomenological treatments of pdfs in real QCD at moderate $Q$ frequently do parametrize such ``higher twist'' contributions in pdf extractions~\cite{Accardi:2009br,Accardi:2016qay}.  

\section{Working in transverse coordinate space}
\label{s.coord}

To further extend the analogy with TMD factorization as it is used in QCD, we now consider the $W$-terms in \erefs{hadrotensappF1}{hadrotensappF2}, but in transverse coordinate space,
\begin{align}
F_1^W(\xbj,Q,\T{q}{}) &{}= \frac{1}{2} \int \frac{\diff[2]{\T{b}{}}}{(2 \pi)^2}
    ~ e^{-i\T{q}{}\cdot \T{b}{} } \tilde{f}_{q/p}(\xbj,\T{b}{};\mu) \, , \no
F_2^W(\xbj,Q,\T{q}{}) &{}= \xbj \int \frac{\diff[2]{\T{b}{}}}{(2 \pi)^2}
    ~ e^{-i\T{q}{}\cdot \T{b}{} } \tilde{f}_{q/p}(\xbj,\T{b}{};\mu) \, .
\end{align}
In the low order Yukawa theory, we can write down the explicit transverse coordinate space version of the quark TMD pdf in terms of Bessel functions. It is
\begin{align}
   \tilde{f}_{q/p}(\xi,\T{b}{};\mu) &{}= 
2 a_\lambda(\mu) (1 - \xi)\int\diff{\Tsc{k}{}}{}\Tsc{k}{}\frac{\Tscsq{k}{}+\chi^2(\xi)}{\left[\Tscsq{k}{}+\Delta(\xi)^2\right]^2}J_0(\Tsc{b}{}\Tsc{k}{}) \no
&{}= 2 a_\lambda(\mu) (1 - \xi) \left\{ K_0\parz{\Tsc{b}{} \Delta(\xi) } - \frac{\Tsc{b}{} \left[ \Delta(\xi)^2 - \chi(\xi)^2\right]}{2 \Delta(\xi)} K_1\parz{\Tsc{b}{} \Delta(\xi) } \right\} \, . \label{e.exactbspace}
\end{align}
A sample of the $\Tsc{b}{}$-space TMD pdfs is shown in \fref{btspaceTMD} for several values of the momentum fraction $\xi$.  \\
\begin{figure}[!h]
	\centering
	\begin{subfigure}[b]{0.48\textwidth}
		\includegraphics[width=\textwidth]{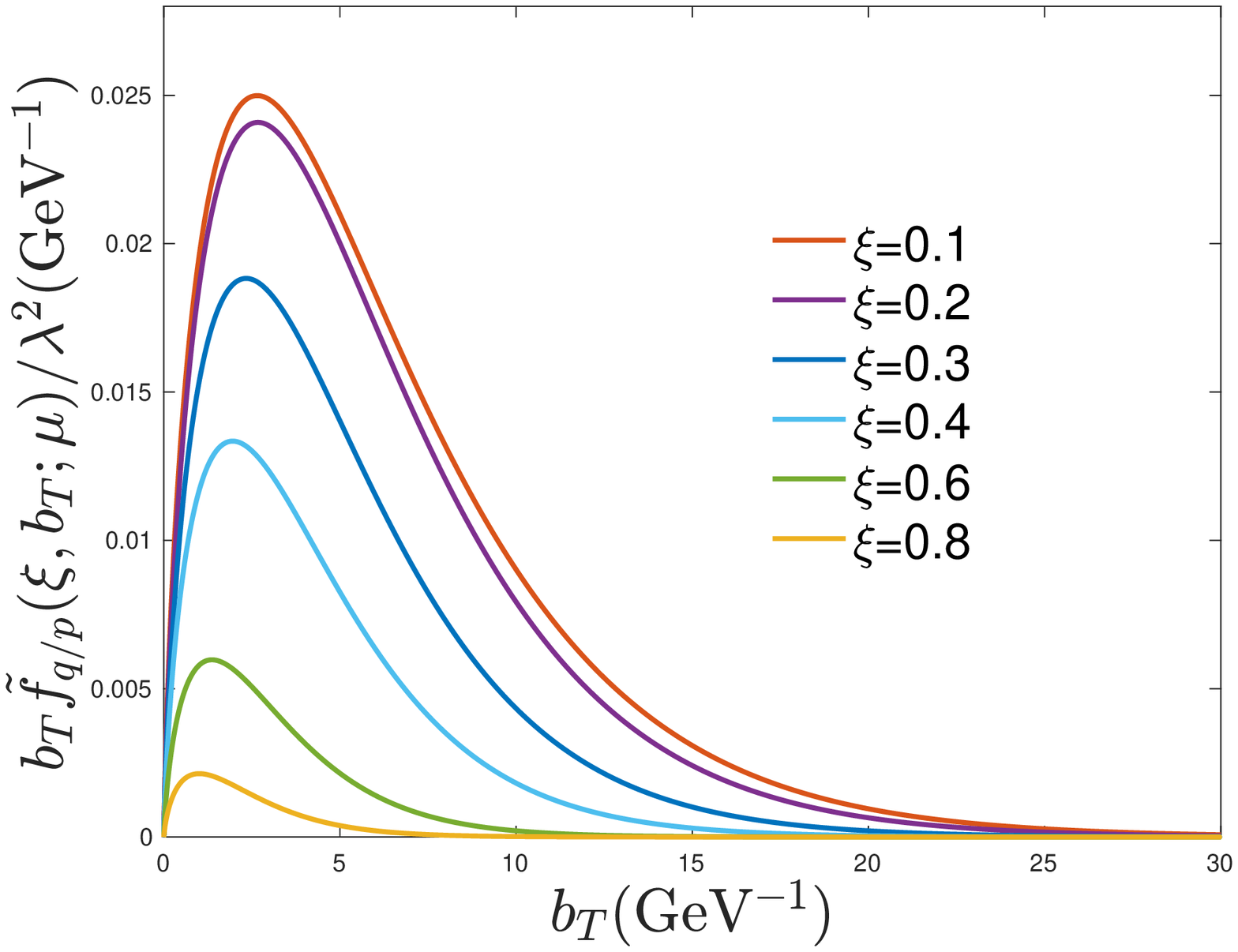}
		\caption{}
		\label{fTms_1}
	\end{subfigure}
	\hspace{1mm}
	\begin{subfigure}[b]{0.48\textwidth}
		\includegraphics[width=\textwidth]{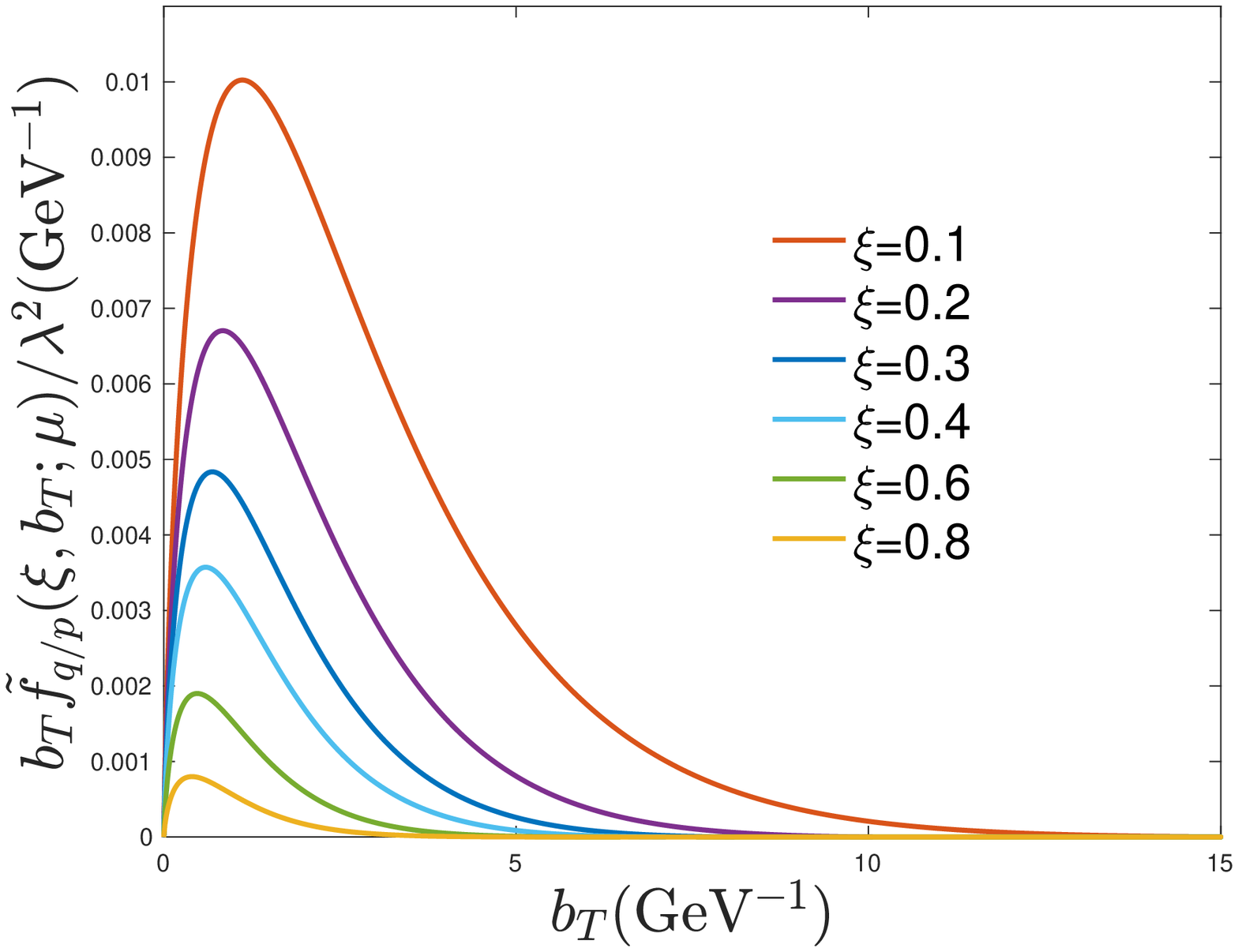}
		\caption{}
		\label{fTms_15}
	\end{subfigure}
	\caption{The coordinate space quark TMD pdf for several values of momentum fraction $\xi$. Mass scales are (a) $\mquark = 0.3$~GeV, $\tarmass = 1.0$~GeV and $\mgluon = 1.0$~GeV (b) $\mquark = 0.3$~GeV, $\tarmass = 1.0$~GeV and $\mgluon = 1.5$~GeV. Plots with this general shape are familiar from work with the CSS formalism~\cite[Fig.~4]{Konychev:2005iy}.}
	\label{f.btspaceTMD}
\end{figure}
The coordinate space TMD pdf satisfies an operator product expansion (OPE) in the limit of small $\Tsc{b}{}$, 
\begin{align}
\label{e.basicope}
\tilde{f}_{q/p}(\xbj,\T{b}{};\mu) = 
\sum_j \int_{\xbj}^1 \frac{\diff{\xi}{}}{\xi} \mathcal{\tilde{C}}_{q/j}(\xbj/\xi,\T{b}{};\mu) \tilde{f}_{j/p}(\xi;\mu) + \order{m^2 \Tscsq{b}{}}{} \, , 
\end{align}
with 
\begin{equation}
\label{e.OPE_coeff}
\mathcal{\tilde{C}}_{q/j}(\hat{x},\T{b}{};\mu) = \delta(1-\hat{x}) \delta_{jq} - a_\lambda(\mu) (1- \hat{x}) \ln \parz{\frac{\mu^2 \Tscsq{b}{} e^{2 \gamma_E}}{4}}  \delta_{jp} + \cdots \, .
\end{equation}
There is a term for $\mathcal{C}_{q/s}$ as well, but it does not contribute at the order of the graphs in \fref{basicmodel}, so we do not write it here explicitly. See \aref{opecalc} for a discussion of these formulas.
Thus, 
\begin{align}
&\tilde{f}_{q/p}(\xbj,\T{b}{};\mu) = f_{q/p}(\xbj;\mu) - a_\lambda(\mu) (1- \xbj) \ln \parz{\frac{\mu^2 \Tscsq{b}{} e^{2 \gamma_E}}{4}} + \cdots + \order{m^2 \Tscsq{b}{}} \, , \no
&= \tilde{f}^\text{OPE}_{q/p}(\xbj,\T{b}{};\mu) + \order{m^2 \Tscsq{b}{}} \, .
\label{e.ope}
\end{align} 
where $f_{q/p}(\xbj;\mu)$ is the $\order{a_\lambda(\mu)}$ quark-in-hadron collinear pdf from \eref{finP} and the second term uses 
\eref{PinP}. We will call the approximation wherein the $\order{m^2 \Tscsq{b}{}}$ terms in \eref{ope} are dropped $\tilde{f}^\text{OPE}_{q/p}(\xbj,\T{b}{};\mu)$. Figure~\ref{f.OPEbtspaceTMD} compares the OPE approximation with the unapproximated calculation in \eref{exactbspace}, and confirms that the two agree in the small $\Tsc{b}{}$ limit where the $\order{m^2 \Tscsq{b}{}}$ contributions are negligible. In the $\Tsc{b}{} \to \infty$ IR limit, the OPE calculation has a (negative) divergence. 
\begin{figure}[!h]
	\centering
	\begin{subfigure}[b]{0.48\textwidth}
		\includegraphics[width=\textwidth]{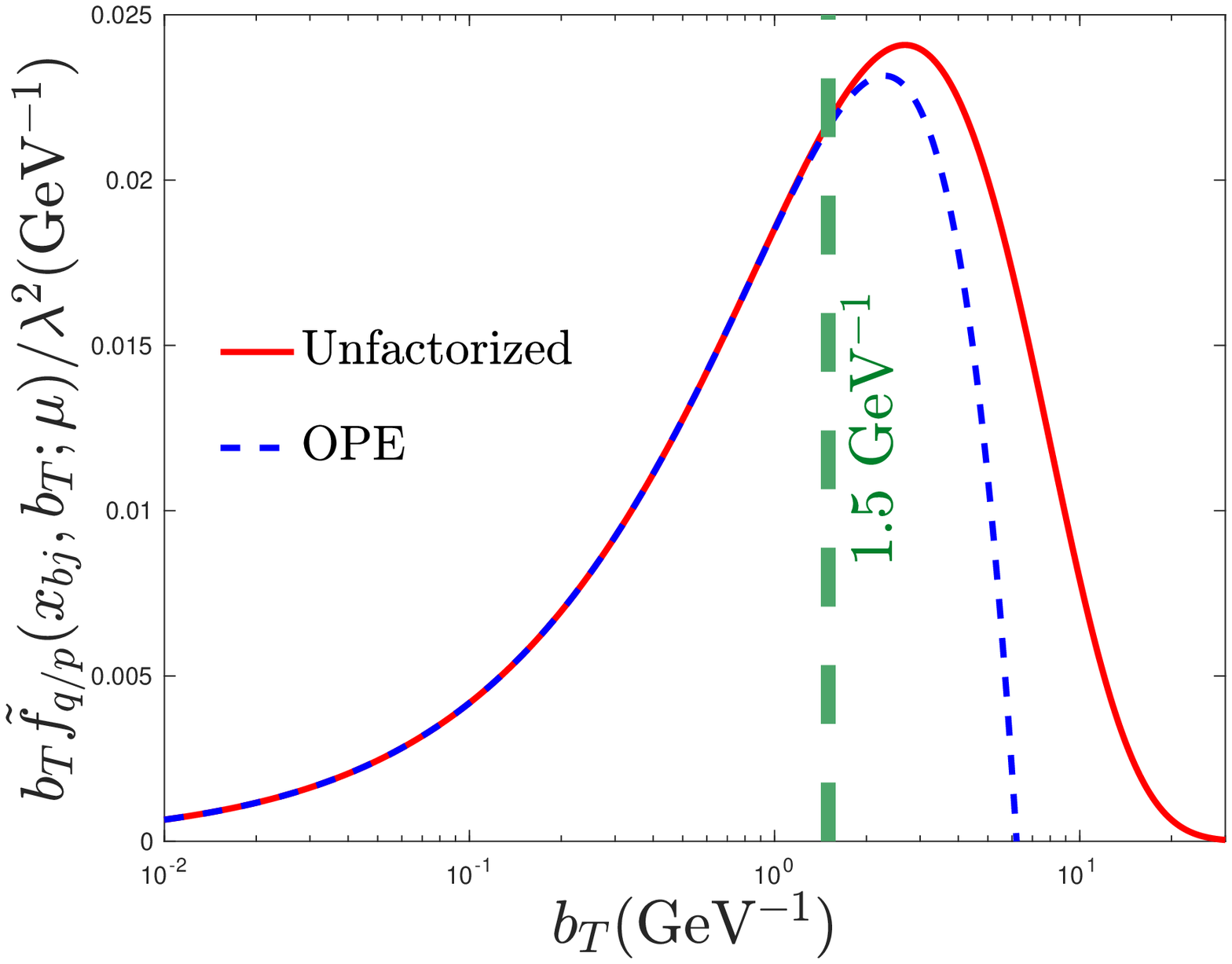}
		\caption{}
		\label{fTOPEms_1}
	\end{subfigure}
	\hspace{1mm}
	\begin{subfigure}[b]{0.48\textwidth}
		\includegraphics[width=\textwidth]{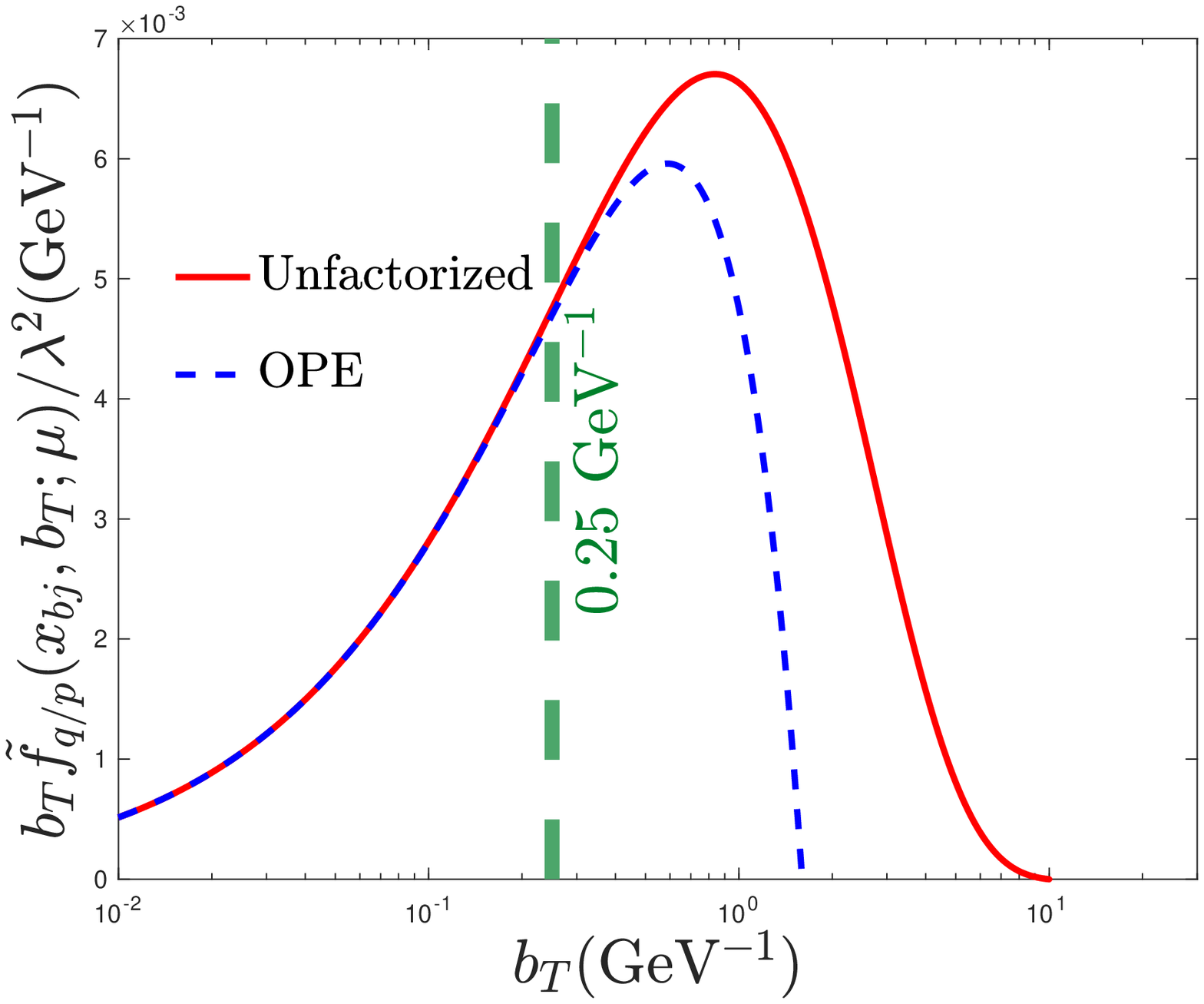}
		\caption{}
		\label{fTOPEms_15}
	\end{subfigure}
	\caption{The unfactorized coordinate space quark TMD pdf  and $\tilde{f}^\text{OPE}_{q/p}(\xbj,\T{b}{};\mu)$ for $x_{bj}=0.2$.The green line shows the approximate maximum allowable value of $\bmax$. Any value of $\bmax$ smaller than this approximate maximum allowable value is justified.  (a) $\mquark = 0.3$~GeV, $\tarmass = 1.0$~GeV, $\mgluon = 1.0$~GeV and $\bmax \lesssim 1.5$ GeV$^{-1}$ (b) $\mquark = 0.3$~GeV, $\tarmass = 1.0$~GeV $\mgluon = 1.5$~GeV and $\bmax \lesssim 0.25$ GeV$^{-1}$}
	\label{f.OPEbtspaceTMD}
\end{figure}

In QCD versions of this, one is motivated to isolate the contributions from the small $\Tsc{b}{}$ region, which is insensitive to soft, large-distance mass scales, from the $m$-dependent large $\Tsc{b}{}$ contributions. Then the small $\Tsc{b}{}$ part can be calculated perturbatively in QCD using the OPE and collinear factorization. If the remaining large $\Tsc{b}{}$ contribution is sequestered from the perturbative part, it can be treated as a universal nonperturbative contribution and parametrized phenomenologically. 

A standard scheme~\cite{Collins:1981va} for separating out the $m$-dependent portion of the TMD pdf (what would be the nonperturbative part in QCD) is the ``$b_*$-method.''
There, one demarcates the regions of large and small $\Tsc{b}{}$ by replacing $\T{b}{}$ with a different transverse coordinate variable $\bstar$ with the property that 
\begin{equation}
\bstar(\Tsc{b}{}) = 
\begin{dcases}
\T{b}{} & b_T \ll b_{\rm max} \\
\vect{b}_{\rm max} & b_T \gg b_{\rm max} \,  \label{e.bdefold}
\end{dcases}\, ,
\end{equation}
where $\bmax$ is a transverse size that is chosen to demarcate the boundary between what are considered large and small transverse coordinate regions. The most commonly used functional form is
\begin{align}
\label{e.bstar}
  \bstar(\Tsc{b}{}) = \frac{ \T{b}{} }{ \sqrt{ 1 + \Tscsq{b}{}/\bmax^2} } \, . 
\end{align}
The only requirement on $\bmax$ is that it should be small enough that $\Tsc{b}{} \lesssim \bmax$ contributions to the W term are small enough that the $\order{m^2 \Tscsq{b}{}}$ in \eref{ope} are negligible. Since the evolution factor in \eref{evolved} is $\Tsc{b}{}$-independent, we can write 
\begin{equation}
\label{e.f_g_separate}
\tilde{f}_{q/p}(\xbj,\T{b}{};\mu) = \tilde{f}_{q/p}(\xbj,\bstar;\mu) \frac{\tilde{f}_{q/p}(\xbj,\T{b}{};\mu)}{\tilde{f}_{q/p}(\xbj,\bstar;\mu)} = 
\tilde{f}_{q/p}(\xbj,\bstar;\mu) \exp \{-g_{q/p}(\xbj,\T{b}{})\} \, , 
\end{equation}
and the function we have defined as 
\begin{equation}
\label{e.g_def}
g_{q/p}(\xbj,\T{b}{}) \equiv -\ln \parz{\frac{\tilde{f}_{q/p}(\xbj,\T{b}{};\mu)}{\tilde{f}_{q/p}(\xbj,\bstar;\mu)}} \, .
\end{equation}
will be exactly scale-independent because the $\mu$-dependence is an overall $\Tsc{b}{}$-independent factor. Equation~\eqref{e.exactbspace} substituted in \eref{g_def} gives the explicit $g_{q/p}(\xbj,\T{b}{})$ for the Yukawa theory example. Note carefully that $g_{q/p}(\xbj,\T{b}{})$ depends on the detailed choice of $\bstar$ and the value of $\bmax$. 

If $\bmax$ is small compared to $\sim 1/m$, then we can use the OPE approximation and write 
\begin{equation}
\label{e.OPE_express}
\tilde{f}_{q/p}(\xbj,\T{b}{};\mu) = \tilde{f}^\text{OPE}_{q/p}(\xbj,\bstar;\mu) \exp \{-g_{q/p}(\xbj,\T{b}{})\} + \order{m^2 \bmax^2} \, , 
\end{equation}
and, if $\bmax$ is small enough, we can just drop the 
$\order{m^2 \bmax^2}$ errors. The maximum allowable $\bmax$ before which the $\order{m^2 \bmax^2}$ errors start to be important depends, of course, on the ``nonperturbative'' scales like the masses in \eref{masses}. Comparing plots (a) and (b) in \fref{OPEbtspaceTMD} shows that the $\bmax$ dependence on those masses is quite strong. 

In applications to QCD at high energies, it is often the hope that expressions analogous to \eref{OPE_express} can be used to exploit the OPE part $\tilde{f}^\text{OPE}_{q/p}(\xbj,\bstar;\mu)$ for the widest possible range of $\Tsc{b}{}$, thereby minimizing the importance of the $m$-dependent $g_{q/p}(\xbj,\T{b}{})$ functions and maximally exploiting the predictive power in collinear pdfs with collinear factorization alone. Thus, one chooses $\bmax$ to be as large as possible while still guaranteeing that it is reasonably justified to drop the powers of $m^2 \bmax^2$ in \eref{OPE_express}. In earlier sections, we defined $Q_0$ to be the scale below which it is no longer justified to neglect powers of $m/Q_0$, so we should set
\begin{equation}
\bmax \approx 1/Q_0 \, .
\end{equation}
\begin{figure}[!h]
	\centering
	\begin{subfigure}[b]{0.48\textwidth}
		\includegraphics[width=\textwidth]{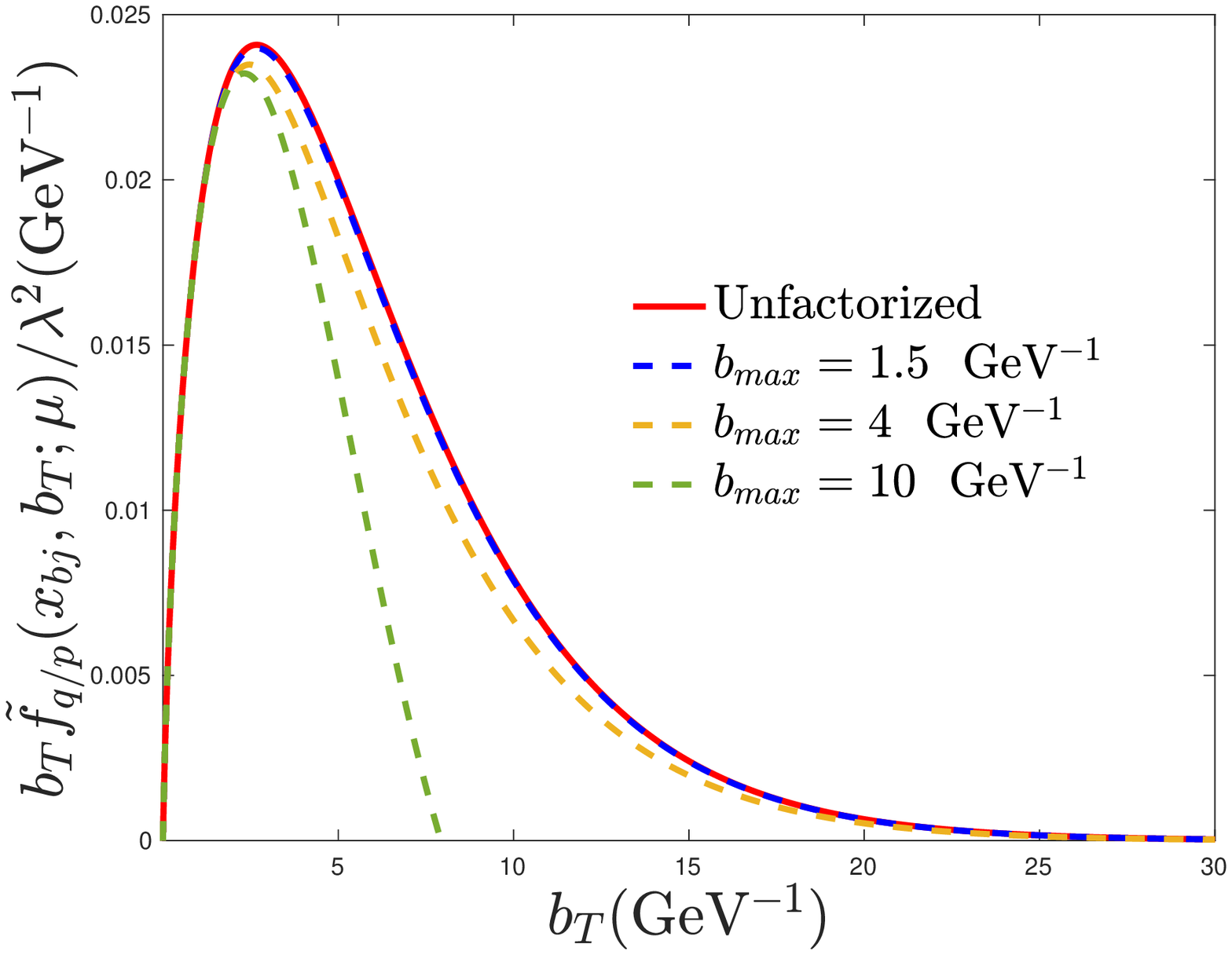}
		\caption{}
		\label{fTbmaxms_1}
	\end{subfigure}
	\hspace{1mm}
	\begin{subfigure}[b]{0.48\textwidth}
		\includegraphics[width=\textwidth]{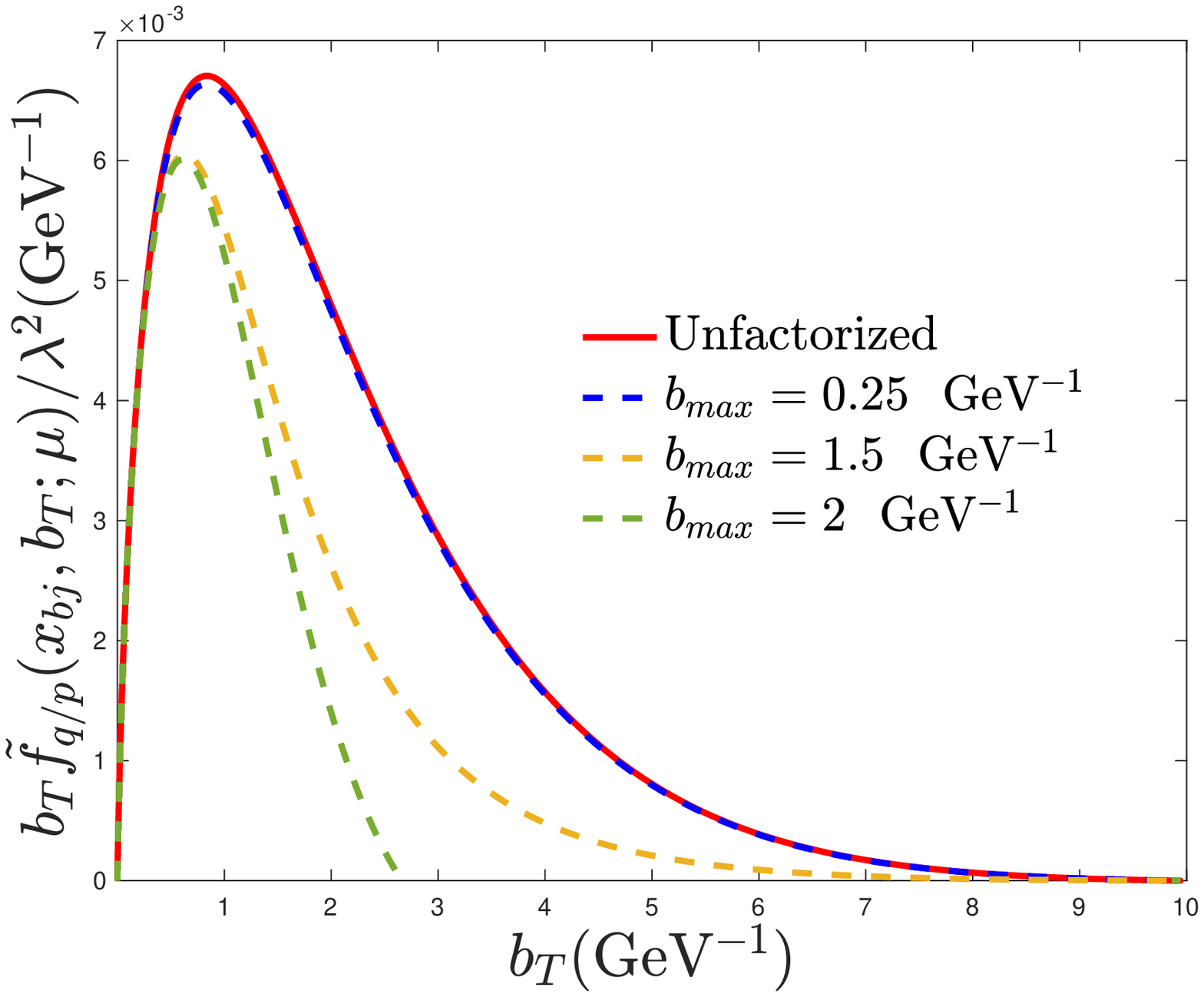}
		\caption{}
		\label{fTbmaxms_15}
	\end{subfigure}
	\caption{The unfactorized coordinate space quark TMD pdf  and the coordinate space quark TMD pdf $\tilde{f}^{\text{Evol}}_{q/p}(\xbj,\T{b}{};Q)$ for several values of $\bmax$. The mass scales are (a) $\mquark = 0.3$~GeV, $\tarmass = 1.0$~GeV and $\mgluon = 1.0$~GeV (b) $\mquark = 0.3$~GeV, $\tarmass = 1.0$~GeV and $\mgluon = 1.5$~GeV. (Note the change in the horizontal axis.)}
	\label{f.fbmax}
\end{figure}
In analogous situations in QCD, the strategy would be to minimize contributions from higher orders in the hard part of the OPE of \eref{OPE_coeff} as $\Tsc{b}{} \to 0$ in the $\tilde{f}_{q/p}(\xbj,\bstarsc;\mu)$ factor of \eref{f_g_separate}, so that $\tilde{f}^\text{OPE}_{q/p}(\xbj,\bstar;\mu) \approx f_{q/p}(\xbj;\mu)$ up to a fixed number of calculable higher orders. To this end, one chooses the scale $\mu$ to be order $1/\bstarsc$. Let us thus define,
\begin{equation}
\mu_{\bstar} = b_0/\bstarsc \, , 
\end{equation}
where $b_0 \equiv 2 e^{-\gamma_E}$. Using $\mu_{\bstar}$ in \eref{OPE_coeff} eliminates the logarithmic term in \eref{ope} (or, rather, moves it into the collinear pdf). 
In QCD factorization, however, calculations of the overall cross sections require $\mu \sim Q$. Therefore, there are two steps to the scale-setting. In the Yukawa theory example, we first apply the TMD evolution equation (\erefs{tmd_evol}{evolved}) and write
\begin{equation}
\tilde{f}_{q/p}(\xbj,\bstar;\mu) = \tilde{f}_{q/p}(\xbj,\bstar;\mubstar) \exp\left\{-2 \int_{\mubstar}^\mu \frac{\diff{\mu'}{}}{\mu'} \gamma_2(a_\lambda(\mu')) \right\} \, , \label{e.evolved2}
\end{equation}
to relate a generic scale $\mu$ to the choice $\mu = \mu_{\bstarsc}$. 
Then, substituting \eref{evolved2} into \eref{f_g_separate}, repeating the step of approximating with the OPE, and finally setting $\mu = Q$ gives
\begin{align}
\label{e.f_g_separate2}
\tilde{f}_{q/p}(\xbj,\T{b}{};Q) &{}= 
\tilde{f}_{q/p}(\xbj,\bstar;\mubstar) \exp \left\{-2 \int_{\mubstar}^Q \frac{\diff{\mu}{}}{\mu} \gamma_2(a_\lambda(\mu))
-g_{q/p}(\xbj,\T{b}{})\right\} \,  \no
&{}= 
\tilde{f}^\text{OPE}_{q/p}(\xbj,\bstar;\mubstar) \exp \left\{-2 \int_{\mubstar}^Q \frac{\diff{\mu}{}}{\mu} \gamma_2(a_\lambda(\mu))
-g_{q/p}(\xbj,\T{b}{})\right\} + \order{m^2 \bmax^2} \, . 
\end{align}
From its definition in \eref{g_def}, $g_{q/p}(\xbj,\T{b}{})$ vanishes like a power of $\Tscsq{b}{}$ as $\Tsc{b}{} \to 0$. Therefore, it mainly affects the low transverse momentum behavior. Dropping the errors on the second line of \eref{f_g_separate2} gives an approximation reminiscent to what is done in QCD,
\begin{equation}
\tilde{f}^{\text{Evol}}_{q/p}(\xbj,\T{b}{};Q) \equiv \tilde{f}^\text{OPE}_{q/p}(\xbj,\bstar;\mubstar) \exp \left\{-2 \int_{\mubstar}^Q \frac{\diff{\mu}{}}{\mu} \gamma_2(a_\lambda(\mu))
-g_{q/p}(\xbj,\T{b}{})\right\} \, . \label{e.approxevol}
\end{equation}
The ``$\text{Evol}$'' superscript here marks this as  another approximation to the exact operator definition of the TMD pdf in \eref{tmdpdfdef}. It indicates that this is the ``evolved'' $\Tsc{b}{}$-space TMD pdf after the OPE is applied and the error terms in \eref{f_g_separate2} are dropped. Compare this form of the TMD pdf to Eq.~(33) of Ref.~\cite{Rogers:2015sqa}. Thus, 
\begin{equation}
\tilde{f}_{q/p}(\xbj,\T{b}{};Q) = \tilde{f}^{\text{Evol}}_{q/p}(\xbj,\T{b}{};Q) + \order{m^2 \bmax^2} \, .
\end{equation}
Hence, the standard separation of a TMD pdf into an OPE part and an exponential of g-functions is accurate in the limit that the arbitrary $\bmax$ is chosen to be very small. 

Notice that,
\begin{equation}
\tilde{f}^{\text{Evol}}_{q/p}(\xbj,\T{b}{};Q) = f_{q/p}(\xbj;\mubstar) \exp \left\{
-g_{q/p}(\xbj,\T{b}{})\right\} + \order{a_\lambda^2, m^2 \bmax^2} \, ,  \label{e.approxevolapprox}
\end{equation}
If we restrict consideration to the graphs in \fref{basicmodel}, as we do throughout this paper, then 
$\tilde{f}^{\text{Evol}}_{q/p}(\xbj,\T{b}{};Q)$ is just 
$f_{q/p}(\xbj;\mubstar) \exp \left\{
-g_{q/p}(\xbj,\T{b}{})\right\}$. We will use this approximation in all figures below.

Transforming the W terms in the Yukawa theory example of \erefs{hadrotensappF1}{hadrotensappF2} into coordinate space allows them to be reexpressed in terms of $\tilde{f}^{\text{Evol}}_{q/p}(\xbj,\T{b}{};Q)$,
\begin{align}
\tilde{F}_1(\xbj,Q,\T{b}{}) &= \frac{1}{2} \tilde{f}_{q/p}(\xbj,\T{b}{};Q) + \tilde{Y}_1 + \order{\frac{m^2}{Q^2}} = \frac{1}{2} \tilde{f}^{\text{Evol}}_{q/p}(\xbj,\T{b}{};Q) + \tilde{Y}_1 + \order{\frac{m^2}{Q^2}} \, , \label{e.hadrotensappF1tilde} \\
\tilde{F}_2(\xbj,Q,\T{b}{}) &= \xbj \tilde{f}_{q/p}(\xbj,\T{b}{};Q) + \tilde{Y}_2  + \order{\frac{m^2}{Q^2}}  = \xbj \tilde{f}^{\text{Evol}}_{q/p}(\xbj,\T{b}{};Q) + \tilde{Y}_2 + \order{\frac{m^2}{Q^2}} \, .
\label{e.hadrotensappF2tilde}
\end{align}

Restricting to the graphs in \fref{basicmodel}, we may examine the effect of a non-zero $\bmax$. 
Figure~\ref{f.fbmax} compares $\tilde{f}^{\text{Evol}}_{q/p}(\xbj,\T{b}{};Q)$ for several values of $\bmax$ with the original unapproximated $\tilde{f}_{q/p}(\xi,\T{b}{};\mu)$ of \eref{exactbspace}. The deviation of the $\tilde{f}^{\text{Evol}}_{q/p}(\xbj,\T{b}{};Q)$ curves from the unapproximated curve 
is a measure of the error induced by neglecting the $\order{m^2 \Tscsq{b}{}}$ terms in \eref{f_g_separate2}. For any set of intrinsic mass scales, there exists a value of $\bmax$ below which the curves are no longer distinguishable by eye. We can, for instance, identify $\bmax \lesssim 1.5$~GeV$^{-1}$ and $\bmax \lesssim 0.25$~GeV$^{-1}$ for the case with $\mgluon = 1$~GeV and $\mgluon = 1.5$~GeV respectively.
In each case, the $\order{m^2 \bmax^2}$ terms in \eref{OPE_express} are negligible so long as $\bmax$ is chosen to be smaller than these values. Figure~\ref{f.fbmax} also shows that once $\bmax$ is made acceptably small, the $\bmax$-dependence in $\tilde{f}^{\text{Evol}}_{q/p}(\xbj,\T{b}{};Q)$ vanishes:
\begin{equation}
\label{e.derivcrit}
\frac{\diff{}{}}{\diff{\bmax}{}} \tilde{f}^{\text{Evol}}_{q/p}(\xbj,\T{b}{};Q) \to 0 \, .
\end{equation} 
When $\bmax$ is small, changing it amounts to simply transferring $m$-independent contributions between the first (OPE) factor and the second (``nonperturbaive'' $e^{-g}$) factor in \eref{OPE_express}. A significant $\bmax$-dependence indicates either that $\bmax$ is too large or that the model of $g_{q/p}(\xbj,\T{b}{})$ is not complete in the small $\Tsc{b}{}$ region. In a theory, like QCD, where explicit calculations over large time and distance scales are nontrivial, \eref{derivcrit} is the appropriate criterion for checking if $\bmax$ is sufficiently small. For an example, see Sec.~IX of \cite{Gonzalez-Hernandez:2022ifv}.  

While the steps above are not helpful for calculating in the specific case of the Yukawa theory, they are nonetheless very useful for illustrating how the procedure works. 
Equation~\eqref{e.approxevol} with \erefs{hadrotensappF1tilde}{hadrotensappF2tilde} is analogous to the way cross sections in QCD are often expressed when one uses the CSS or similar formalisms in $\Tsc{b}{}$-space. As just emphasized, the $\bstarsc$ strategy for isolating $m$-dependent (``nonperturbative'') and massless OPE (``perturbative'') contributions in the two separate factors of \eref{f_g_separate} is only reliable if $\bmax$ is chosen small enough that it is justfiable to neglect the $\order{m^2 \Tscsq{b}{}}$ terms in \eref{f_g_separate2}. 

In QCD, the functions that correspond to $g_{q/p}(\xbj,\T{b}{})$ contain information about the large distance physics, so they are nonperturbative. In phenomenology,
see for instance Refs.~\cite{Nadolsky:2000ky,Taghavi:2016tse,Scimemi:2019cmh,Bacchetta:2022awv}
, the usual strategy is to replace them with ansatz parametrizations that are fit to experimental data.\footnote{Although there are rapidly improving lattice based methods for calculating them. See for instance Refs.~\cite{Constantinou:2020hdm,LPC:2022zci} and references therein.}
In the Yukawa theory example, we know the explicit expression for $g_{q/p}(\xbj,\T{b}{})$ through \eref{exactbspace} and \eref{g_def}, so it is possible to directly examine the effect of replacing it with an ansatz approximation. Typical parametrizations of $g_{q/p}(\xbj,\T{b}{})$ are linear or quadratic in $\Tsc{b}{}$:
\begin{equation}
g_{q/p}(\xbj,\T{b}{}) \approx g_1 \Tsc{b}{} \qquad \text{or} \qquad g_{q/p}(\xbj,\T{b}{}) \approx g_1 \Tscsq{b}{} \, . \label{e.ansatzes}
\end{equation}
In the Yukawa theory, one expects correlation functions to vary roughly like $\sim e^{-m \Tsc{b}{}}/\Tsc{b}{}$ over large distances, and this is reflected in the approximately linear behavior of $g_{q/p}(\xbj,\T{b}{})$ at large $\Tsc{b}{}$, 
in agreement with Refs.~\cite{Schweitzer:2012hh,Collins:2014jpa}. 

To ensure that the ``perturbative'' and ``nonperturbative'' contributions are completely separated, we use $\bmax = 1.5$~GeV$^{-1}$ (for $\mgluon = 1$~GeV) and $\bmax = 0.25$~GeV$^{-1}$ (for $\mgluon = 1.5$~GeV), in accordance with the observations of \fref{fbmax}. 

With an appropriately chosen $g_1$, the linear ansatz can be made to give reasonable agreement with the true $g_{q/p}(\xbj,\T{b}{})$ over a wide range of $\Tsc{b}{}$, but it produces significant errors in the tail region in transverse momentum space. The quadratic ansatz can also be made to reproduce the qualitative behavior at large $\Tsc{q}{}$, but overall it performs much worse than the linear ansatz. This is shown in \fref{ktplots} for two sets of values for masses. The corresponding coordinate space TMD pdfs are shown in \fref{ansatz}.
\begin{figure}[!h]
	\centering
	\begin{subfigure}[b]{0.49\textwidth}
		\includegraphics[width=\textwidth]{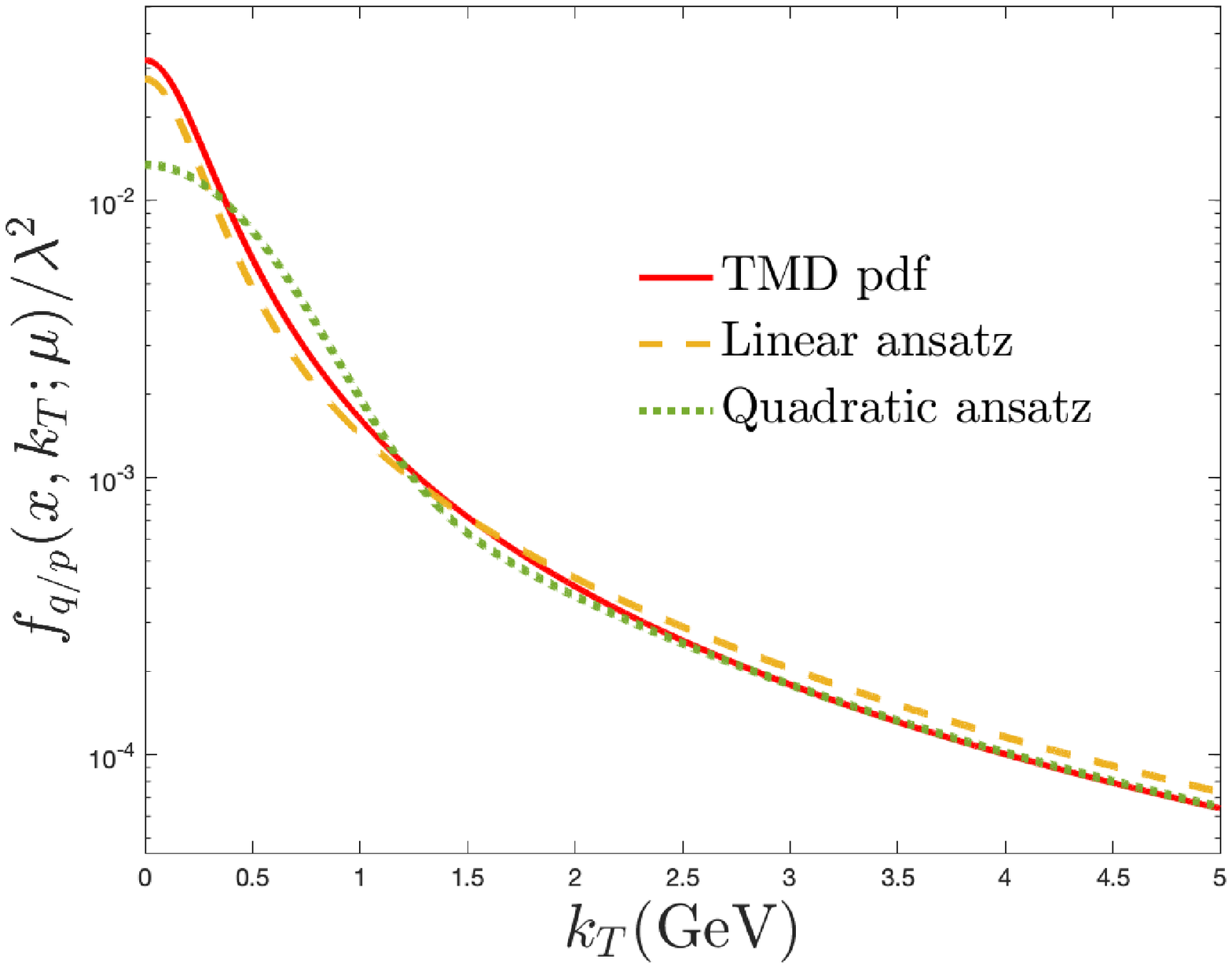}
		\caption{}
	\end{subfigure}
	\hspace{1mm}
	\begin{subfigure}[b]{0.49\textwidth}
		\includegraphics[width=\textwidth]{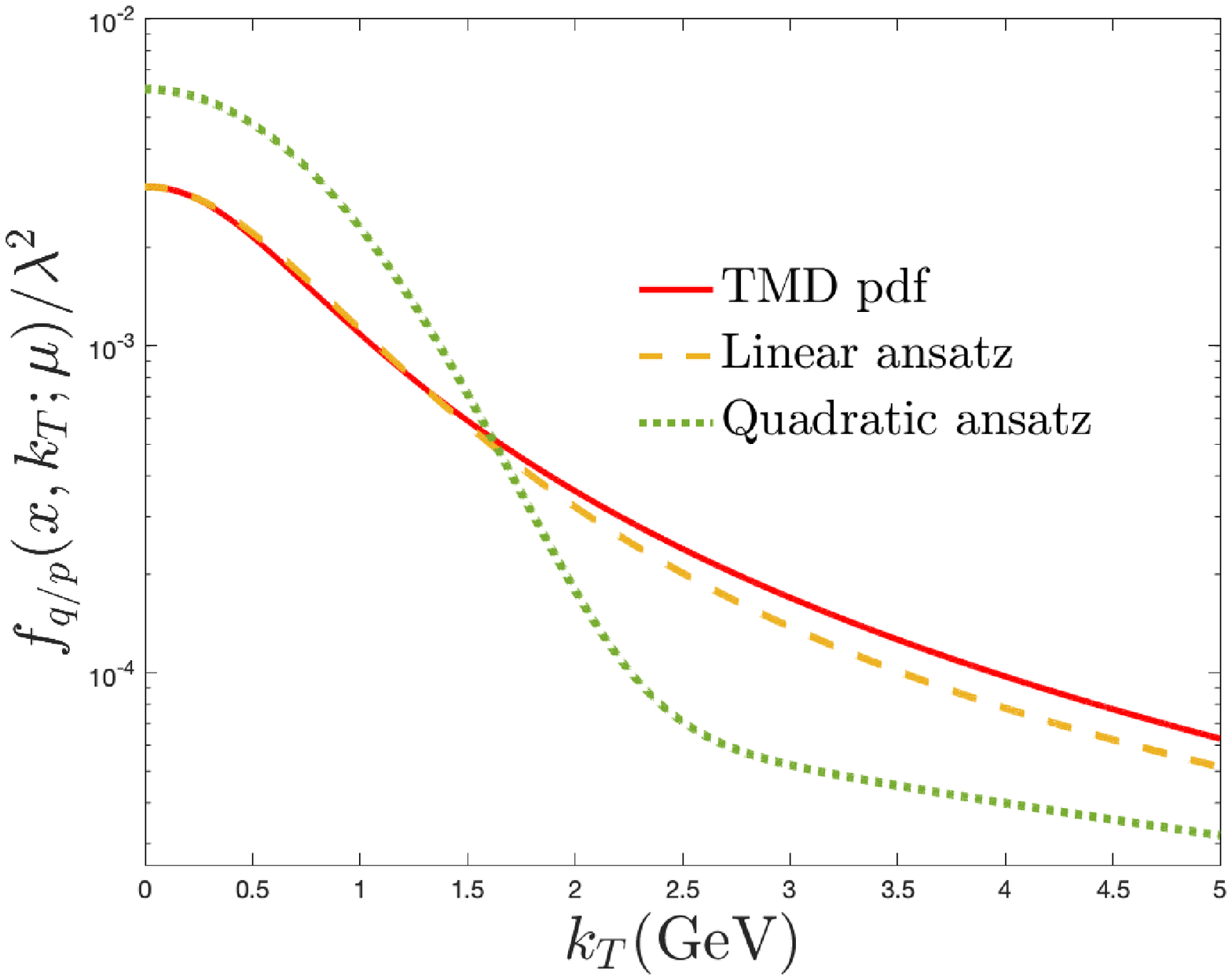}
		\caption{}
	\end{subfigure}
  \caption{The original unfactorized TMD pdf from \eref{TMDpdfresult} (solid red line) compared with the TMD pdfs in the $\tilde{f}^{\text{Evol}}_{q/p}(\xbj,\T{b}{};Q)$ approximation from \eref{approxevol} and using the ansatzes in \eref{ansatzes}. These curves are the Fourier transforms of those in \fref{ansatz}.
  In plot (a), $\tarmass = \mgluon = 1$~GeV, $\mquark = 0.3$~GeV with $g_1 = 0.3$~GeV for the linear case and $g_1 = 0.1$~GeV$^2$ for the quadratic case. In plot (b), $\tarmass = 1$~GeV, $\mquark = 0.3$~GeV and $\mgluon = 1.5$~GeV with $g_1 = 1$~GeV for the linear ansatz and $g_1 = 0.25$~GeV$^2$ for the quadratic case.
 }
\label{f.ktplots}
\end{figure}
\begin{figure}[!h]
	\centering
	\begin{subfigure}[b]{0.49\textwidth}
		\includegraphics[width=\textwidth]{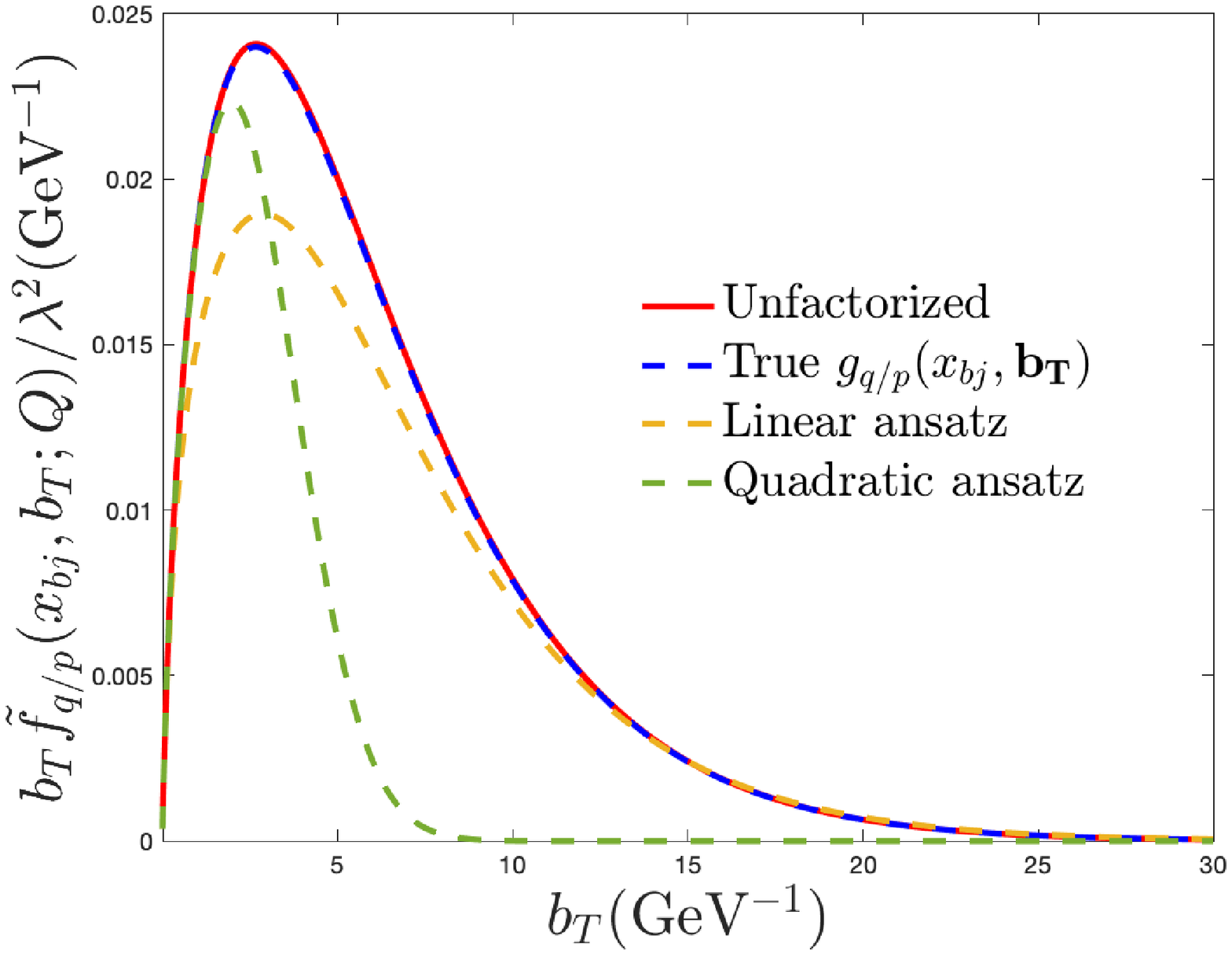}
		\caption{}
		\label{f.linear}
	\end{subfigure}
	\hspace{1mm}
	\begin{subfigure}[b]{0.49\textwidth}
		\includegraphics[width=\textwidth]{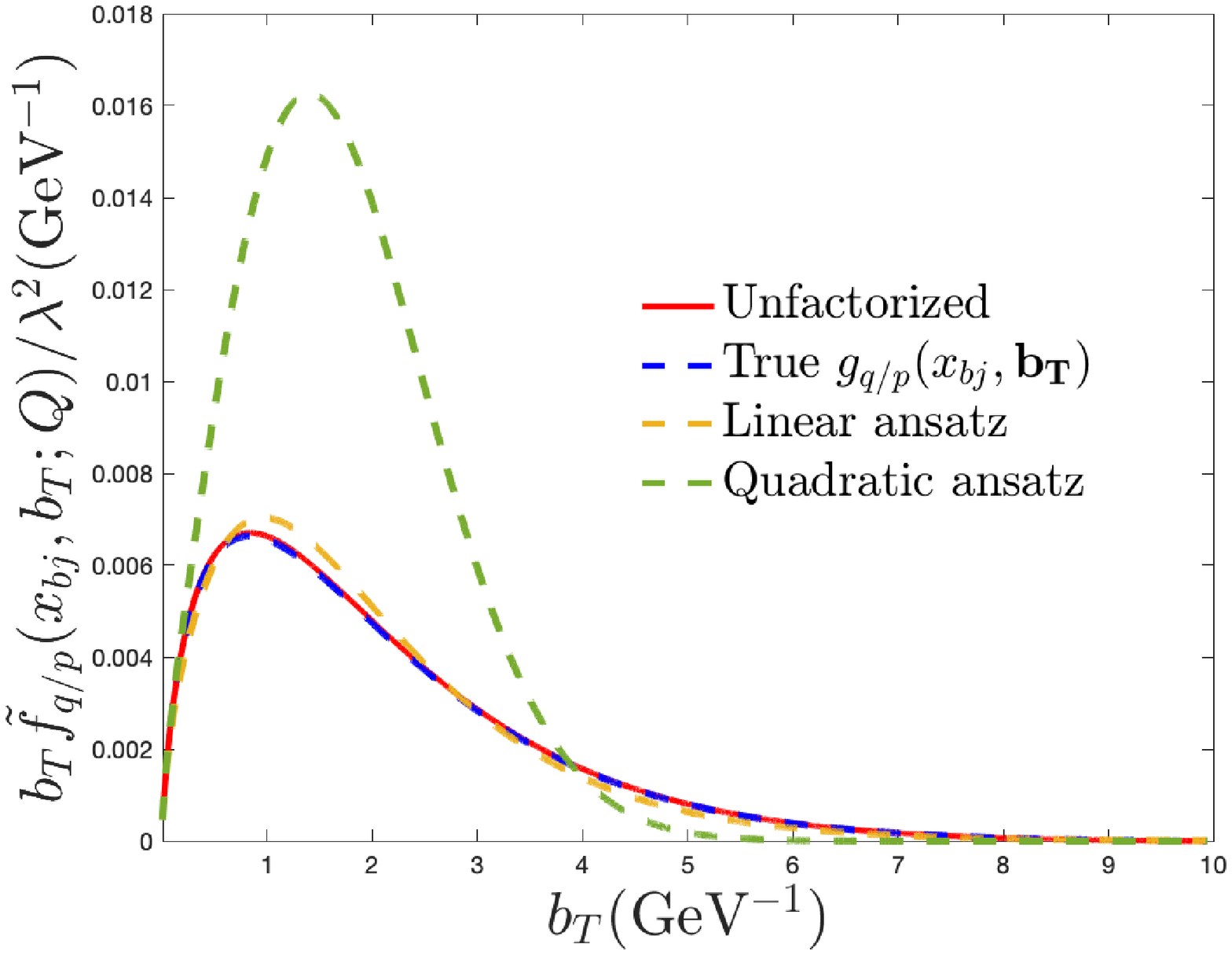}
		\caption{}
		\label{f.quadratic}
	\end{subfigure}
	\caption{The TMD pdfs of \fref{ktplots} before the Fourier transform into momentum space.  In (a), the chosen mass scales are $\mquark = 0.3$~GeV, $\tarmass = \mgluon=  1.0$~GeV. The $g_{q/p}(\xbj,\T{b}{})$ function uses $\bmax = 1.5$~GeV$^{-1}$ while $g_1 = 0.3$~GeV (linear ansatz) and $g_1 = 0.1$~GeV$^2$ (quadradtic ansatz).
 In (b), $\mquark = 0.3$~GeV, $\tarmass =  1.0$~GeV and $\mgluon =  1.5$~GeV. The $g_{q/p}(\xbj,\T{b}{})$ function uses $\bmax = 0.25$~GeV$^{-1}$ while $g_1 = 1$~GeV (linear ansatz) and $g_1 = 0.25$~GeV$^2$ (quadradtic ansatz).
 }
	\label{f.ansatz}
\end{figure}

\begin{figure}[!h]
	\centering
	\begin{subfigure}[b]{0.49\textwidth}
		\includegraphics[width=\textwidth]{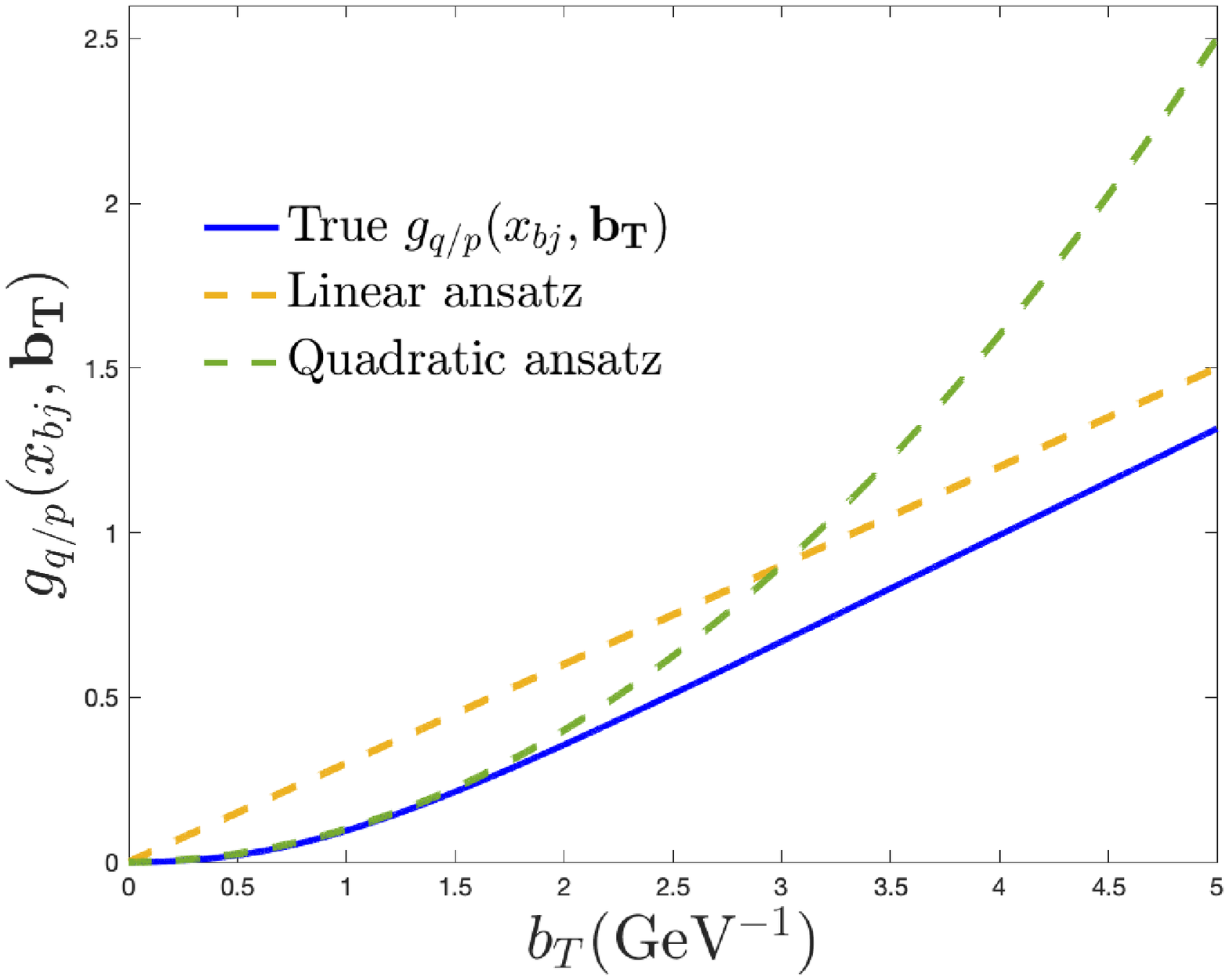}
		\caption{}
	\end{subfigure}
	\hspace{1mm}
	\begin{subfigure}[b]{0.49\textwidth}
		\includegraphics[width=\textwidth]{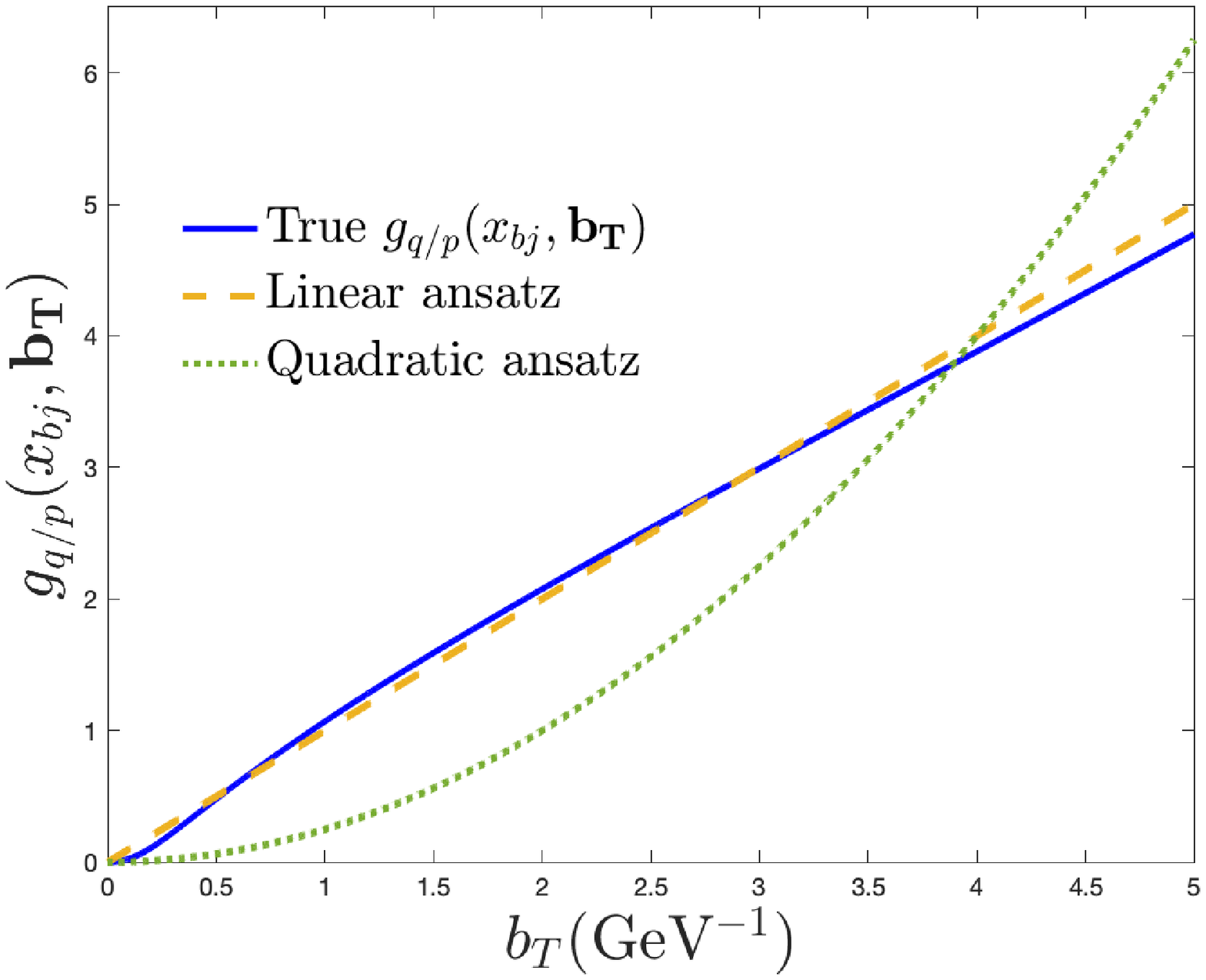}
		\caption{}
	\end{subfigure}
  \caption{
  Plots of the $g$-functions used to obtain \fref{ansatz}(a) and \fref{ansatz}(b). The solid blue curves are the exact $g(\xbj,\T{b}{})$ from \eref{g_def}.
 }
\label{f.gfun}
\end{figure}

Figure~\ref{f.gfun} shows the linear and quadratic versions of $g_{q/p}(\xbj,\T{b}{})$ that were used to obtain the $\tilde{f}^{\text{Evol}}_{q/p}(\xbj,\T{b}{};Q)$ approximations in \frefs{ktplots}{ansatz}, compared with the true $g_{q/p}(\xbj,\T{b}{})$ obtained directly from \eref{exactbspace} and \eref{g_def}.

By increasing $\bmax$ well above $0.25$~GeV$^{-1}$ in the right-hand panel of \fref{ktplots}, it is possible to improve the matching between $\tilde{f}^{\text{Evol}}_{q/p}(\xbj,\T{b}{};Q)$ and the true $\tilde{f}_{q/p}(\xbj,\T{b}{};Q)$ while continuing to use linear and/or quadratic ansatzes. 
However, when the $\order{m^2 \bmax^2}$ errors in \eref{f_g_separate2} are no longer negligible, and the clear separation between a large-$\Tsc{b}{}$ region and a small-$\Tsc{b}{}$ region fails. Thus, while it may be tempting, in applications, to fit $\tilde{f}^{\text{Evol}}_{q/p}(\xbj,\T{b}{};Q)$ by using a large $\bmax$ as a parameter, doing so undermines the original motivation for the steps leading from \eref{basicope} to \eref{approxevol}. Specifically, $\tilde{f}^\text{OPE}_{q/p}(\xbj,\bstar;\mubstar)$ is no longer an accurate approximation to $\tilde{f}_{q/p}(\xbj,\bstar;\mubstar)$. The quality of such a fit is just an artifact of the choice of $\bmax$ and $b_*$, and not a successful application of the OPE (see also, for instance, discussions after Fig.~5 in Ref.~\cite{Grewal:2020hoc}).  
It is preferable, therefore, to keep $\bmax$ small and instead refine the parametrization of $g_{q/p}(\xbj,\T{b}{})$ in a way that ensures that it interpolates smoothly into the small-$\Tsc{b}{}$ region.
  
In a literal parton model where the pdfs are exactly number densities, the TMD and collinear pdfs are related through the integral,
\begin{equation}
\pi \int_0^\infty \diff{\Tscsq{k}{}} f_{q/p}(\xi,\Tsc{k}{}) = f_{q/p}(\xi) \, . \label{e.naiveint}
\end{equation}
As exact relations, integrals like these fail in theories that require renormalization like QCD and the Yukawa toy theory because the integral over $\Tscsq{k}{}$ is UV divergent. After the integral is regulated, \eref{naiveint} only holds in general in an approximate sense. We saw an example of this with the cutoff definition of the collinear pdf in \eref{cuttoff_def}; the integral of the TMD pdf over $\Tscsq{k}{}$ up to a cutoff $k_c^2$ only equaled the standard $\msbar$ pdf up to subleading power corrections. 

It is straightforward to recover the approximate collinear-TMD pdf correspondence of \eref{naiveint} in the coordinate space treatment of the TMD pdf and the $\tilde{f}^{\text{Evol}}_{q/p}(\xbj,\T{b}{};Q)$ approximation, but by regulating the small-$\Tsc{b}{}$ rather than the large $\Tsc{k}{}$ behavior. 
First, one writes the Fourier transform of $\tilde{f}^{\text{Evol}}_{q/p}(x,\T{b}{};Q)$ back to transverse momentum space, 
\begin{equation}
\label{e.bspaceTMDpdf}
f^{\text{Evol}}_{q/p}(x,\T{k}{};Q) = 
\frac{1}{(2 \pi)^2} \int \diff{^2\T{b}{}}{} e^{i \T{k}{} \T{b}{}}
\tilde{f}^{\text{Evol}}_{q/p}(x,\T{b}{};Q) \, .
\end{equation}
The divergence that comes from integrating over all transverse momentum now appears as the divergence at $\Tsc{b}{} = 0$ in the integrand of \eref{bspaceTMDpdf}. To regulate it, one may replace $\Tsc{b}{}$ inside $\tilde{f}_{q/p}(x,\T{b}{};Q)$ with a function that levels off at a constant lower bound $\bmin$ as $\Tsc{b}{} \to 0$. An example is~\cite{Collins:2016hqq},
\begin{equation}
b_c(\Tsc{b}{}) = \sqrt{\Tscsq{b}{} + \bmin^2} \, ,
\end{equation}
but any well-behaved interpolating function is acceptable. 
Replacing $\tilde{f}^{\text{Evol}}_{q/p}(x,\T{b}{};Q)$ in \eref{bspaceTMDpdf} by $\tilde{f}^{\text{Evol}}_{q/p}(x,b_c(\Tsc{b}{});Q)$ and integrating over all transverse momentum gives
\begin{align}
\pi \int_0^{\infty} \diff{\Tscsq{k}{}}{} f^{\text{Evol}}_{q/p}(\xbj,\T{k}{};Q) &{}\to \tilde{f}^{\text{Evol}}_{q/p}(\xbj,\bmin;Q) \no
= f_{q/p}(\xbj;b_0/\bstarsc(\bmin)) & \exp \left\{-2 \int_{b_0/\bstarsc(\bmin)}^Q \frac{\diff{\mu}{}}{\mu} \gamma_2(a_\lambda(\mu))
-g_{q/p}(\xbj,\bmin)\right\} \, \no
= f_{q/p}(\xbj;b_0/\bmin) & + \order{\frac{\bmin^2}{\bmax^2},m^2 \bmax^2,m^2 \bmin^2,a_\lambda^2}  \, . \label{e.intrel}
\end{align}
When $m \ll 1/\bmax \ll Q$, and $\bmin \approx b_0/Q$, the error terms on the last line are negligible, and the result is the expected 
\begin{equation}
\pi \int_0^{\infty} \diff{\Tscsq{k}{}}{} f^{\text{Evol}}_{q/p}(\xbj,\T{k}{};Q) \approx f_{q/p}(\xbj;Q) \, . \label{e.naive2}
\end{equation}
One possibly misleading aspect of with this way of reconstructing the integral relation in \eref{naiveint} is that it might seem to suggest that there is no role for $g_{q/p}(\xbj,\boldsymbol{b}_\text{min})$ after the transverse momentum integration. 
However, recall that $Q  \approx Q_0$ is, by definition, an acceptably large $Q$ to use with factorization.  But when $Q \approx Q_0$, we have both $\bmax \approx 1/Q_0$ and $\bmin \approx b_0/Q_0$. Therefore, near $Q \approx Q_0$, the ratio $\bmin/\bmax$ does not give a power suppression, and dropping the errors in the last line of \eref{intrel} is unjustified. We can see this problem directly in the Yukawa theory calculation. For example, factorization at $\xbj \approx 0.1$ describes the unfactorized cross section very well even for $Q \approx 2.0$~GeV. However, with $\bmax = 0.25$~GeV and $\bmin = b_0/(2~\text{GeV})$, 
\begin{equation}
\frac{f_{q/p}(\xbj = .1;b_0/\bstarsc(\bmin))}{f_{q/p}(\xbj = .1;b_0/\bmin)} \approx 1.4 \, . \label{e.tmdcutcompare1}
\end{equation}
While it is true that the approximate integral relation in \eref{naiveint} does hold very well even for these scales, it requires that one keep an accurate $g_{q/p}(\xbj,\boldsymbol{b}_\text{min})$ in the exponent on the second line of \eref{intrel},
\begin{equation}
\frac{f_{q/p}^{\text{Evol}}(\xbj = .1;\bmin;2~\text{GeV})}{f_{q/p}(\xbj = .1;b_0/\bmin)} \approx 1.0 \, .
\label{e.tmdcutcompare2}
\end{equation}
In these calculations, we have used $\mquark = 0.3$~GeV, and $\tarmass = \mgluon=  1.0$~GeV.
Dropping the $g_{q/p}(\xbj,\boldsymbol{b}_\text{min})$ is only acceptable in the limit where $Q \gg Q_0$. In other words, near the input scale where hadronic structure effects are most likely to be relevant, and where it is desirable to preserve the parton structure interpretation embodied by \eref{naiveint} as closely as possible, the $g$-functions cannot be neglected. 

\section{Discussion}
\label{s.conclusion}

We have shown that full calculations of DIS cross sections in QFTs that are simpler than QCD, but which nonetheless require renormalization, are useful for highlighting general but subtle properties of parton densities and for examining the limits of factorization while side stepping issues like confinement, large coupling, gauge invariance, and other complicating features of QCD. We have illustrated several examples in this paper, including the demonstration in \sref{TMDandSIDIS} of how the large-$\Tsc{q}{}$ ``$Y$-term'' contribution to SIDIS is necessary to maintain a reasonably accurate description after the inclusive integral over all $\Tsc{q}{}$, and the importance of the $g_{q/p}(\xbj,\T{b}{})$ function in \eref{OPE_express} for maintaining the usual parton model relationship, \eref{naiveint}, between the collinear and TMD pdfs near the input scale $Q \approx Q_0$. (Compare \eref{tmdcutcompare1} and \eref{tmdcutcompare2}.) Our discussion of the GPM following \eref{hadrotens2} emphasizes the importance of taking into account the UV divergences in TMD functions when they are integrated over transverse momentum, especially in applications that require higher precision than a leading power parton model can provide. See \cite{Rogers:2020tfs} for more on this.

The steps for factorizing the cross section, reviewed in \srefs{pdfs}{collinearsteps}, exemplify the difference between what in \cite{Collins:2021vke} are called ``track-A'' and ``track-B'' approaches. The steps in this paper are in the track-A approach; they begin with the bare pdf in \eref{pdfdef}, and no actual collinear divergences ever arise. 

Of course, a detailed description of the transition between ``perturbative'' ($m$-independent) and ``nonperturbative'' ($m$-sensitive) behavior depends entirely on the nature of intrinsic mass scales like $\mquark$, $\mgluon$ and $\tarmass$.  

Although many basic features of factorization are present in both QCD and in the Yukawa toy theory, there are, of course, major differences between the two theories, and attempts to draw any conclusions about one from the other should be made only with extreme caution. In the Yukawa theory, the target ``hadron'' state we have considered is not a bound state, all interactions are point-like, and there is no confinement or asymptotic freedom.  In some sense, the Yukawa theory is a version of the scalar diquark model that is sometimes used in phenomenology (e.g.~\cite{Bacchetta:2008af,Kang:2010hg,Guerrero:2020hom}) in the limit that the quark-hadron coupling is point-like. One particularly noticeable difference comes from absence of soft gluons and lightcone divergences in the Yukawa theory as compared to the gauge theories. These soft gluons are responsible for, among other things, the well-known nonperturbative Collins-Soper evolution factor $e^{-g_K \ln(Q/Q_0)}$ that is present in QCD but not in \eref{approxevol}. (See, e.g., Eq.~(114) of \cite{Gonzalez-Hernandez:2022ifv}.) They also lead to the TMD pdf turning negative at large transverse momentum. See, for example, Fig.~5 of \cite{Gonzalez-Hernandez:2022ifv}. Contrast this with the plots in \fref{ktplots} for the Yukawa theory, which are strictly positive.
Indeed, in the Yukawa theory, the transition to the small transverse momentum asymptote happen rather quickly as $\Tsc{k}{}$ decreases below $Q$ -- see~\fref{TMD_WY_vs_unfact}, especially for smaller $\xbj$. In QCD, one generally must consider very small $\Tsc{k}{}$ before the asymptotic limit is reached. See, for example, \cite{Boglione:2014oea}. 

Despite these cautionary remarks, we nevertheless hope that calculations like these will be useful for stress testing other general assertions about factorization and the properties of pdfs in the future. To this end, we have made a convenient Wolfram Mathematica package for generating the cross sections and pdfs of this paper available at~\cite{yuknb}. 

\vskip 0.3in
\acknowledgments
We thank Markus Diefenthaler for useful comments. 
Ted Rogers and Tommaso Rainaldi were supported by the U.S. Department of Energy, Office of Science, Office of Nuclear Physics, under Award Number DE-SC0018106. 
Fatma Aslan was supported by NSF 
under the Award No.~1812423 and  Award No.\ 2111490,
by the U.S.~Department of Energy.
L.G. is supported by the U.S. Department of Energy, Office of Science, Office of Nuclear Physics, under Contract No. DE-FG02-07ER41460. This work 
was also supported by the DOE Contract No. DE- AC05-06OR23177, under which 
Jefferson Science Associates, LLC operates Jefferson Lab.

\begin{appendix}

\section{Details of the unapproximated calculation}
\label{a.exact}

\begin{figure}[h!]
\centering
\includegraphics[width=8cm]{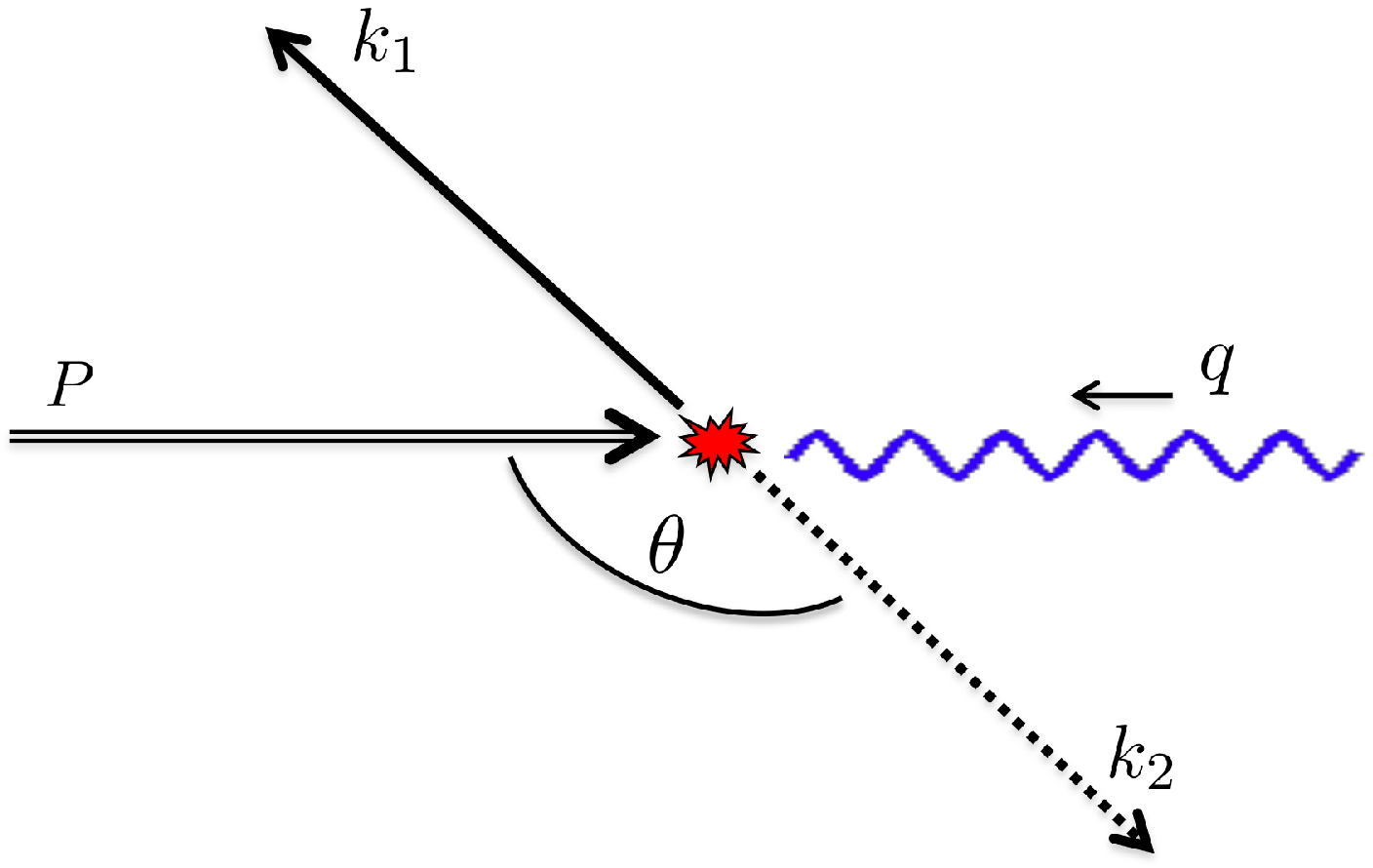}
\caption{Center-of-mass kinematics of semi-inclusive DIS at the order shown in \fref{basicmodel}.}
\label{f.kindiag}
\end{figure}

\subsection{Center-of-mass kinematics}

Following \fref{kindiag}, define the momenta
\begin{align}
k_1 &{}= k + q\, , \\
k_2 &{}= p - k\, , \\
k_1^2 &{}= \mquark^2 \, , \\
k_2^2 &{}= \mgluon^2 \, .
\end{align}
Then, in the center-of-mass
\begin{align}
s &{}= \parz{p + q}^2 = \tarmass^2 + \frac{Q^2 (1-\xbj)}{\xbj} \\
P &{}= \parz{p^0,0,0,\sqrt{p^{0 \, 2} - \tarmass^2}} \\
q &{}= \parz{q^0,0,0,-\sqrt{p^{0 \, 2} - \tarmass^2}} \\
k_1^0 + k_2^0 &{}= \sqrt{s} \equiv 2 E \\
\sqrt{k_1^{0 \, 2} - \mquark^2} &{}= -\sqrt{k_2^{0 \, 2} - \mgluon^2} 
\end{align}
\begin{equation}
p^0 = \frac{\tarmass^2 + Q^2/(2 \xbj)}{2 E}
\end{equation}
\begin{align}
k_1^0 &{}= E - \frac{\mgluon^2 - \mquark^2}{4 E} \, , \\
k_2^0 &{}= E + \frac{\mgluon^2 - \mquark^2}{4 E} \, . \\
\end{align}
\begin{equation}
q^0 = \begin{cases}
	+\sqrt{p^{0\, 2} - \tarmass^2 - Q^2} \,& \text{if} \;\;  x < 0.5 \\
	-\sqrt{p^{0\, 2} - \tarmass^2 - Q^2} \, & \text{if} \;\; x > 0.5 \,  \\
\end{cases} \, .
\end{equation}
For $\phi = 0$,
\begin{align}
k_1 &{}= \parz{k_1^0,\ktmax \sin \theta,0,\ktmax \cos \theta} \, \\
k_2 &{}= \parz{k_2^0,-\ktmax \sin \theta,0,-\ktmax \cos \theta}
\end{align}
where

\begin{equation}
\label{e.ktmaxval}
\ktmax = \sqrt{E^2 - \frac{\mgluon^2 + \mquark^2}{2} + \frac{\parz{\mgluon^2 - \mquark^2}^2}{16 E^2}}
\end{equation}

\begin{equation}
\Tscsq{k}{}= \ktmax^2 \parz{1 - \cos^2 \theta} \, .
\end{equation}
\begin{equation}
\cos \theta = 
\begin{cases}
	-\sqrt{1 - \frac{\Tscsq{k}{}}{\ktmax^2}} \,& \text{if} \;\;  \pi/2 \leq \theta \leq \pi \\
	\\
	+\sqrt{1 - \frac{\Tscsq{k}{}}{\ktmax^2}} \, & \text{if} \;\; 0 \leq \theta \leq \pi/2 \,  \\
\end{cases} \, .
\end{equation}
The Mandelstam variables are
\begin{align}
t &{}= k^2 = \tarmass^2 + \mgluon^2 - 2 p^{0} k_2^0 
\parz{1 + \frac{\ktmax p_z}{p^0 k_2^0} \cos \theta } \, \\
u &{}= \tarmass^2 + \mquark^2 - 2 p^{0} k_1^0 
\parz{1 - \frac{\ktmax p_z}{p^0 k_1^0} \cos \theta } \, ,
\end{align}
Note that
\begin{align}
p \cdot k &{}= \frac{k^2 - \mgluon^2 + \tarmass^2}{2} , \\
q \cdot k &{}= \frac{-k^2 + \mquark^2 + Q^2}{2}\, , \\
p \cdot q &{}= \frac{Q^2}{2 \xbj} \, .
\end{align}
Also, there is a kinematical upper bound on the value of $\xbj$:
\begin{equation}
\label{e.xmax}
x_\text{max} = \frac{Q^2}{\left[ (\mquark + \mgluon)^2 - \tarmass^2 + Q^2 \right]} \, .
\end{equation}
\subsection{Organization of the calculation}

We organize each graphical contribution to the hadronic tensor by expressing it as
\begin{equation}
W^{\mu \nu} = \frac{1}{2} \frac{1}{4 \pi} \int \frac{\diff{^4 k}{}}{(2 \pi)^4} \frac{N^{\mu \nu}}{D} (2 \pi) \delta_+ ((P-k)^2 - \mgluon^2) (2 \pi) \delta_+ ((k+q)^2 - \mquark^2) 
 \, .
 \label{e.mod_hadtens}
\end{equation}
$N^{\mu \nu}$ is the collection of all numerator factors from Dirac traces, etc. $D$ is the collection of all propagator denominators. We can also write
\begin{equation}
W^{\mu \nu} = \frac{1}{2} \frac{1}{4 \pi} \int \frac{\diff{^2\T{k}{}}{}}{(2 \pi)^2} W^{\mu \nu}_\text{INT} \, 
\end{equation}
where 
\begin{align}
W^{\mu \nu}_\text{INT} &{}\equiv \int \diff{k^+}{} \diff{k^-}{} \frac{N^{\mu \nu}}{D} \delta_+ ((P-k)^2 - \mgluon^2) \delta_+ ((k+q)^2 - \mquark^2)  \, .
\end{align}
To express $W^{\mu \nu}$ in terms of the SIDIS hadronic tensor, recall that
\begin{align}
 \langle N \rangle W^{\mu \nu} &{}= \sum_B \int \frac{\diff{^2 \T{k}{}}{} \diff{\zn}{}}{4 \zn} W_\text{SIDIS}^{\mu \nu}
 \label{e.mod_hadtens2} \, .
\end{align}
In all the graphs of \fref{basicmodel}, $\langle N \rangle = 2$. But the computation is identical when $B$ is the quark and when it is the scalar, so the $\sum_B$ also gives a factor of $2$. Thus, we can write simply 
\begin{align}
 W^{\mu \nu} &{}= \int \frac{\diff{^2 \T{k}{}}{} \diff{\zn}{}}{4 \zn} W_\text{SIDIS}^{\mu \nu}
 \label{e.mod_hadtens3} \, .
\end{align}
so
\begin{align}
\frac{W^{\mu \nu}_\text{INT}}{32 \pi^3} =\int \frac{\diff{\zn}{}}{4 \zn} W_\text{SIDIS}^{\mu \nu} \, .
\label{e.mod_relation}
\end{align}
Since $z$ is always fixed by final state kinematics in the low order graphs of \fref{basicmodel}, we will find it more convenient to work with $\frac{W^{\mu \nu}_\text{INT}}{32 \pi^3}$ than with $W_\text{SIDIS}^{\mu \nu}$ directly. 

Switching variables from $k$ to $k_2$, \eref{mod_hadtens} becomes
\begin{align}
W^{\mu \nu} 
&{}= \frac{1}{2} \frac{1}{4 \pi} \int \frac{\diff{k_2^0}{} \diff{\left| {\bf k}_2 \right|}{} \left| {\bf k}_2 \right|^2 \diff{\Omega}{} }{(2 \pi)^4} \frac{N^{\mu \nu}}{D} (2 \pi) \delta_+ (k_2^2 - \mgluon^2) (2 \pi) \delta_+ ((p + q - k_2)^2 - \mquark^2) \no
 &{}= \frac{1}{2} \frac{1}{4 \pi} \int \frac{\diff{k_2^0}{} \diff{\left| {\bf k}_2 \right|}{} \left| {\bf k}_2 \right|^2 \diff{\Omega}{} }{(2 \pi)^2} \frac{N^{\mu \nu}}{D}  \frac{1}{2 k_2^0} \delta \parz{k_2^0 - \sqrt{\left| {\bf k}_2 \right|^2 +\mgluon^2}}  \delta_+ ((p + q - k_2)^2 - \mquark^2) 
 \, \no
 &{}= \frac{1}{2} \frac{1}{4 \pi} \int_0^{\infty} \frac{\diff{\left| {\bf k}_2 \right|}{} \left| {\bf k}_2 \right|^2}{2 k_2^0 (2 \pi)^2} \int \diff{\Omega}{} \frac{N^{\mu \nu}}{D} \frac{k_2^0}{4 E \left| {\bf k}_2 \right|} \delta (\left| {\bf k}_2 \right| - \ktmax) \no
 &{}=\frac{1}{128 \pi^2} \int_{-1}^{1} \diff{\parz{\cos \theta}}{} \;  \frac{\ktmax}{E} \frac{N^{\mu \nu}}{D} \, .
\end{align}
where in the last line we have kept only azimuthally symmetric contributions. $N^{\mu \nu}$ and $D$ for each graph can be simplified by writing
\begin{align}
k^2 - \mquark^2 &{}= \tarmass^2 + \mgluon^2 - \mquark^2 - 2 p^{0} k_2^0 
\parz{1 + \frac{\ktmax p^z}{p^0 k_2^0} \cos \theta } \nonumber \\
&{}=A - B\parz{1 \pm C \sqrt{1 - \frac{\Tscsq{k}{}}{\ktmax^2}}} \, ,
\label{e.densimp}
\end{align} 
where
\begin{align}
A &= M_p^2 + \mgluon^2 - \mquark^2 \\
B &= 2 p^0 k_2^0 \\ 
C &= \frac{\ktmax p^z}{p^0 k_2^0} \, .
\end{align}
The minus sign in \eref{densimp} is for the $\theta > \pi/2$ contribution and the plus sign is for $\theta < \pi/2$. 
Changing variables back to transverse momentum and using $\diff{\theta} \sin \theta = \frac{\diff{\Tscsq{k}{}}{}}{2 k_m^2 \cos \theta}$,
\begin{equation}
\label{e.Wcoll}
W^{\mu \nu} = \left. \frac{1}{32 \pi^2} \int_{0}^{\ktmax^2} \; \frac{\diff{\Tscsq{k}{}}{}}{8 E \ktmax\sqrt{1 - \frac{\Tscsq{k}{}}{\ktmax^2}}}   \left( \frac{N^{\mu \nu}}{D} \right|_{\theta > \pi/2} + \left.  \frac{N^{\mu \nu}}{D} \right|_{\theta < \pi/2} \right) \, .
\end{equation}
So, 
\begin{equation}
W^{\mu \nu}_\text{INT} = \frac{1}{8 E \ktmax \sqrt{1 - \frac{\Tscsq{k}{}}{\ktmax^2}}}
\parz{ \left. \frac{N^{\mu \nu}}{D} \right|_{\theta > \pi/2} +
\left. \frac{N^{\mu \nu}}{D} \right|_{\theta < \pi/2} } \, .
\end{equation}
and
\begin{equation}
\text{Jac} = \frac{1}{8 E \ktmax \sqrt{1 - \frac{\Tscsq{k}{}}{\ktmax^2}}} \, .
\end{equation}
In general we may also account for the azimuthal angle $\varphi$ so that it is
\begin{equation*}
   \diff{^2\T{k}{}}{} = \Tsc{k}{}\diff{\Tsc{k}{}}{}\diff{\varphi}{} = \frac{\diff{\varphi}{}}{2}\diff{\Tscsq{k}{}}{} = -\diff{\varphi}{}k_m^2\cos\theta\diff{\cos\theta}{}
\end{equation*}
So,
\begin{equation*}
\begin{split}
     W^{\mu\nu}&\equiv \frac{1}{2}\frac{1}{4\pi}\int\frac{\diff{^2\T{k}{}}{}}{(2\pi)^2}W^{\mu\nu}_{\rm INT}\\
     & = \frac{1}{128\pi^2}\int_{-1}^1 \diff{\cos\theta}{}\int_0^{2\pi}\frac{\diff{\varphi}{}}{(2\pi)}\frac{k_m}{E}\frac{N^{\mu\nu}}{D}\\
     & = \frac{1}{256\pi^2}\int_0^{2\pi}\frac{\diff{\varphi}{}}{(2\pi)}\int_0^{k^2_m}\frac{\diff{\Tscsq{k}{}}{}}{k_mE\,|\kappa(\Tscsq{k}{})|}\left(\left.\frac{N^{\mu \nu}}{D} \right|_{+\kappa} +
\left. \frac{N^{\mu \nu}}{D} \right|_{-\kappa}\right) \\
& = \frac{1}{32\pi^3}\int \frac{\diff{\varphi}{}}{2}\diff{\Tscsq{k}{}}{}\frac{1}{8k_mE|\kappa(\Tscsq{k}{})|}\left(\left.\frac{N^{\mu \nu}}{D} \right|_{+\kappa} +
\left. \frac{N^{\mu \nu}}{D} \right|_{-\kappa}\right)
\end{split}
\end{equation*}
where
\begin{equation*}
    \kappa(\Tscsq{k}{})\equiv \sqrt{1-\frac{\Tscsq{k}{}}{k_m^2}}
\end{equation*}
In the end it is thus
\begin{equation*}
    W^{\mu\nu}_{\rm INT}(\Tscsq{k}{},\varphi) = \frac{1}{8k_mE|\kappa(\Tscsq{k}{})|}\left(\left.\frac{N^{\mu \nu}}{D} \right|_{+\kappa} +
\left. \frac{N^{\mu \nu}}{D} \right|_{-\kappa}\right).
\end{equation*}

\subsection{Structure functions}

For dealing with specific structure functions, we first write another abbreviation, 
\begin{equation}
I^{\mu \nu} = \frac{N^{\mu \nu}}{D} \, .
\end{equation}
And 
\begin{equation}
\contractortot{j}^{\mu \nu} I_{\mu \nu} = I_j \, .
\end{equation}
where $j$ labels a projection tensor. For any $j$, the $I_j$ for the sum of graphs may be expanded in powers of the $t$-channel propagator, 
\begin{equation}
\label{e.expansion}
I_j = I_{j,2} (k^2 - \mquark^2)^2 + I_{j,1} (k^2 - \mquark^2) + I_{j,0} + I_{j,-1} \frac{1}{k^2 - \mquark^2} + I_{j,-2} \frac{1}{(k^2 - \mquark^2)^2} \, ,
\end{equation}
where the $I_{j,n}$'s are independent of $\Tsc{k}{}$ and depend only on masses, $\xbj$, and $Q$.
All the unpolarized and azimuthally symmetric structure functions can be expressed in terms of linear combinations of the projection tensors
\begin{align}
\contractortot{g}^{\mu\nu} = g^{\mu\nu} ,\;\quad
\contractortot{PP}^{\mu\nu} = P^\mu P^\nu \, .
\label{e.PgPPtot}
\end{align}
So a natural next step is to determine 
\begin{equation}
W_g = \contractortot{g}^{\mu \nu} W_{\mu \nu} \, \quad \& \quad 
W_{PP} = \contractortot{PP}^{\mu \nu} W_{\mu \nu} \, .
\end{equation}

First, write projections of \eref{Wcoll} as 
\begin{equation}
\label{e.Wcoll2}
W_j = \contractortot{j}^{\mu \nu} W_{\mu \nu} = \frac{1}{32 \pi^2} \int_{0}^{\ktmax^2} \; \frac{\diff{\Tscsq{k}{}}{}}{8 E \ktmax\sqrt{1 - \frac{\Tscsq{k}{}}{\ktmax^2}}}   \left( I_j |_{\theta > \pi/2} + \left. I_j \right|_{\theta < \pi/2} \right) = 
\pi \int_{0}^{\ktmax^2} \diff{\Tscsq{k}{}}{} \, \frac{W_{\text{INT},j}}{32 \pi^3} \, .
\end{equation}
Then all the integrals over $\Tscsq{k}{}$ take one of the following forms
\begin{align}
&{}\Gamma^+_2 = \int_{0}^{\ktmax^2} \; \left. \frac{\diff{\Tscsq{k}{}}{}}{\sqrt{1 - \frac{\Tscsq{k}{}}{\ktmax^2}}} (k^2 - \mquark^2)^2 \right|_{\theta < \pi/2} = 2 \ktmax^2 \left(A^2-A B (C+2)+\frac{1}{3} B^2 (C
   (C+3)+3)\right) \label{e.gam2plus} \\
&{}\Gamma^+_1 = \left. \int_{0}^{\ktmax^2} \; \frac{\diff{\Tscsq{k}{}}{}}{\sqrt{1 - \frac{\Tscsq{k}{}}{\ktmax^2}}} (k^2 - \mquark^2) \right|_{\theta < \pi/2} = 2 \ktmax^2 \left(A-\frac{1}{2} B (C+2)\right) \\
&{}\Gamma^+_0 = \left. \int_{0}^{\ktmax^2} \; \frac{\diff{\Tscsq{k}{}}{}}{\sqrt{1 - \frac{\Tscsq{k}{}}{\ktmax^2}}}  \right|_{\theta < \pi/2} = 2 \ktmax^2 \\
&{}\Gamma^+_{-1} = \left. \int_{0}^{\ktmax^2} \; \frac{\diff{\Tscsq{k}{}}{}}{\sqrt{1 - \frac{\Tscsq{k}{}}{\ktmax^2}}} \frac{1}{(k^2 - \mquark^2)} \right|_{\theta < \pi/2} = \frac{2 \ktmax^2}{BC} \ln \left( \frac{B-A}{B(C+1) -A} \right) \\
&{}\Gamma^+_{-2} = \left. \int_{0}^{\ktmax^2} \; \frac{\diff{\Tscsq{k}{}}{}}{\sqrt{1 - \frac{\Tscsq{k}{}}{\ktmax^2}}} \frac{1}{(k^2 - \mquark^2)^2} \right|_{\theta < \pi/2} = 2 \ktmax^2\frac{1}{(B-A) (-A+B C+B)} \, . \label{e.gammin2plus}
\end{align}
The $+$ superscript here means these are the integrals for $\theta < \pi /2$ and so the $C$ comes with a $+$ sign. Identical expressions apply to $\Gamma^-_n$ but with $C \to -C$. The subscripts refer to the power on $(k^2 - \mquark^2)$.

Substituting~\eref{expansion}~ into~\eref{Wcoll2} and using \erefs{gam2plus}{gammin2plus} gives an analytic expression 
\begin{equation}
\label{e.analytic}
W_j = \frac{1}{256 E \ktmax \pi^2} \left[ I_{j,2} (\Gamma_2^+ + \Gamma_2^-) + I_{j,1} (\Gamma_1^+ + \Gamma_1^-) + 2 \Gamma_0^+ I_{j,0} + I_{j,-1} (\Gamma_{-1}^+ + \Gamma_{-1}^-) + I_{j,-2} (\Gamma_{-2}^+ + \Gamma_{-2}^-) \right] \, ,
\end{equation}
where 
\begin{align}
\Gamma_2^+ + \Gamma_2^- &{}= 4 \ktmax^2 \left((A-B)^2+ \frac{B^2C^2}{3}   \right) \\
\Gamma_1^+ + \Gamma_1^- &{}= 4 \ktmax^2 \left( A - B \right) \\
\Gamma_0^+ + \Gamma_0^- &{}= 4 \ktmax^2 \\
\Gamma_{-1}^+ + \Gamma_{-1}^- &{}= -\frac{2 \ktmax^2}{BC} \ln \left( \frac{B(1+C) -A}{B(1-C) -A} \right) \\
\Gamma_{-2}^+ + \Gamma_{-2}^- &{}= 4 \ktmax^2 
\left( \frac{1}{(A - B)^2 - B^2 C^2} \right) \, ,
\end{align}
All that is needed to get explicit expressions for structure functions is to get the $I_{j,n}$ for $\contractortot{g}^{\mu \nu}$ and $\contractortot{PP}^{\mu \nu}$ and specify numerical values for the masses $Q^2$ and $\xbj$. 

\subsection{Simple case}

The explicit expressions for the $I$'s, $A$, $B$, and $C$ are cumbersome for general combinations of masses, so we will write them here only for a special case. The expressions for $A$, $B$, $C$, $\ktmax$, $E$ and the $I$'s are especially simple for  
$\pmass = m$, $\mgluon = 2 m$, and $\mquark = m$. The results are
\begin{align}
\ktmax &{} = \frac{1}{2} \sqrt{-\frac{Q^2 (\xbj-1) \left(8 m^2 \xbj+Q^2 (\xbj-1)\right)}{\xbj \left(Q^2 (\xbj-1)-m^2
   \xbj\right)}} \label{e.kmexample} \\
E &{}= \frac{1}{2} \sqrt{m^2-\frac{Q^2 (\xbj-1)}{\xbj}}  \\
A &{}= 4 m^2 \\
B &{}= \frac{\left(Q^2 (\xbj-1)-4 m^2 \xbj\right) \left(2 m^2 \xbj+Q^2\right)}{2 \xbj \left(Q^2
   (\xbj-1)-m^2 \xbj\right)} \\
C &{}= -\frac{Q^2 \sqrt{(\xbj-1) \left(8 m^2 \xbj+Q^2 (\xbj-1)\right) \left(4 m^2
   \xbj^2+Q^2\right)}}{\left(Q^2 (\xbj-1)-4 m^2 \xbj\right) \left(2 m^2 \xbj+Q^2\right)} \\
I_{g,-2} &{}= 0 \\
I_{g,-1} &{}= -\frac{4 \left(Q^2 (\xbj-1)^2-8 m^2 \xbj^2\right)}{(\xbj-1) \xbj} \\
I_{g,0} &{}= 8 \\
I_{g,1} &{}= \frac{4 \xbj}{Q^2 (1-\xbj)} \\
I_{g,2} &{}= 0 \,  \\
I_{pp,-2} &{}= 0  \\
I_{pp,-1} &{} = \frac{32 m^4 \xbj}{\xbj-1}-\frac{2 m^2 Q^2 (\xbj+3)}{\xbj}\\
I_{pp,0} &{}= \frac{4 m^2 (1-3 \xbj)}{\xbj-1}+\frac{2 Q^2}{\xbj} \\
I_{pp,1} &{}= 2-\frac{2 m^2 \xbj}{Q^2 (\xbj-1)} \\ 
I_{pp,2} &{}= 0
\label{e.Ippexample} \, \, . 
\end{align}
Therefore, in this case we may write \eref{analytic} as 
\begin{equation}
\label{e.analytic2}
W_j = \frac{\ktmax}{64 E \pi^2} \left[ I_{j,2} \left(A^2-2 A B +\frac{1}{3} B^2 (1 + C^2)  \right) + I_{j,1} \left( A - B \right) + I_{j,0} - I_{j,-1} \frac{1}{2BC} \ln \left( \frac{B(1+C) -A}{B(1-C) -A} \right) \right] \, .
\end{equation}
and substitute \erefs{kmexample}{Ippexample} to get numerical values for $W_g$ and $W_{PP}$.
The standard structure functions $F_1$ and $F_2$ are
\begin{align}
F_1 {}& \equiv 
 -\frac{1}{2} W_g
   +\frac{2 Q^2 \xn^2}{(\pmass^2 \xn^2 + Q^2)^2} W_{PP} \label{e.Con1} \, ,
                          \\
F_2 {}& \equiv \frac{12 Q^4 \xn^3 \left(Q^2-\pmass^2 \xn^2\right)}
	 {\left(Q^2 + \pmass^2 \xn^2\right)^4}
    \left( W_{PP}
	 -\frac{\left(\pmass^2 \xn^2+Q^2\right)^2}{12 Q^2 \xn^2}
	   W_g \right) \, . \label{e.Con2}
\end{align}
Here, $\xn$ is Nachtmann $x$:
\begin{equation}
\xn = \frac{2 \xbj}{1 + \sqrt{1 + \frac{4 \xbj^2 \pmass^2}{Q^2}}} \, .
\end{equation}


\section{The small-$\Tsc{b}{}$ operator product expansion}
\label{a.opecalc}

In this appendix, we extract the small-$\Tsc{b}{}$ limit of the TMD pdf (or ff) in collinear factorization. We start with the quark pdf in \eref{fbspace} and consider the limit where $\Tsc{b}{} \approx 1/\mu$ with $\mu/m \to \infty$, 
In the region of the integrand where $\Tsc{k}{} \approx m$,  we can expand it as, 
\begin{equation}
e^{-i \T{k}{} \cdot \T{b}{} } f_{q/p}(x,\T{k}{};\mu) = f_{i/p}(x,\T{k}{};\mu) + \order{\Tscsq{k}{} \Tscsq{b}{} }
= f_{i/p}(x,\T{k}{};\mu) + \order{\frac{\Tscsq{k}{}}{\mu^2} } 
\, . \label{e.opeAP1}
\end{equation}
Thus, we define the small-$\Tsc{k}{}$ part of the approximation to \eref{fbspace} as 
\begin{equation}
\tilde{f}_{q/p}(x,\T{b}{}\approx 1/\mu;\mu) \approx \int_\text{UVR} \diff{^2 \T{k}{}}{} f_{q/p}(x,\T{k}{};\mu) \equiv f_{q/p}(x;\mu) \, . \label{e.firstapprox}
\end{equation}
Dropping the $e^{-i \T{k}{} \cdot \T{b}{} }$ factor causes the integral over $\Tsc{k}{}$ to become UV divergent, so we must introduce a UV regulator and/or UV renormalization scheme to have a well-defined $f(x;\mu)$. We notate this with the ``$\text{UVR}$,'' for ``UV regulator,'' on the integral sign. The specific choice of UV regulator is arbitrary.  Equation~\eqref{e.firstapprox} is a poor approximation because of the UV divergence and the $\order{\Tscsq{k}{}/\mu^2}$ correction in \eref{opeAP1}, which is unsuppressed when $\Tsc{k}{} \approx \mu$. However, it can be corrected by using collinear perturbation theory for large enough $\mu$. To implement this, we rewrite \eref{fbspace} in the small $\Tsc{b}{}$ region as 
\begin{equation}
\label{e.opePM}
\tilde{f}_{q/p}(x,\T{b}{}\approx 1/\mu;\mu) = f_{q/p}(x;\mu) + 
\int \diff{^2 \T{k}{}}{}  \left\{ e^{-i \T{k}{} \cdot \T{b}{} } f_{q/p}(x,\T{k}{};\mu) - 
\delta^{(2)}(\T{k}{}) f_{q/p}(x;\mu) \right\} \, .
\end{equation}
The integrand of the second term in \eref{opePM} is the contribution from the $\order{\Tscsq{k}{}/\mu^2} \sim \order{\Tscsq{k}{}\Tscsq{b}{}}$ error in \eref{opeAP1} if we transform into $\Tsc{b}{}$-space and use the definition in \eref{firstapprox}. The full integral in \eref{opePM} is both IR and collinear finite, so there is no need for a $\text{UVR}$ in the integral. By construction, its only unsuppressed contribution comes form the region where $\Tsc{k}{}$ is comparable to $\mu$. However, the $\Tsc{k}{} \approx \mu$ behavior of $f_{q/p}(x,\T{k}{};\mu)$ is expressible in collinear factorization, 
\begin{equation}
\label{e.asyexp}
\Tsc{k}{} f_{q/p}(x,\T{k}{} \approx \mu;\mu) = \sum_i \Tsc{k}{} \frac{1}{\Tscsq{k}{}} \left[ \mathcal{C}_{q/i} \otimes f_{i/p}\right](x,\Tsc{k}{}/\mu) + \order{\frac{m^2}{\Tscsq{k}{}}} \, .
\end{equation}
See \eref{asym_pdf} for the Yukawa theory version of this.
The hard coefficient $\mathcal{C}_{q/i}$ can be expanded in collinear perturbation theory. It is dimensionless, but it depends on logarithms of $\Tsc{k}{}/\mu$. 
The $\order{m^2/\Tscsq{k}{}}$ in \eref{asyexp} combines with the $\order{\Tscsq{k}{}\Tscsq{b}{}}$ error from \eref{opeAP1} to give an overall error that is suppressed by $\order{m^2 \Tscsq{b}{}}$. 
Up to $\order{m^2 \Tscsq{b}{}}$ terms, therefore, the right side of \eref{opePM} only involves collinear pdfs. This is the small-$\Tsc{b}{}$ OPE for the unpolarized case. Notice that the unsuppressed term in \eref{opePM} is not completely unique because it depends upon the scheme used in \eref{firstapprox} for treating the UV region. 

The full statement of the transverse coordinate space OPE for the quark TMD pdf is 
\begin{align}
\label{e.ope2}
\tilde{f}_{q/p}(x,\T{b}{};\mu) = 
\sum_j \int_{x}^1 \frac{\diff{\xi}{}}{\xi} \mathcal{\tilde{C}}_{q/j}(x/\xi,\T{b}{};\mu) \tilde{f}_{j/p}(\xi;\mu) + \order{m^2 \Tscsq{b}{}}{} \, , 
\end{align}
Reading off the first term from \eref{opePM} gives the zeroth order hard coefficient 
\begin{equation}
\mathcal{\tilde{C}}_{q/j}^{(0)} = \delta(1-x/\xi) \delta_{qj} \, .
\end{equation}
The higher order $\mathcal{\tilde{C}}_{q/j}^{(n)}$ with $n \geq 1$ are calculable from the second term in \eref{opePM}. The steps are to: i.) calculate the integral, ii.) extract the small-$\Tsc{b}{}$ limit, iii.)  drop any $\Tsc{b}{}$-suppressed corrections, and iv.) identify the $\mathcal{\tilde{C}}_{q/j}^{(n)}$ coefficients. The coefficients are independent of the target, so one  generally uses a massless parton target to calculate. Thus, one calculates
\begin{equation}
\label{e.opePM2}
\int \diff{^2 \T{k}{}}{}  \left\{ e^{-i \T{k}{} \cdot \T{b}{} } f^\text{partonic}_{q/i}(x,\T{k}{};\mu) - 
\delta^{(2)}(\T{k}{}) f^\text{partonic}_{q/i}(x;\mu) \right\} \, ,
\end{equation}
where $i$ is a massless parton. With a massless parton target, $f^\text{partonic}_{q/i}(x,\T{k}{};\mu)$ can, in general, involve terms proportional to $\delta^{(2)}(\T{k}{})$. 

In the Yukawa theory example, it is easiest to do the calculation of $\mathcal{\tilde{C}}_{q/j}^{(1)}$ by dealing with each term in the integral in \eref{opePM2} separately. We will also make the replacement $\Tscsq{k}{} \to \Tscsq{k}{} + m^2$ in propagator denominators to regulate any collinear divergences at intermediate steps. Then we may combine the two terms in \eref{opePM2} and set $m \to 0$. The second term in \eref{opePM2} is just the negative of the collinear pdf (see \eref{unint}), and with $\msbar$ renormalization it is
\begin{equation}
-\frac{a_\lambda(\mu)}{\pi} (2 \pi \mu)^{2 \epsilon} (1-\xi)
\int \diff{^{2 - 2 \epsilon} \T{k}{}}{}  \frac{ \Tscsq{k}{} 
}
	  {\big[\Tscsq{k}{} + m^2
	   \big]^2} - \overline{{\rm MS}} \;\; \text{C.T.} \,  \stackrel{\epsilon \to 0}{=} 
    -a_\lambda(\mu) (1 - \xi) \left[ -1 + \ln\parz{\frac{\mu^2}{m^2}} \right] \label{e.2ndterm} \, .
\end{equation}
We may calculate the first term in \eref{opePM2} in four dimensions since it is finite. It is
\begin{align}
\frac{a_\lambda(\mu)}{\pi} (1-\xi)
\int \diff{^2 \T{k}{}}{} e^{-i \T{k}{} \cdot \T{b}{} } \frac{ \Tscsq{k}{} 
}
	  {\big[\Tscsq{k}{} + m^2
	   \big]^2}  &{}= 
    2 a_\lambda(\mu) (1 - \xi) \left[ K_0\parz{\Tsc{b}{} m} - \frac{\Tsc{b}{} m}{2} K_1\parz{\Tsc{b}{} m}  \right]  \, \no
&{}=a_\lambda(\mu) (1 - \xi) \left[-\ln\parz{\frac{\Tscsq{b}{} m^2 e^{2 \gamma_E}}{4}} - 1\right] + \order{m^2 \Tscsq{b}{}} \, . \label{e.1stterm}
\end{align}
Adding \eref{1stterm} and \eref{2ndterm} and setting $m = 0$ gives for \eref{opePM2}
\begin{align}
-a_\lambda(\mu) (1-\xi) \ln\parz{\frac{\Tscsq{b}{} \mu^2 e^{2 \gamma_E}}{4}} &{}= \sum_j \int_{\xi}^1 \frac{\diff{z}{}}{z} \left[ -a_\lambda(\mu) (1-z) \ln\parz{\frac{\Tscsq{b}{} \mu^2 e^{2 \gamma_E}}{4}} \delta_{jp} \right] \delta(1 - \xi/z) \delta_{pp} \no
&{}= \sum_{j\in p} \int_{\xi}^1 \frac{\diff{z}{}}{z} \mathcal{\tilde{C}}^{(1)}_{q/j}(z,\T{b}{};\mu) \tilde{f}^{(0)}_{j/p}(\xi/z;\mu) \, .
\end{align}
So we read off
\begin{equation}
\mathcal{\tilde{C}}^{(1)}_{q/j}(z,\T{b}{};\mu) = -a_\lambda(\mu) (1-z) \ln\parz{\frac{\Tscsq{b}{} \mu^2 e^{2 \gamma_E}}{4}} \delta_{jp} \, , \label{e.C1final}
\end{equation}
which corresponds to \eref{OPE_coeff}.

The manipulations above work equally well if, instead of the $\Tscsq{k}{} \to \Tscsq{k}{} + m^2$ replacement in \eref{2ndterm}, we used dimensional regularization to handle the collinear divergences, as is more standard in QCD. But the use of $m^2$ makes the separate rolls of UV and collinear behavior very transparent. The full result for the OPE contains, in addition to the term in \eref{C1final}, another term for $j = s$, but we do not write it here since it does enter explicitly in the calculations in the main body of the text.

\end{appendix}

\bibliography{bibliography}

\end{document}